# SOEN6471

# Advanced System Architecture

## PM4 – Refactoring

| | | |
|---|---|---|
| *TEAM 1* | Akhilesh Masna<br>Computer Science and Software Engineering Dept.<br>**Concordia University**<br>**Montreal, Canada** | Anil Ganesh<br>Computer Science and Software Engineering Dept.<br>**Concordia University**<br>**Montreal, Canada** |
| Prakash Tirunampalli<br>Computer Science and Software Engineering Dept.<br>**Concordia University**<br>**Montreal, Canada** | Sai Ganesh Gaddam<br>Computer Science and Software Engineering Dept.<br>**Concordia University**<br>**Montreal, Canada** | Katam Raju<br>Computer Science and Software Engineering Dept.<br>**Concordia University**<br>**Montreal, Canada** |
| Avinash Mandapaka<br>Computer Science and Software Engineering Dept.<br>**Concordia University**<br>**Montreal, Canada** | Bharath Reddy Gujjula<br>Computer Science and Software Engineering Dept.<br>**Concordia University**<br>**Montreal, Canada** | Iustin-Daniel Iacob<br>Computer Science and Software Engineering Dept.<br>**Concordia University**<br>**Montreal, Canada** |

# I. ABSTRACT

In the current scenario, many organizations invest on open-source systems which are becoming popular and result in rapid growth, where in many of them have not met the quality standards which resulted in need for assessing quality. Initially we represent our work by analyzing the two open source case studies which are

(1) Distributed Modular Audio Recognition Framework (DMARF) is an open-source framework which consists of Natural Language Processing (NLP) implemented using Java which facilitates extensibility by adding new algorithms,

(2) General Intensional Programming System (GIPSY) is a platform designed to support intensional programming languages which are built using intensional logic and their imperative counter-parts for the intensional execution model.

During this background study we identified few metrics which are used to assess the quality characteristics of a software product defined by ISO standards. Among the metrics, we identified the number of the java classes and methods using SonarQube. Followed by that, the actors and stakeholders have been categorized and focused on the evolution of fully dressed use cases. Besides, we analyzed the requirements and compiled the conceptual UML domain model diagrams with the responsibilities and relationships based on the functionalities, which leads to the creation of the class diagrams. Later the analysis and interpretation of results has been done using the metric tools to verify results which have been implemented and to identify the code smells accordingly. Finally the implication is towards performing the system level refactoring by applying appropriate refactoring methods to enhance the quality and performance of the open source systems. Besides, the respective test cases have been portrayed to ensure that there is not much behavioral change with the existing architecture.

# II. INTRODUCTION

Software engineering is a systematic process for design, implementation and maintenance of the software systems not only from the product perspective but from the process perspective as well. The quality of the process (such as design and implementation) is reflected in the intrinsic product quality. The scope of the study is to express and depict important quality attributes of the system architecture, while targeting mainly the domain model of the system. First iteration of the study presents a set of background cases study from the architectural perspective of the conceptual system design, and tries to identify, based upon selected scientific studies, architectural patterns, and requirements and explain thoroughly the basic concepts of the architecture system design.

The study is based on the assumption that the background case studies and the scientific research included as reference of this study are analyzed from the software architectural perspective and certain metrics, patterns and concepts are depicted from the mentioned references.

In the second iteration of the document, we looked into the problem domain of the two background studies, with the scope of identifying the certain characteristics while remaining in the conceptual area.

The third iteration of the study, the focus is on the architectural design of the cases studies and the design pattern used across the system. Automated tools were used for construction of the actual UML class diagram of the system. Further, several patterns were identified in the source code and background and description of their usage is depicted in the design section.

In the final iteration, the system level refactoring is implemented to overcome the code smells which are identified in the modular conceptual components of the systems.

# III. BACKGROUND

## A. *Background research: OSS case studies*

This current research analysis depicts some of the architectural patterns and concepts drawn from the selected scientific research studies. Two software systems are described in the study:

- DMARF, open source Distributed Modular Audio Recognition Framework;

- GIPSY, open source General Intensional Programming System.

Both systems source code are available online [1] [2].

We introduce the conceptual model and the high-level requirements gradually as they are identified along the studies presented into the current study. Although some of the components might be described repetitively, they are analyzed from different perspective. The scope is to identify as many as possible and depict domain model in a clear fashion of each of the systems under analysis.

Further, after the basic analysis of the systems, the architectural model and main requirements identified are summarized along a dedicated section. Some of the enterprise architectural design patterns used have been depicted from the case studies.

Second iteration of the study, the problem domain of the background studies is analyzed and characteristics are depicted and further described in details. The purpose is to gather concepts and ideas from the conceptual representation of the studies and describe them herein. These will layout the basic foundation of the architectural representation of the two systems under study.

The third iteration of the study is targeting the actual system design of the cases studies and on the best practices in the architecture and implementation. Hence, in order to reduce the effort of depicting the actual UML Class diagram, several tools were evaluated and then selected to perform reverse engineering of the system's source code. The advantage is these type of tools dramatically reduce the reengineering effort and the results are consistently accurate. The tool used (and described in the future section), can identify the associations, relationships and many other perspective of the system under study. Yet, none is capable of identifying pattern design.

Next, the design of the systems was analyzed and code "bad smells" were identified and described, then later a few refactoring suggestions were shown. Many of these "bad smells" software code were identified by using again automated tools such as McCabe, JDeodorant, Logiscope. These tools perform a systematic process analyzing the source code, running complex measurements and being capable of identifying the "problems" using an engineering approach.

<u>Case study: DMARF</u>

*Managing Distributed MARF with SNMP* [3]

The Modular Audio Recognition Framework (MARF) is an open-source research platform and collection of modules for natural language processing (NLP) algorithms written in Java and packaged as an extensible framework, which would facilitate addition of new algorithms. The Distributed MARF (DMARF) can run distributed over the network in a distributed architecture (whereas the intercommunication uses CORBA, XML-RPC and Java RMI). The system architecture has started shifting from one centralized system (such as IBM mainframe, prior to 90's) towards a large geographically dispersed architecture, with many benefits and improvements. Hence, such benefits introduced further more complexity and new requirements in terms of stability and reliability of the systems, so the need for better management, configuration, monitoring and testing played a crucial role in the viability of the system.

In essence, DMARF system required an enterprise distributed framework to be able to survive the new distributed architecture, yet a standard monitoring and management technology needed to be implemented. The most known and supported protocol, by both commercial and open-source multiplatform protocol is Simple Network Management Protocol (SNMP). The application nodes deploy SNMP agents across the entire distributed network and these agents collect, transmit, service and instrument tasks, to a centralized server manager. The protocol is supported by almost every application and hardware vendor.

DMARF system implementation has agent's proxy and manager that communicate over the SNMP protocol. The individual nodes can be added and removed, along with their agent proxies. The manager receives information from the agents, collects and displays the data in a meaningful format (and eventually sends commands to the agents).

In a real critical systems implementation, this is the architected design of the successful systems. Even more, usually the manager is duplicated such that every single point of failure is eliminated from the network of nodes. Implementations support that nodes enter and exit network in a dynamic fashion, nodes might fail, and managers might fail and so on. Yet the business continuity, for such critical systems is 100% availability. On the other hand, such complexity introduces additional cost and effort. Similarly, DMARF system aims at enterprise grade architecture system, where the nodes deploy agent proxies and transmit data to the manager, which in turn stores and displays the data in a meaningful format. Although some difficulties were encountered due to the underlying framework used (AdventNet and SimpleWeb), the proof of concept has proven that DMARF is capable of an enterprise class architectural design.

*Self-optimization property in autonomic specification of Distributed MARF with ASSL* [6]

The classic Modular Audio Recognition Framework (MARF) is an open source research platform and a collection of various algorithm implementations for pattern recognition, signal processing, natural language processing etc. written in java [6]. MARF can run distributive or standalone systems. Sometimes it may just act as a library in applications.

This paper in the reference [6] explains about the self-optimization features in the DMARF. To specify the self-optimization policy, Autonomic System Specification language (ASSL) is used.

The classical MARF was expanded to permit the phases of the pipeline to run as distributed nodes illustrated in the below figure.

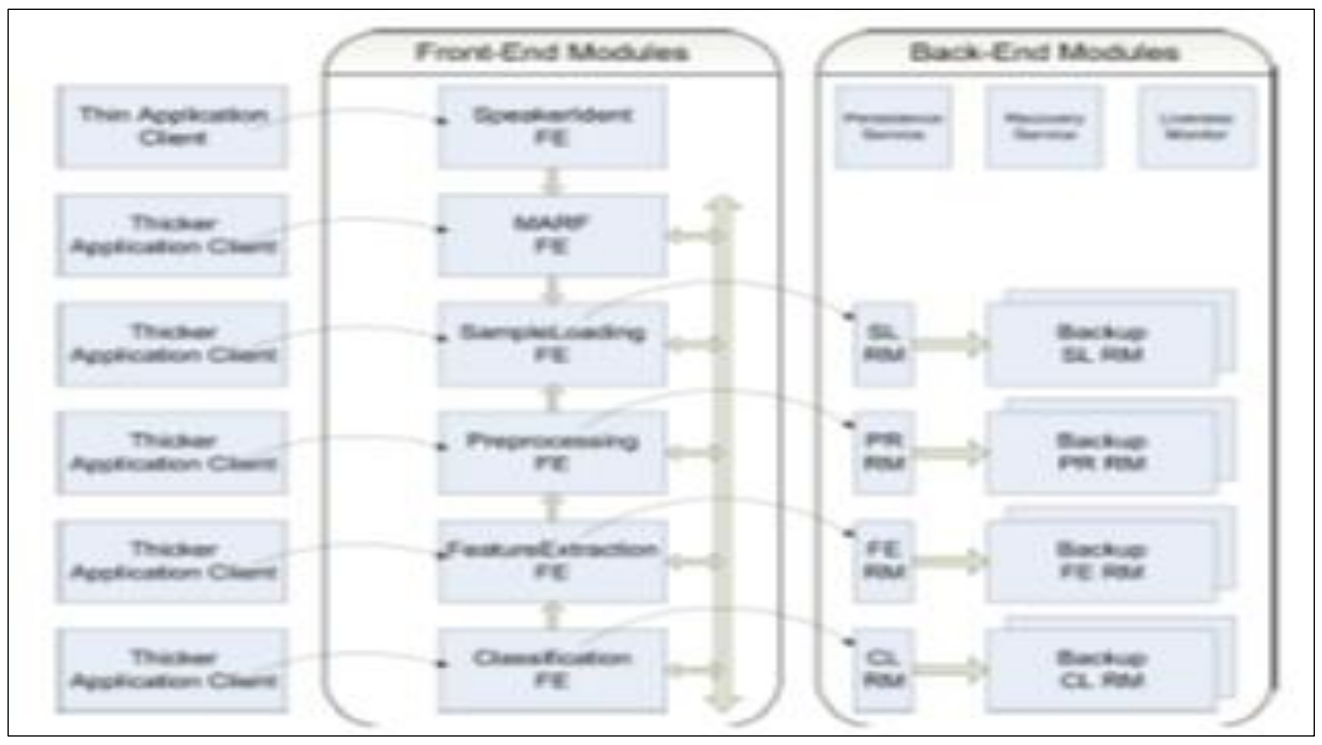

Figure 1: The Distributed MARF pipeline [6].

In order to understand the DMARF self-optimization mechanism, initially the requirements should be comprehended.

There are two functional requirements for DMARF installations related to self-optimization:

1. *Training set classification data replication*. The DMARF based application does extensive data processing through a pipeline which comes under sample loading and classification stage where they do a lot of operations and potentially heavy computations. The stand-alone MARF employs a dynamic programming to store the immediate results in the same stage. So, a lot of data is absorbed by the classification stage. This may results in storage of data in different locations causing re-computations of already computed data. So it is better to communicate with the members of that stage. Such replication would optimize a lot of computational effort on the end users.

2. *Dynamic communication protocol selection*. This protocol is used to automatically select most efficient communication protocol in the current environment.

After analyzing the DMARF architecture, an understanding of ASSL is also needed. ASSL considers autonomic systems composed of autonomic elements communicating over interaction protocols [6]. To know briefly, they are specified in different tiers namely AS tier, AS interaction protocol and AE tier.

The ASSL Self optimization for DMARF, it is made autonomic by adding an autonomic manager layer to the architecture. This layer implies autonomic behavior over the entire system by implementing the self-management policies. As the DMARF classification is improved with self-optimization policy, ASSL is used to specify the policy and generate the implementation. So it is known that, while constructing a self-optimization specification model for ADMARF, an algorithm with ASSL is devised for the classification stage of the DMARF's pattern recognition pipeline. This algorithm when it is fully implemented, ADMARF system will be fully functional to autonomic environments.

*Towards security hardening of scientific distributed demand-driven and pipelined computing systems* [9]

The security aspects with respect to the case studies GIPSY and DMARF are data/demand integrity, data/demand origin authentication, confidentiality, availability and malicious code detection. To overcome the above security aspects in the both case studies the Java data security Framework (JDSF), proxy certificates can be implemented as the security layer.

In GIPSY the components are related to development, compilation and execution. GIPSY compilation is collection of GIPC framework and eduction execution engine (GEE). These two components undergo major amendments to support the security aspects. As the GIPSY is more flexible for the use of external languages and resources, the security has become a major challenge as it is vulnerable to unsigned code and also grants privileges to the outsiders.

The classical MARF is a collection of pattern recognition, signal processing and natural language processing (NLP). The main component in the MARF is pipeline which helps to communicate with each other to get the data. This pipeline contains four basic stages: sample loading, preprocessing, feature extraction and training/classification. This MARF is extended to support the front-end by allowing the pipeline in all the stages. And later it's further extended with SNMP by implementing the proxy SNMPv2. The security relies on the underlying protocols of DMARF like Java RMI, COBRA, XML-RPC and SNMP.

The differences between the GIPSY and DMARF are GIPSY adopts the demand-driven eductive execution model (GEE) and DMARF is pipelined. Both are written in Java languages. GIPSY deals with the programming languages and so it contains malicious code, but DMARF does not deal with code execution, it only processes the data.

JDSF is a proposed framework for the security in the databases in Java. It mainly deals with the data security aspects like confidentiality, integrity, authentication and availability. The detection of malicious-code, high-availability is sub-topics among them.

The confidentiality in both the systems is more about correctness, accuracy, computation, results and their availability. It is more relevant to the application. The JDSF's confidentiality framework integration into the security later will invoke the JDSF calls whenever the data is entering or leaving the system. It can be source based or proxy based.

The integrity of the data can be achieved by Java virtual machine and also by the JDSF's integrity sub-framework. The proxy certificates are better used in the integrity aspect.

The authentication is related to the integrity and the correctness of the data coming from the untrusted source. This also deals with the code singing in the procedural demands and make sure that it is coming from the trusted source. The JDSF authentication sub-frame work can be used similar to the confidentiality and Integrity and can also achieved by DNSsec methodology and can be hierarchical.

Availability is generally very hard to achieve in the distributed systems and JDSF has no requirements with this availability the detection of malicious code will require more research and will be achieved in the future.

*Towards Autonomic Specification of Distributed MARF with ASSL: Self Healing* [10]

The self-healing property in the Distributed Modular Audio Recognition Frame (DMARF) works with assistance of Autonomic System Specification Language. The ASSL enhances the DMARF using autonomic middleware layer which is added to DMARF which organizes the four stages of DMARF pattern recognition pipeline, where these autonomic elements are managed by distinct autonomic manager.

The aim of autonomic computing (AC) is to apply the principles of self-regulation and complexity hiding and also used for transforming complex systems into self-managing autonomic systems, so that the software system can manage and deal with requirements automatically. This feature enables by using the AC Tools which helps in problem formation, system design, system analysis and system implementations.

The ASSL is used to integrate autonomic features into DMARF, which cannot be used in autonomous systems due to lack of provision among the applications. Hence, ASSL uses three core autonomic requirements:

- Self-healing
- Self optimization
- Self protection

DMARF is the extended version of classical MARF where the pipeline stages are made into distributed nodes.
Common applications of DMARF systems are:

- High volume processing of recorded audio, textual etc.

DMARF Self-healing Requirements:

- In DMARF, the system should be able to recover by itself in the form of replication for keeping at least one route of pipeline available. Here replication is providing two different pre-processing services.

ASSL:

ASSL approaches the issue of formal specification and code generation of ASs with in framework. ASSL considers autonomic systems that are composed of autonomic elements (AE), which interacts over interaction protocols.

To specify these, ASSL uses three major tiers.

- AS tier: It forms general and global AS perspective.
- AS Interaction Protocol (ASIP) tier: It forms a communication protocol perspective.
- AE tier: It forms a unit level perspective.

Sub Tiers in ASSL Specification Model:

- AS/AE Self-Management Policies: ASSL model specifies four AC self-management policies: self-configuring, self-healing, self-optimizing, self-protecting.
- Autonomic element interaction protocol is a private communication protocol used by ASSL, which specifies same constructs as ASIP does.
- Managed Resource interface elements are controlled by AE.
- AS/AE Metrics are said to be set of parameters and observables, which are controlled by AE.
- AE's Architecture is specified as a correlation among AE's and group of AE's.

Self-Healing Model for DMARF:

- In Autonomic DMARF (ADMARF), the atomicity is to be added to DMARF behavior. ADMARF addresses the self-healing model where the ASSL specifies the self-healing behavior of ADMARF by addressing node replacement and node recovery.

AS Tier Specification:

· AS tier specifies the global ADMARF self-healing behavior. Other than AS tier, ASSL SELF_HEALING self-management policy structure is used.

AE Tier Specification:

· It specifies the self-healing policy for each AE in the ADMARF where DMARF stages are specified with special autonomic manager.

*Autonomic specification of self-protection for DMARF with ASSL* [16]

This paper mainly focuses on an autonomic computing layer covering DMARF by specific autonomic properties at each of pattern recognition stages of the same. Today both software and hardware facing challenges about complexity and this will be biggest threat to the continuous progress in the IT industry. Autonomic computing (AC) focus on the reduction of workload needed to maintain complex system by transforming them to the self-managing autonomic system. By applying the AC principles it can reduce the complexity problems like security.

The AC applies the principles of self-regulation and complexity hiding. The AC tool is ASSL framework which helps the researchers with problem specification, system design and implementation, system analysis and evaluation.

Distributed MARF is based on MARF whose pipeline stages were made into distributed nodes. MARF is an open source research platform and it's a collection of algorithms written in java and arranged into a modular. In the below figure implemented algorithms are grouped in white boxes and progress algorithms are grouped in grey boxes.

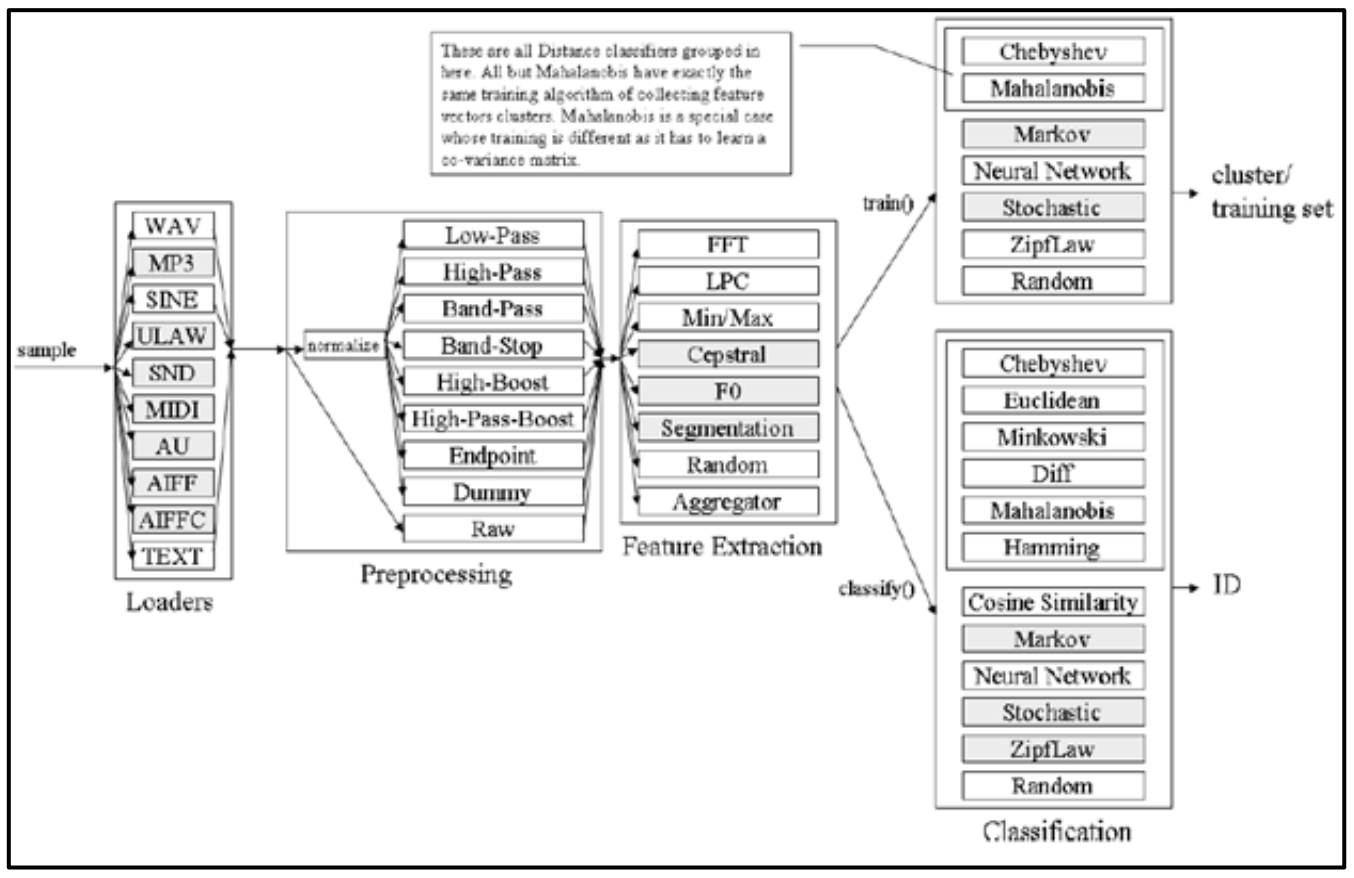

Figure 2: MARF's pattern recognition pipeline

DMARF allows the stages of the pipeline to run as distributed nodes as well as front-end is shown in the below figure. The basic stages and front-end were implemented without backup recovery.

.

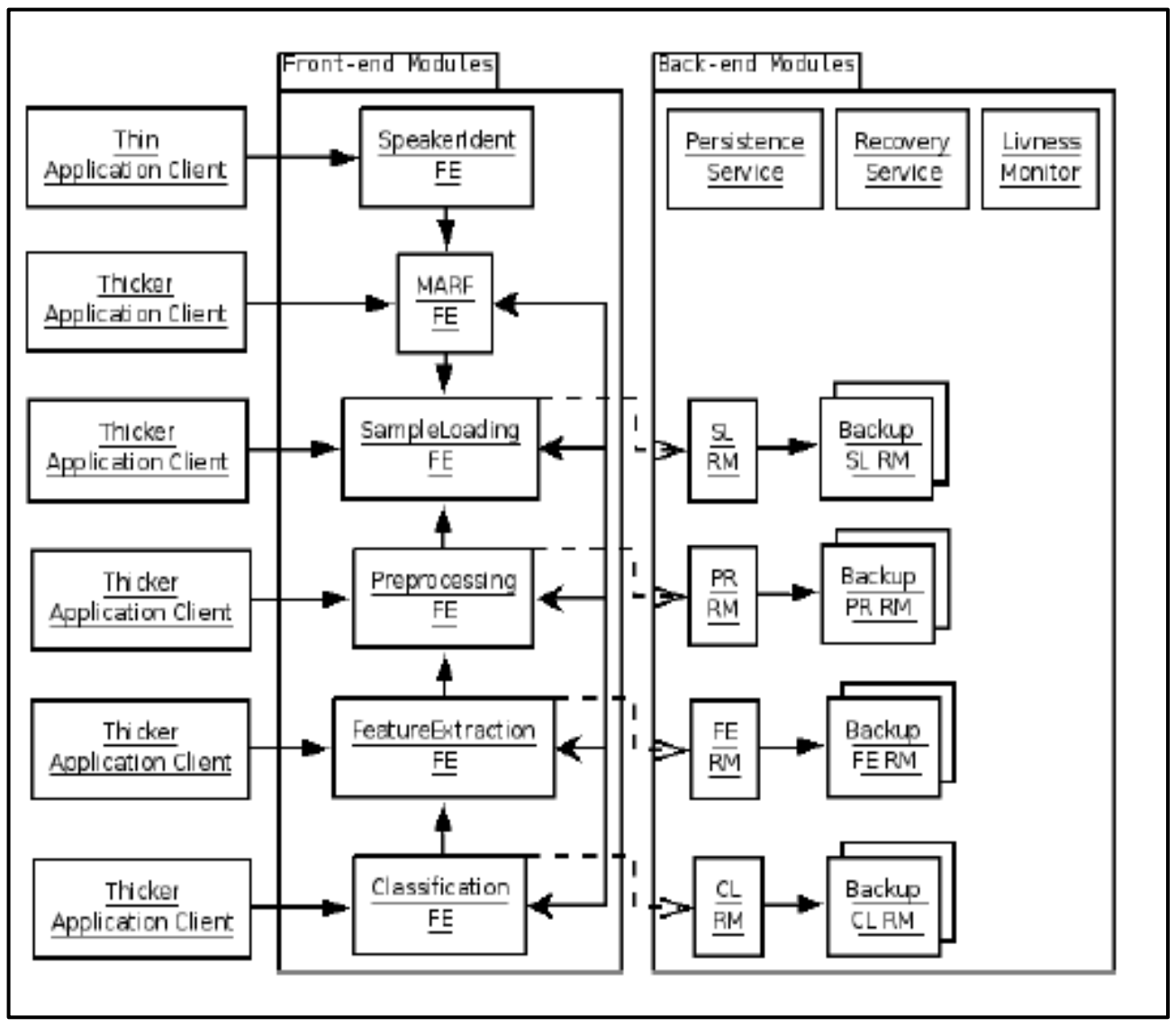

Figure 3: Distributed MARF pipeline

DMARF self-protection requirement captures as an autonomic system primarily covers the autonomic functioning of distributed pattern recognition pipeline But self-protection of DMARF based system as is less important in localized scientific environment and is more important in global environment. DMARF based system should stick to the specification where each node proves its identity to the other nodes participating in the pipelines. This will insure the data authentication and also protect against from spoofing of data. Autonomic system specification language (ASSL) approaches the problem of formal specification and code generation of autonomic systems (ASs) within a framework. In genera ASSL considers autonomic systems (AS) as composed of autonomic elements (AE) communication over interaction protocols and in order to specify those ASSL defines through formulation of tiers.

ASSL self-protection model for DMARF: DMARF must incorporate special autonomic computing behavior. Self-protecting is one of the self-management properties that must be addressed by ADMARF. If the message is private or public it should receive in AS followed by AM of stage AE level for private message and AM of AS level for public message.

In IP Tier specification communication activities, channels, entities must be in order to allow both internal and external entities to communicate. At ASIP tier a single public message called (publicMessage), a single sequential bi-directional public communication channel called (publicLink), public communication functions (recievePublicMessages and sendPublicMessages) and if operational value returns false in case of public message received is insecure (thereIsInsecurePublicMessage).

```
//receive public messages if the message is secure
FUNCTION receivePublicMessages {
  DOES {
    IF ( AS.METRICS.thereIsInsecurePublicMessage ) THEN
      MESSAGES.publicMessage << CHANNELS.publicLink
    END
  }
}
```

Figure 4: ASSL specification of receive public message

AS Tier Specifications: To protect AS from the insecure public message specifying a SELF-PROTECTING policy. In specification model, the publicMessageSecure event will be activated when a publicMessage is about to be received by the AS.

```
SELF_PROTECTING {
  // a new incoming message has been detected
  FLUENT inSecurityCheck {
    INITIATED_BY { EVENTS.publicMessageIsComing }
    TERMINATED_BY { EVENTS.publicMessageSecure,
                    EVENTS.publicMessageInsecure }
  }
  MAPPING {
    CONDITIONS { inSecurityCheck }
    DO_ACTIONS { ACTIONS.checkPublicMessage }
  }
}
```

Figure 5: AS Tier self-protecting policy

AE Tier specification is a tier, which specifies that SELF-PROTECTING mechanism for private messages. Similar to the AS tier the AE level SELF-PROTECTING is specified with a single insecurity check mapped to the checkPrivateMessage action. Then initiated by the privateMessageIsComming event and terminated by the privateMessageIsSecure event or by the privateMessageSecure event.

*Conclusion*: Constructed a self-protection specification model for ADMARF. Although this is not a complete autonomic specification model for ADMARF, but shown how ASSL helps us to achieve desired atomicity in DMARF through the self-protecting property specification.

*Distributed Modular Audio Recognition Framework (DMARF) and its Applications over Web Services* [13]

The paper focuses on the architectural design and implementation of DMARF, its applications, advantages and disadvantages. DMARF is a distributed version of Modular Audio Recognition Framework (MARF) implemented in Java. It has RMI and CORBA implementation done in a modular way to achieve the interoperability but now facing a problem of flexibility because of the need to use CORBA-enabled clients outside of MARF's modules. The proposed solution is to use the web services to cover this gap in DMARF.

The classic MARF written in Java consists of pipeline stages to communicate with each other to get the data in a chained manner [13]. Whereas the distributed version of MARF will allow the stages of the pipeline to run as distributed nodes [13]. As MARF allows to process high volume of audio, textual or imagery data using a desktop, with the DMARF this feature turned out as distributed. Introduction of Web Services makes it even more widely available over internet [13]. Based on the applications of the MARF, it is noted that the nodes are communicated in a pipeline manner. The current work is to make this into distributed manner so that parallel usage of services can be possible on distributed computers. In the DMARF, if any process in the pipeline crashes then the information at that stage and the pending transactions will be lost too. Hence to avoid this situation, DMARF has been extended by adding "warm standby". It is a module of DMARF, runs on different machine simply acts as a backup server. Whenever the primary server fails/ crashes then this machine will come into action and take over the system enabling uninterrupted services. The architecture of DMARF is designed such that it can achieve the desired functionalities without fail. The main principles are [13] platform independence, database independent API, simplicity and maintainability.

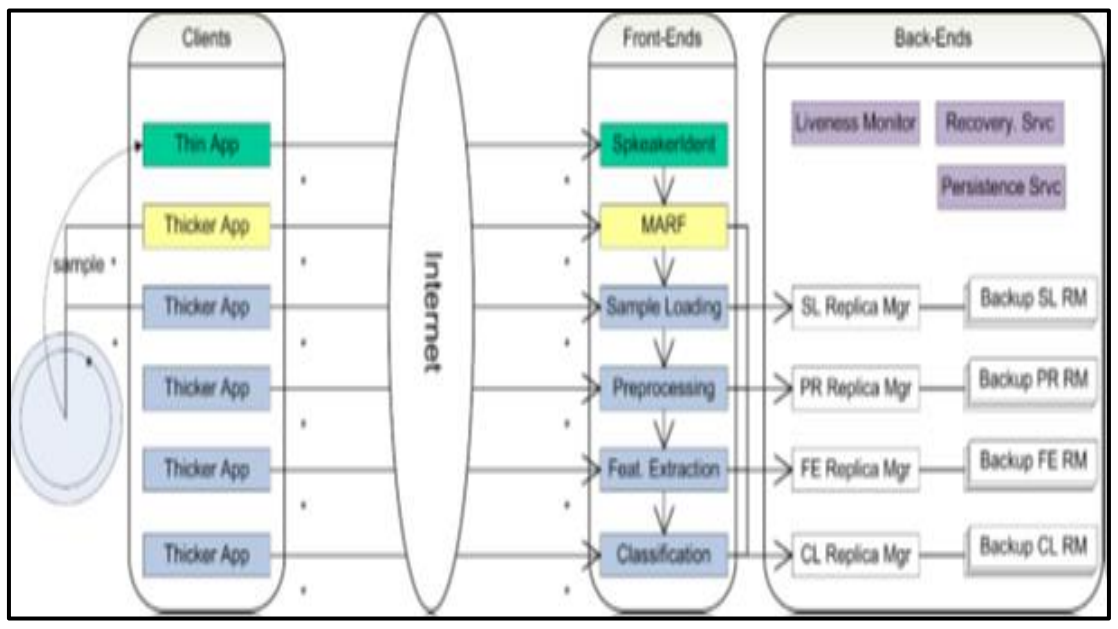

Figure 6: The Distributed MARF Pipeline [13]

Module view shows how the DMARF modules are divided into layers as shown in fig1, front-end and back-end. The application specific front end and pipeline service comprises the front-end of the system. All the stages are associated with the database and other storage sub functions. As per the execution of the system, for the implementation of the WS, DMARF uses Apache Tomcat [14] as servlet container. Along with the WS, to support message passing between methods, a Java XML remote procedure call (JAX-RPC) is used over Simple Object Access Protocol (SOAP) WS.

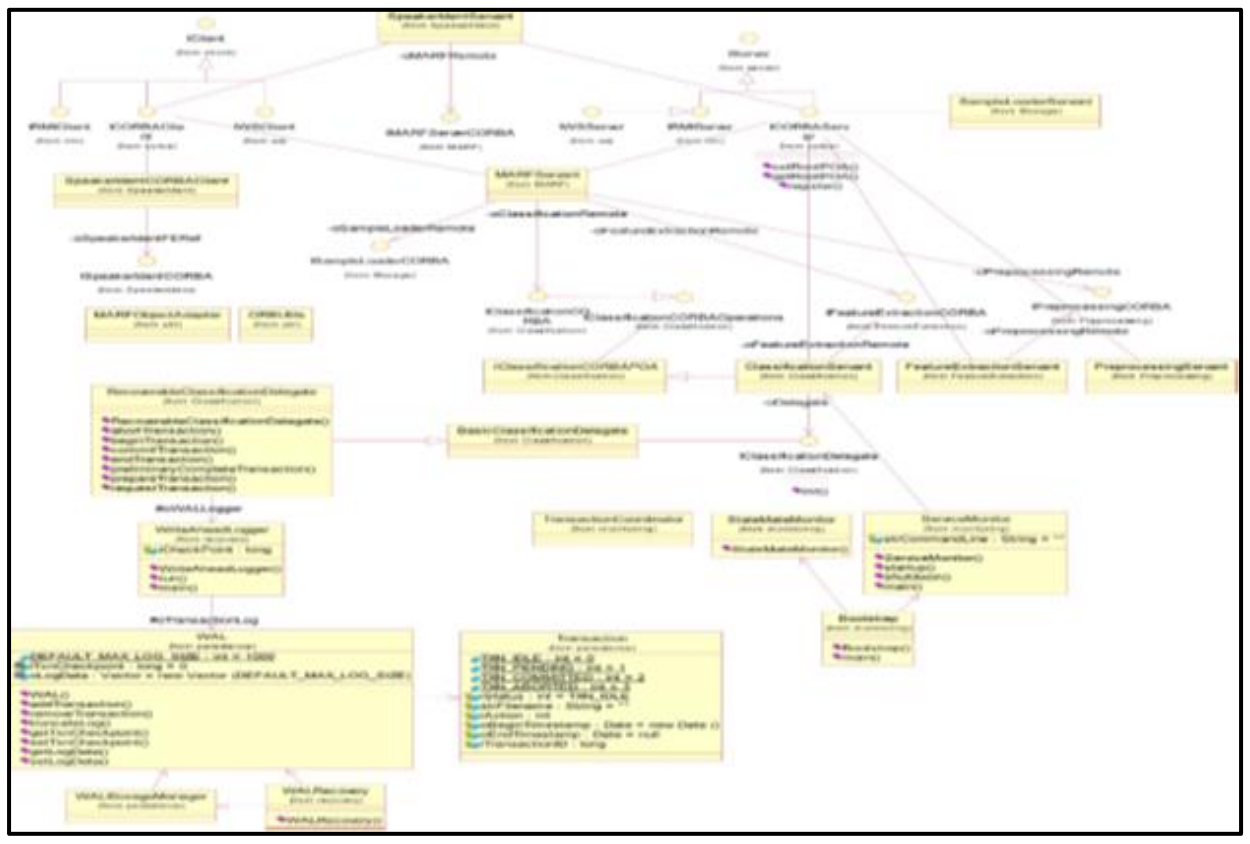

Figure 7: General Architecture Class Diagram of marf.net [13]

As shown in above figure, the architecture of the DMARF is shown in the form of UML diagrams representing the main modules and the relationships among them. To implement WS for greater portability of MARF, mostly USED Java WS with JAX-RPC, Java Servlets, Java Networking, Apache Tomcat, and J2SE are the technologies that are mostly used. However there is work undergoing in order to include other communication technologies (JINI, JMS, DCOM+, .NET Remoting). [13] The main goal is to achieve disaster recovery, fault tolerance, high availability and replication.

*On Design and Implementation of Distributed Modular Audio Recognition Framework [16]*

MARF is designed for pattern recognition processing. It is a pipeline stage procedure where accidents access about pending transactions and processing in module, not just lost if methodology is inaccessible. Some applications are purely sequential with no concurrency when processed in a bulk of voice samples. In fig below MARF's pipeline's goal is to distribute the stages of pipeline as service and stages that are not present i.e. sample loading, front-end application service. In fig below Design of distributed version is presented which indicates different levels of basic front-ends from higher to lower. Some features are implemented, but not all modules work. Besides, some of six services work well in CORBA, RMI and WS modes. Some of design considerations for WS there is no remote object reference, for which a class is created as RemoteObjectReference which is encapsulated with type and URL passed to modules, so it can be used to connect. All communication modules rely on delegates for business and transactions. Basic delegate redirect the business logic and provide basis for transaction, which is not accurate and recoverable is extension of the basic functionality with the transactional on the top of basic operation.

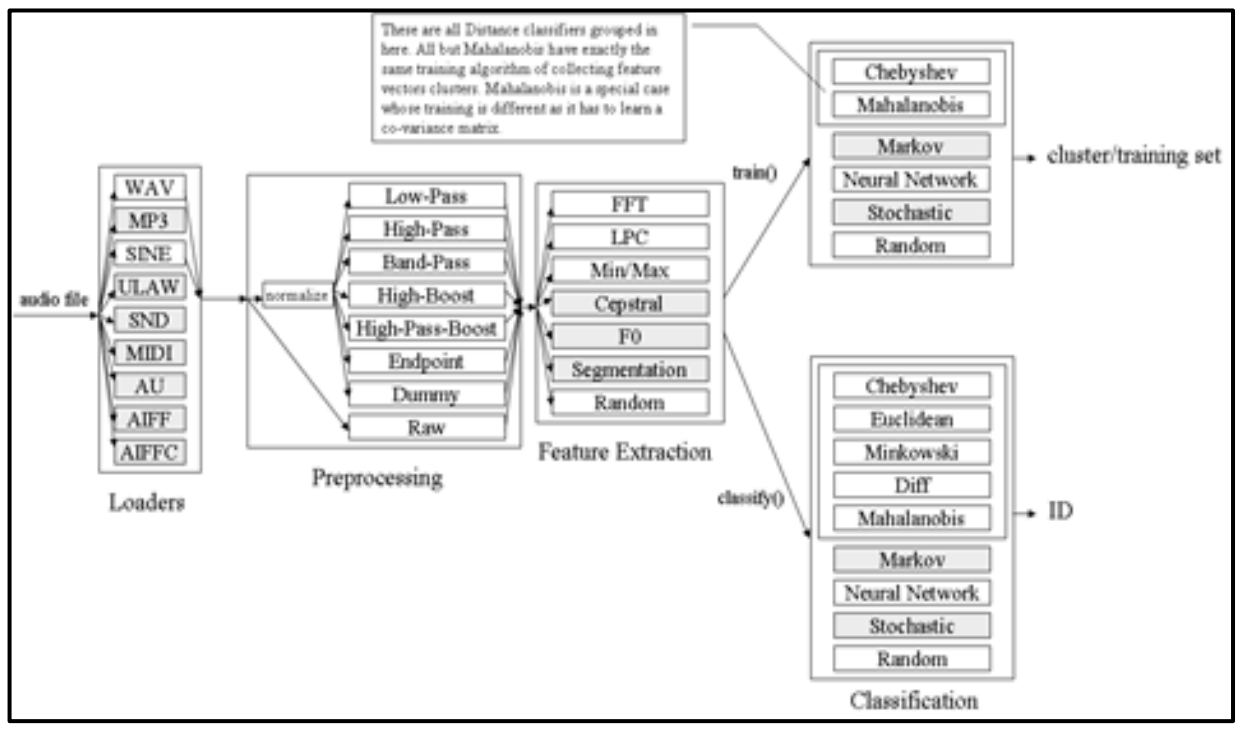

Figure 8: The Core MARF Pipeline Data Flow

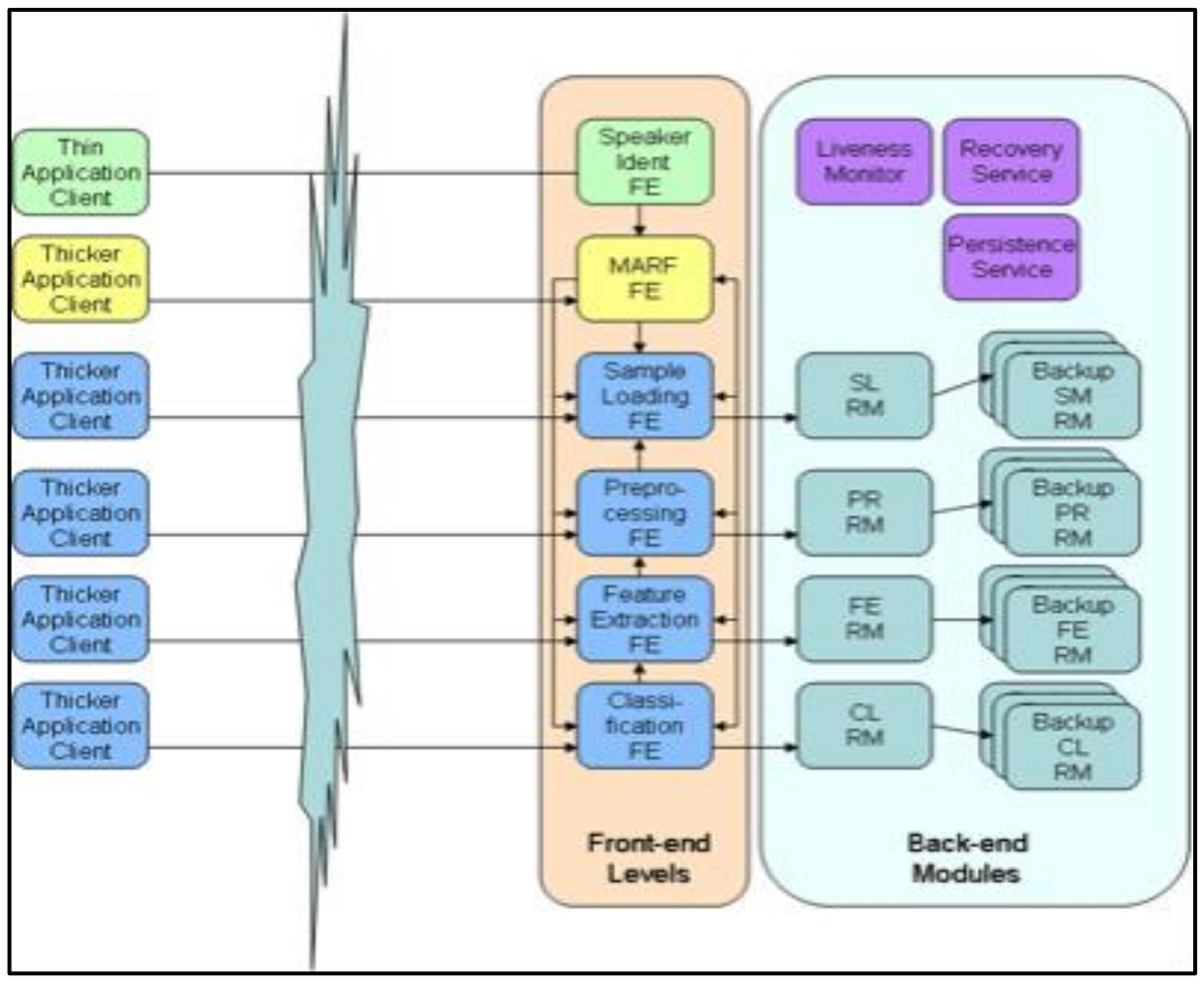

Figure 9: The Distributed MARF Pipeline

WAL consists of entries called Transactions, where transactions are a data structures holding unique ID. WAL is a message-logging protocol designed for DMARF. It is recovery process for unfinished transactions and prevention for data loss. It maintains backups and Logs. WAL max size is 1000 entries, where such entries allow future feature called PITR. In configuration and deployment CORBA, RMI, and WS use dmarf-hosts.properties at beginning to find the service to register and all the scripts of GNU and Ant makefiles. Makefile target for single wave file and shell test for CORBA pipeline contains 295 samples and 31 testing wave samples, which multiplies to 4 configs and 16 test configs.

DMARF is implemented in two views, where the Module view application is divided into two phases, front-end which exists on client side and back-end on server side, which is all about layering. Execution view is all about JVM, which it is about runtime Entities. Class diagrams are used to summarize the major modules and relationships. Log File Format is the file produced with the help of Logger class, which is the classical format "[timestamp]: message" it intercepts to write STDOUT or STDERR for every file and STDOUT and STDERR is also preserved so if the file stuck in runs, the log data is applied.

The issues and limitations after running all the six services on the same machine runs out of files which reaches fault kernel limits. MARF's deals with flaws for rigid and less concurrent Transaction ID "wrap-around" for long running systems and transactions with message passing operations. MARF writes about dumps long run servers, which have potential to have their ID's recycle after overflow and no particular calculation of time.

This confirmation usage of Distributed MARF has demonstrated a probability for the pipeline organizes and not just to be executed in a pipeline and stand-alone modes on a few machines. This could be valuable in add any of the specified administrations to customers that have low computational force or no obliged environment to run the entire pipeline by regional standards or which cannot manage the cost of long-running process. Moreover, they ran across configuration defects in the traditional MARF itself that must be adjusted, related to the storage and parameter passing among modules.

*Towards a self-forensics property in the ASSL toolset* [20]

Forensic lucid was initially proposed for specification and autonomic deduction and event reconstruction in the cybercrime domain of digital forensics. It has been used for other domains such as investigation of incidents in various crash investigations. Forensic lucid is primary experimental platform for compilation and evaluation is GIPSY.

The concept of self-forensics and the idea of its implementation within ASSL and GIPSY are described through their founding core work. The background work of ASSL formal specification ToolSet takes as an input a specification of property of autonomic systems and does formal syntax and semantics checks of the specifications and if the checks pass, it generates a java collection of classes and interfaces corresponding to the specification.

The ASSL framework includes the autonomic multi-tier systems architecture (AS) including formal language constructs to specify service-level objectives (SLOs). Core-self CHOP autonomic properties corresponding architecture, allow actions, events and metrics to aid the self-management aspect of the systems, also specifies the interaction protocols between the AS managed autonomic elements, ASSL formal modeling specification and model checking has been applied to a number open source, academic and research software systems specifications, for E.g., Distributed audio recognition framework (DMARF) and the general intentional programming system (GIPSY).

The forensic autonomic property (SFAP) consists of adding the syntax and semantic checker of port to lexical analyzer, parser and semantic checker of ASSL as well as adding the appropriate code generator for JOOIP and forensic lucid to translate forensic. The future work will be to complete the implementation of the lucid property and export it into the target, like ADMARF, AGIPSY.

<u>Case study: GIPSY</u>

*Towards a multi-tier runtime system for GIPSY* [4]

General Intensional Programming System (aka GIPSY) represents a multi-tier runtime system that further unifies the distributed technologies, which are used for the implementation of the Demand Migration Framework (DMF), used to streamline distributed execution of the hybrid intensional-imperative programs.

Intensional programming implies declarative programming language, which is based on the denotational semantics. GIPSY system includes a compiler (GIPC) based on the notion of the Generic Intensional Programming Language (GIPL) and stands as the runtime of other flavors of the Lucid languages.

As a result of the Lucid's prescriptive semantics, the parallelism of the Lucid programs is intrinsic, and the Lucid application can be understood as producer-consumer networks computing in parallel[1]. Communication between distinct components of the architecture was initially designed using Java RMI that added constraints in the deployment architecture. Furthermore, two branches have been created, one based on JINI[2] and second based on JMS[3]

[1] E. A. Ashcroft and W. W. Wadge. R for semantics. ACM Transactions on Programming Languages and Systems, 4(2):283–294, Apr. 1982
[2] JINI: Network architecture for construction of distributed systems

of the DMF. Yet, these components are not interoperable, which increased the effort even higher.

After in-depth analysis, a new design was applied [5] by constructing wrapper classes for each tier type introduced. Such design, each GIPSY worker node (which usually translates to single physical box), would be registered within the network participating in computation. Finally, the distributed computation prototypes were implemented by using a design strategy, the abstract factory, which will construct the multi-threaded, RMI, JINI or JMS nodes.

Although JINI is not commonly used in the distributed systems, JMS is the industry standard for messaging in distributed systems using Java, with proven stability and reliability (many open source and commercial implementations: IBM MQ, Tibco EMS, ApacheMQ, Oracle AQ, and many others). Lately, the JMS shifted towards implementing in addition API to support non-Java clients, nowadays known as Advanced Queuing Messaging Protocol (AMQP).

In a GIPSY peer-to-peer computing network of nodes and tiers, the design was improved and shifted towards the new distributed systems paradigm (which is similar to the Service Oriented Architecture - SOA): demands are propagated without knowing where the actual computation takes place, nodes might fail without fatally affecting the system, worker nodes and tiers can be dynamically added or removed, and finally the nodes could be used for sharing the computation power by computing different demands for different programs.

The details of the GIPSY architecture nodes and tiers, along with the wrapper and API classes are depicted in details in the design architecture [4]. Further work is required in order to improve the performance of the system and the implementation of a security layer.

There is a semnificative overhead for implementing such architecture, yet, the benefits are known and proven in real systems implementation that makes it feasible for long-lived system implementations.

*Towards Autonomic GIPSY [7]*

Autonomic GIPSY's goal is to make general intensional programming system capable of self-managing to a far greater extent than what it is now [7]. The architecture of autonomic general intensional programming system realizes about various aspects like autonomic computing including goal-driven self-protection, self-healing, self-optimization and self-configuration.

Autonomic computing concentrates mainly on making complex computing systems like general intensional programming system smarter and easier. This complexity is reduced through automation. GIPSY is a complex system as it has a multi-tier architecture. Though its architecture provides high scalability, it cannot be achieved as it does not have the required self-management capacity. It is executed in four different tiers called Demand store tier, Demand worker tier, Demand generator tier and GIPSY manager tier. Its architecture is shown below [7].

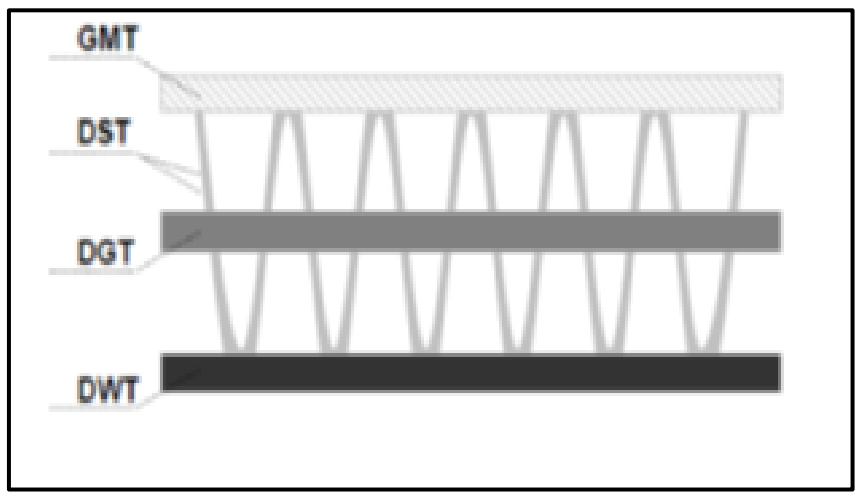

Figure 10: GIPSY multi-tier architecture [6]

Further, AGIPSY architecture uses autonomic computing. Autonomic computing tells that self-management is comprised of four characteristics like self-configuring, self-optimizing, self-healing, self-protecting. These characteristics are used to achieve high level objectives, improve performance, achieve fault tolerance and provide security. While the GIPSY architecture is tier-oriented, the AGIPSY architecture is multi-agent loosely coupled distributed system [7] which makes its architecture look like a grid which is composed of a hierarchical structure. A GIPSY node in order to become an AGIPSY node, it should operate itself without the intervention of external entities. The architecture of AGIPSY is shown below [7].

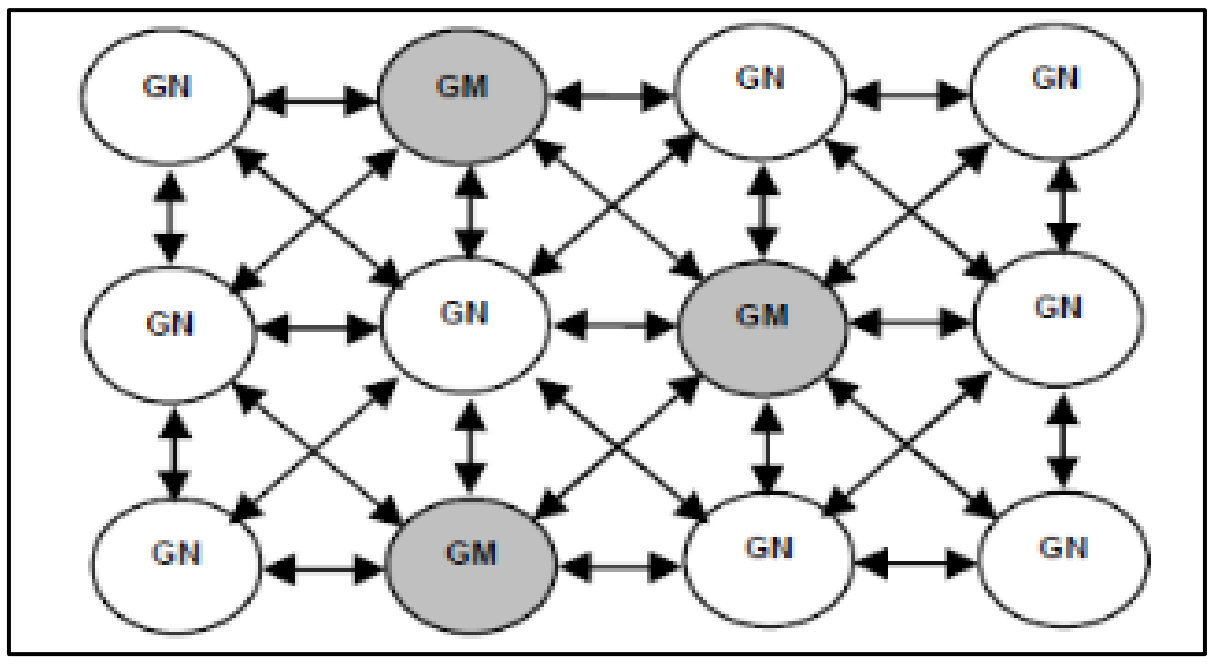

Figure 11: AGIPSY architecture [7]

[3] JMS: Java Messaging System, de-facto messaging system implementation of Java language.

*GIPSY AE*:

The GIPSY AE architecture is based on the architecture of ASSL. ASSL is used to specify the GIPSY AE's. The GIPSY AE architecture is composed of two distinct parts called GIPSY tier controllers and a Node manager (NM). These communicate via an interface to form a loop that directives the workflow of different subcomponents in the node manager. The GIPSY AE architecture is shown below.

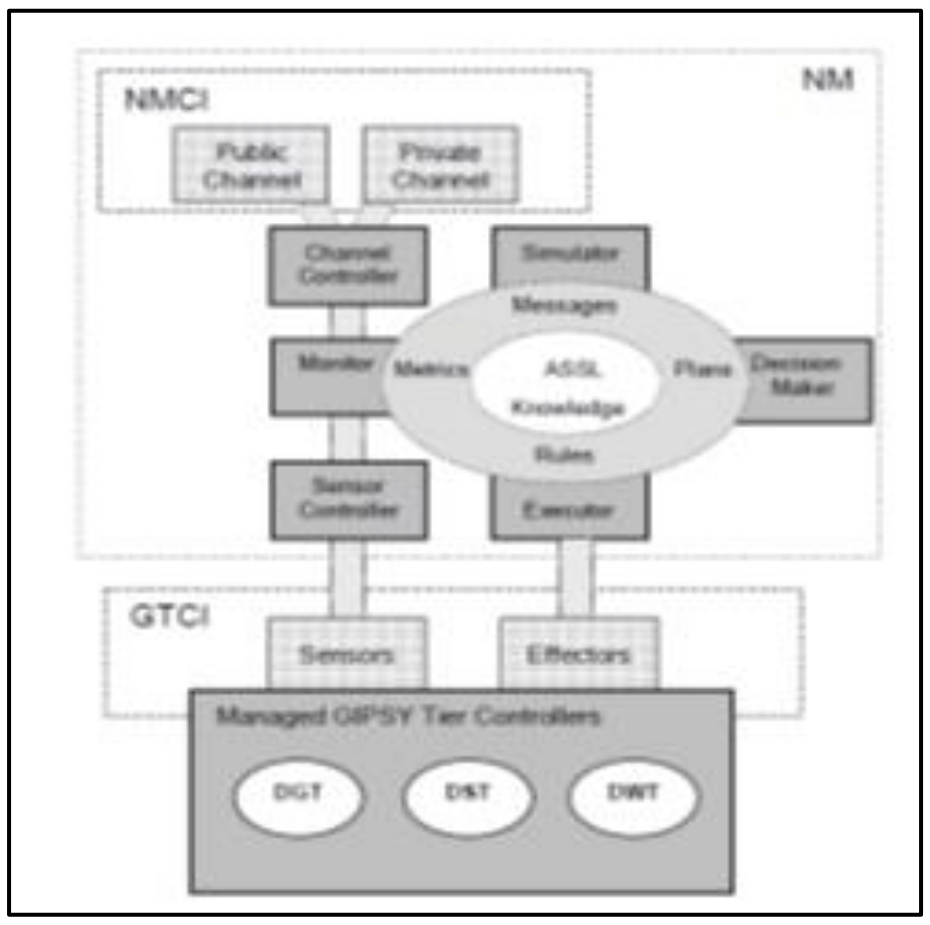

Figure 12: A GIPSY AE architecture [7]

This architecture provides channels to communicate among the node managers and channel controllers to sending and receiving messages. It also contains sensors and sensor controller to tune the adjustment of the sensor's parameters to the appropriate ASSL metric characteristics [7]. In the same way, there are monitors for maintaining the record of all the managed GIPSY tiers with their associated DGs, DWs and DMs.

In the same way, simulator helps with new solutions for unexpected problems, decision maker analyzes the solution, executor uses the effectors to control the execution elaborated by the decision maker and simulator, ASSL knowledge has the complete specification of the GIPSY AE, performance tradeoff controls the performance overhead, scalability-complexity is used to explore the scalability tradeoff.

To conclude, even though GIPSY scales well, its complexity grows rapidly when more GIPSY nodes are needed. So, Autonomic computing solution is needed to make GIPSY Capable of self-management.

*General architecture for demand migration in the GIPSY demand-driven execution engine* [8]

This gives the overview of the GIPSY Demand Migration System (DMS). The GIPSY is a Multi-language programming and Demand-Driven Execution environment. It is implemented to support the intentional programming and family of all programming languages. The DMS connects the GIPSY execution nodes using the generic architecture by using different middleware technologies. The Generators and workers will together form a DMS. The requirements which fulfill the DMS are: it has to run on multiple platforms, it has to be secure in-order to authenticate the identity of the DMS objects, it should be perform asynchronous communication, it should be independent: the DMS must work on different technologies and platform, it should be upgradeable: The DMS should not bound to one technology and should allow the use of the other technologies, in simple it has to be designed as a framework.

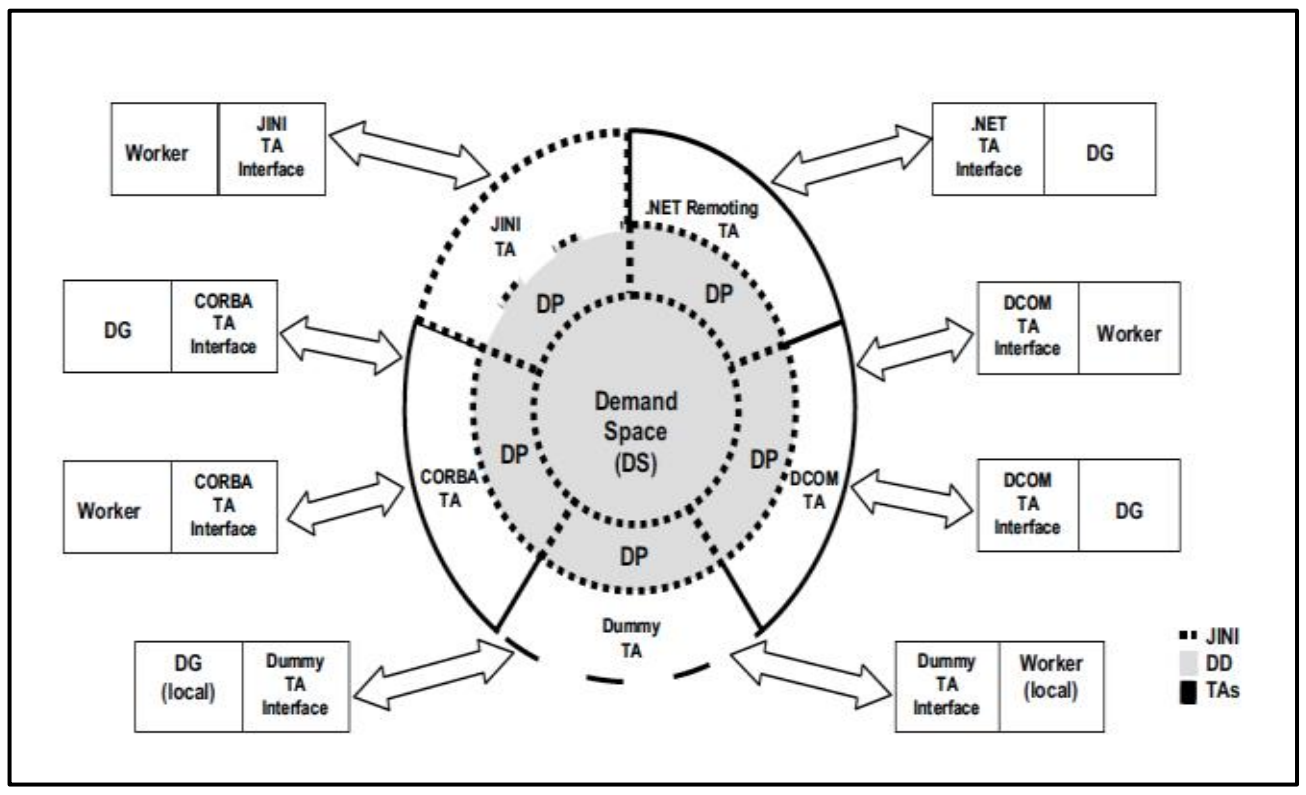

Figure 13: GIPSY Demand Migration System [8]

In DMS architecture it has two major components- Demand Dispatcher (DD) and Transport Agents (TAs). The DD is for the event-driven message storage mechanism and TA is for the delivering the messages. The DD have two entities Demand Space (DS) and Dispatcher proxy (DP). The DD receives the demand from the DG and have a track of the records to be solved. DD acts as bridge between DG's and workers. There are Local DD and Remote DD. In the Local DD, the DD is locally placed in the DG, they won't cross out of the machine boundary and TA's are not needed, as they are local. In the Remote DD the DD and DG are placed on different machines and TA's are required to communicate. The DD categorizes the demands based on the state like process, pending or computed. In process demand is that it has to be sent to worker and executed and if result is returned then it's a computed state and if the result is not populated then it's in pending state. The DS provide space to store all the demands and results. The DD identifies the DG's, workers, TA's by DP's. DP generates global ID for each demand. The TA's are responsible for the carrying the in process demands from DD to workers and vice-versa and also the computed demands from DG to DD and vice-versa. They act like a messengers and also acts as GIPSY transport protocol. One of the TA is based on JINI. It's a java technology and easily supports the GIPSY. It uses the internal protocols called Discovery, join and lookup.

*The GIPSY architecture* [11]

This paper deals with the design and implementation of the GIPSY architecture, which mainly focus on three goals: generality, adaptability and efficiency. Intensional programming is also known as multi-dimensional programming; because expressions involved are indulge to vary in an arbitrary number of dimensions. Lucid is a multi-dimensional intensional programming language. Intensional programming been used to resolve the problems with clear understanding of problems of intensional nature. GIPSY consists of three modular sub-systems they are: General intensional programming compiler (GIPC), General eduction engine (GEE), Intensional programming run-time environment (IPRE). All these components are designed in modular manner to replacement of each of its component at runtime or compile time to improve the efficiency of system.

General Intensional Programming Compiler: Like other functional languages there are many variants of lucid like basic Algebra, functional application, conditional expression, intensional navigation and query. The syntax assumes that identifiers (id) refer as constants, variables, functions or dimensions. Generally functions and dimensions can be first-class values and operational semantics of lucid given as structural operational semantic style.

GIPSY programs are compiled in the two stages. Firstly the program is translated into C and then resulting C program is translated in standard way. Source code consists of two parts the Lucid part which deals with data dependencies between variables and Sequential part that defines the granular sequential computation units.

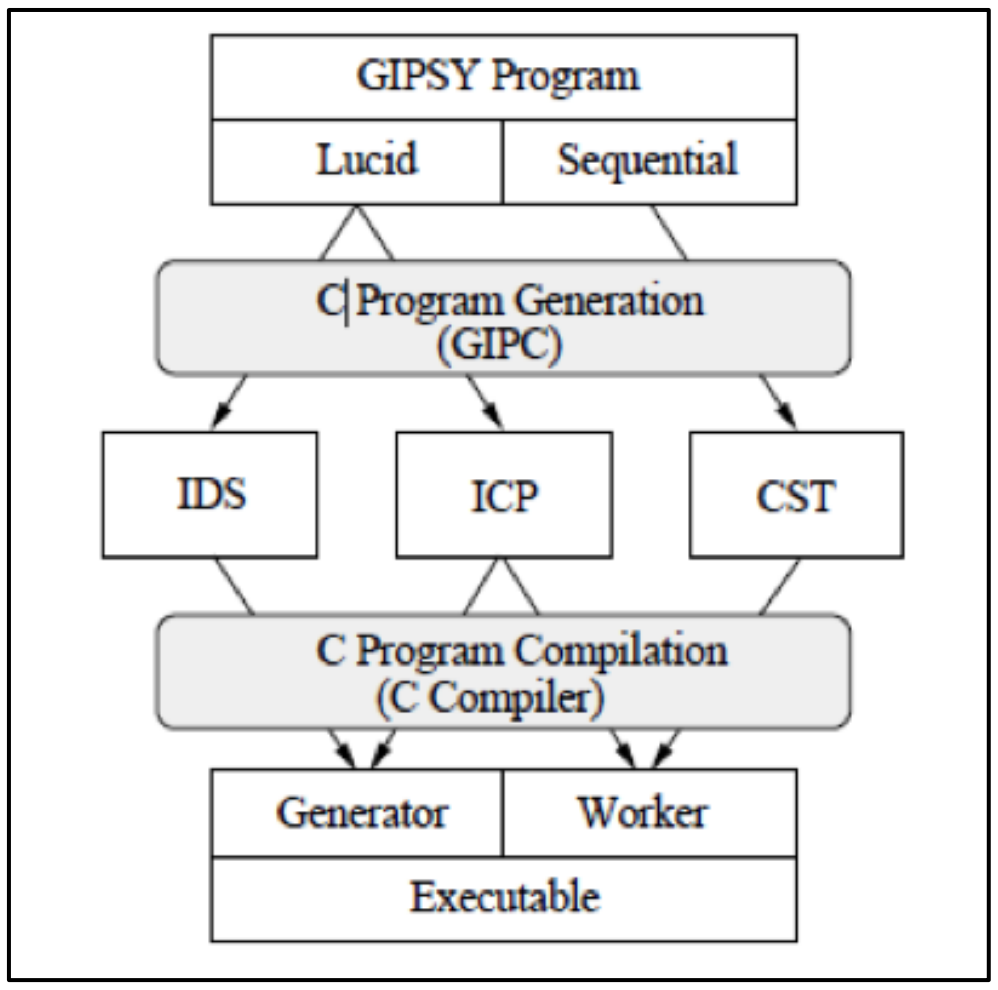

Figure 14: GIPSY program Compilation process

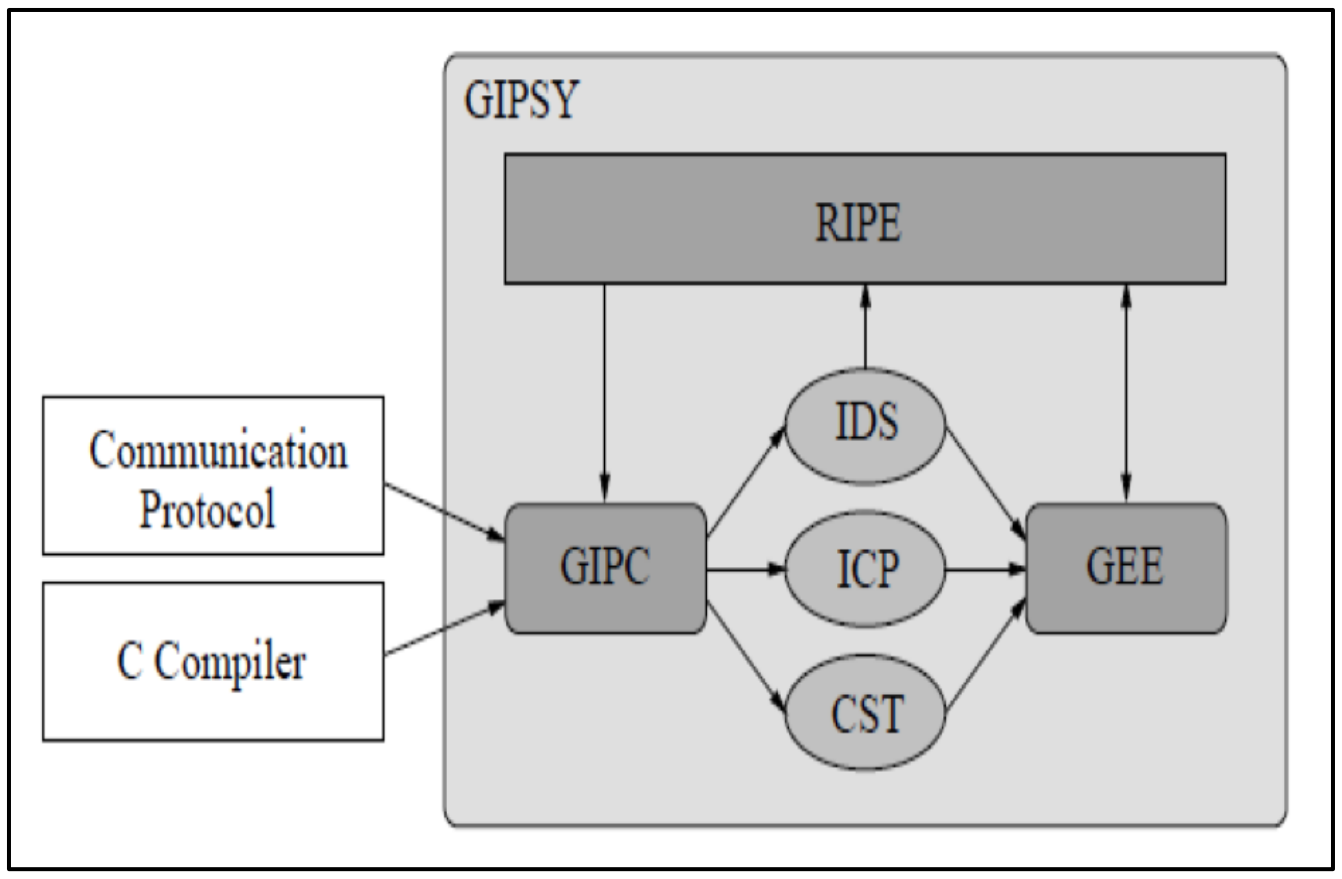

Figure 15: GIPSY Software Architecture

GIPC is modularly designed to allow the individual compilation of IDS, ICP, and CST.

General Eduction Engine: GIPSY uses eduction, which is demand driven computation in conjunction with a value cache called warehouse. Every demand generates a procedure call either remotely or locally and every computed value is stored in warehouse and for every demand an already computed value is extracted from warehouse rather than creating new.

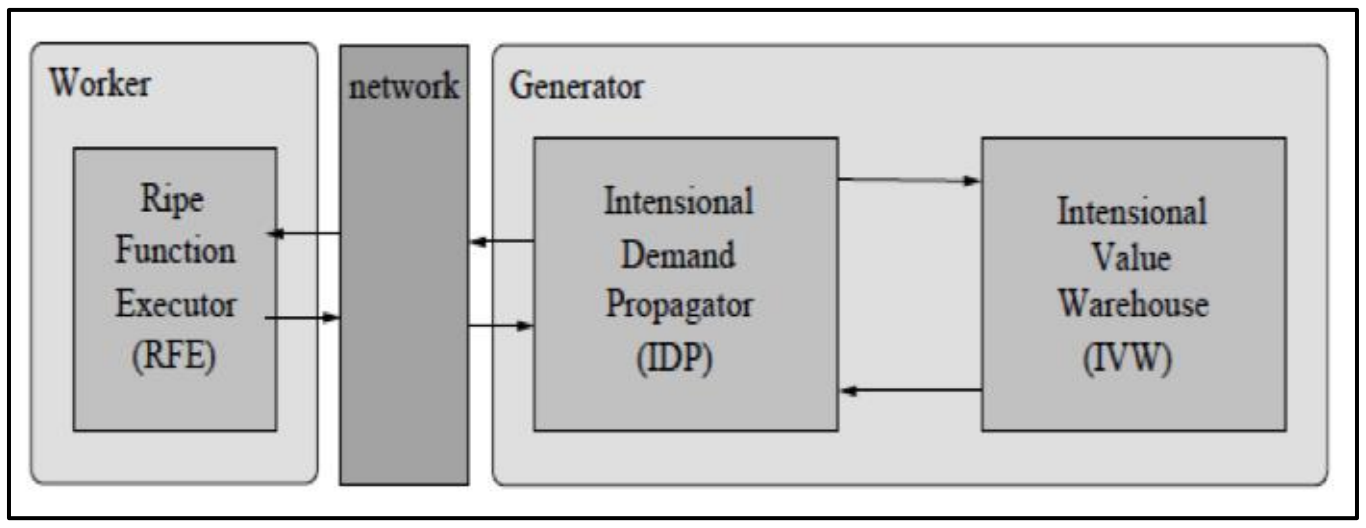

Figure 16: Generator-worker execution Architecture

Generally GIPSY uses the generator-worker execution Architecture. Low charge ripe sequential threads are evaluated by generator and high charge threads by worker. The generator consists of two systems Firstly Intensional demand propagator (IDP) which generates and propagates demands according to the data dependence structure (IDS) by GIPC. Secondly Intensional value warehouse (IVW) which is merely implemented as cache. The cache is the garbage collecting algorithm which improves the performance and enabling the various criteria to identify garbage in warehouse. The worker consists of Ripe Function Executor (RFE) responsible for ripe sequential threads demanded by generator.

Run-time interactive programming environment: RIPE is enabling the visualization of the data flow diagram corresponding to the lucid part of the GIPSY. Lucid programs as multidimensional dataflow graphs had been devised. For instance in hamming problem generating the stream of numbers of 2i 3j 5k in ascending order without repetition.

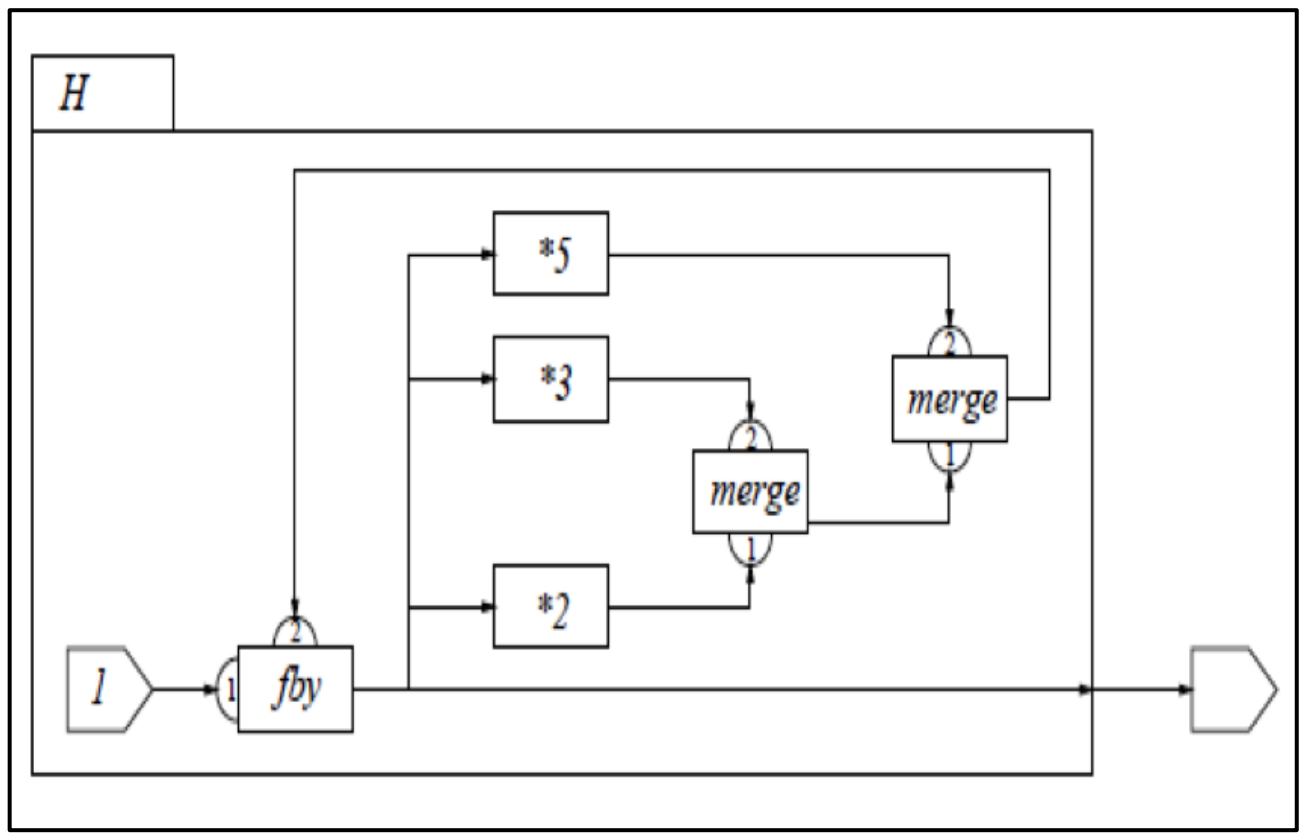

Figure 17: Dataflow graph for Hamming problem

The above figure reveals the dataflow diagram of merge function. RIPE even enable the graphic development of Lucid programs translating the graphic version of program into a textual version can also be compiled into operational version, however these development raises king of problems but solution is not yet settled.

It's proved that intensional programming can be used to solve the problem like distributed systems. This intensional programming framework has proven to provide an excellent solution (in terms of expressiveness) compare to other techniques. Finally the implementation of GIPSY enables us to realize the better solutions in a unified framework and reveals usefulness of approach.

*An Interactive Graph-Based Automation Assistant: A Case Study to Manage the GIPSY's Distributed Multi-tier Run-Time System* [15]

The General Intensional Programming System (GIPSY), is a research project being developed at Concordia University. The primary goal of the system is to evaluate the programs written in Lucid Intensional programming languages using a distributed demand-driven evaluation model. The design of GIPSY is itself integrated with the Lucid compiler framework and demand driven run-time framework. The system primarily includes of the demand generators and demand workers, via a communication node these demand are collected by a demand worker, compute it and send it back to the generator by the communication node. The GIPSY framework itself has the capability to work on the demand driven distributed evaluation of programs not having Lucid language. The paper focuses on how the GIPSY framework can be extended to automate the configuring and managing the GIPSY components via a graph-based approach. [15]

GIPSY run-time is a distributed multi-tier and demand driven framework. The run time system of GIPSY consists of mainly the following components:

1) GIPSY tier – an abstract and generic entity.

2) GIPSY node- a registered process that hosts GIPSY tier(s).

3) GIPSY instance- a group of tier instances acting together to achieve program execution without fail.

To control the GIPSY network, a manager node is implemented [15]. It enables the new nodes to establish connection with it and get instructions from it. Thus the combination of the generators, workers and communication manager nodes creates a virtual network with the manager nodes being able to set the connectivity among the other three. The multi-tier architecture of the GIPSY has been implemented to attain certain set of principles. It comprises of four distinct tiers each has its own specific functionality:

a) Demand Store Tier (DST)-provides persistent storage for demand and their values.

b) Demand Generator Tier (DGT) -generates demands according to the declarations.

c) Demand Worker Tier (DWT) - processes demands.

d) General Manager Tier (GMT) -manages all the other three tiers.

These tiers are responsible for the demand processing and communication as well. There are three types of demands that are migrated in the system. Intensional, Procedural and System demands. Each of these demands is migrated to other tiers using DST. Demands are possibly in any of these three states: pending, processing or computed. It was proposed to have and integrated tool that it can do the tasks, can create, save/load a GIPSY network and start/stop the nodes and allocate/deallocate the tiers. The main component of the system, GMT is designed graphically and then manages the entire GIPSY framework by translating the graphical interactions into complex message between the components. [15] The main goal of this is to reduce the manual interaction with the system till now in use to run the commands into much better automated tool. Sample commands that are used to interact with the system are:

start GMT GMTConfigFile.config

- to start the bootstrap process [15]

deallocate NodeID TierType TierID1 TierID2 TierIDn

- to deallocate tiers [15]

*Advances in the Design and Implementation of a Multi-Tier Architecture in the GIPSY Environment with JAVA* [17]

GIPSY provides the platform to investigate on the intentional and hybrid programming. The GIPSY's compiler consists of GIPL that can translate all the Lucid programming languages to core runtime language and also solved the language independence of runtime system by GEER. Multi-threaded and distributed architecture using Java RMI is designed first and also holds two branches based on implementation of JINI and JMS but no ability of exchange data and delay in the execution of system i.e. GEE. To overthrow GLU to multi-tier and also building the wrapper classes for all tiers DGT, DST, DWT and GMT. Every GIPSY node is a design that translates to single physical computer and register with GIPSY network, hosts instance for each tier enhanced. Four local and distributed prototypes are implemented namely Multi-threaded and RMI, JINI and JMS. When consolidated together by abstract factory and strategy design patterns with extreme programming and the test driven development methodologies, focusing on developing a framework with high extensibility and maintainability for upcoming iterations.

The main concept of this model is generation, propagation and computation of demands and its results. There are also several demand types like Intentional demand, which contains GEERid, ProgramId and Context. Procedural demand form contains GEERid, programId, Object Params, context and code. Resource demand form contains resourceType Id and resourceId. System demand form contains destinationTierId and systemDemandTypeId.

Dispatch Entry class uses universal unique identifier to identify unique demand within the network. Demand signature identifiers will create the same identifiers for all the requests created so that they could be questioned in DST and the result might be concentrated without re-computation. RMI-runtime helps the storage for GIPSY and multi-tier modeling states the expansion of new innovations at point when it is needed. To support the adaptability of GIPSY joining the JINI-DMS, which is focused around JINI and JavaSpaces where JINI is utilized for the Transport Agents and Java spaces is utilized for Demand store. The JMS-DMS which is focused around Java Messaging Service and JBOSS server is utilized as the JMS supplier and Hypersonic Database, which are implanted in JBOSS.

Design holds the main package gipsy.GEE.multitier and subclasses in it i.e gipsy.GEE.DGTWrapper, gipsy.GEE.DSTWrapper, gipsy.GEE.DWTWrapper, gipsy.GEE.GMTWrapper. All the subclasses are inherited from the conceptual class called Generictierwrapper, which has the normal usefulness of IMultTierWrapper, which is only the bundle gipsy.GEE.multitier. In the GenericTierWrapper, API's are in similar to Configuration and ITransportAgent. Configuration is for run-time administration and ITransportedAgent is for all TA operators, initially it doesn't have the regular super interface, which characterize now to backing the engine and design.

DGT and DWT wrappers have normal oGREEpoll, which holds the objects of GIPSY program. The GIPSY program is accumulation of identifiers and GREEpoll is gathering of GEERs and DST joins the oStorageSubsystem, where an item sort of IVWInterface refers to demand store, furthermore there are some supporting classes like EDMFImplementation, TierFactory and NodeController. Main class of GEE must be re-outlining to backing these new improvements. When these progressions are carried out, unit testing and reconciliation testing will happen to verify that if it is bug free and new changes will not influence the old progressions. Along these lines, the Multi-level GIPSY once deployed will serves as the Scalability and flexibility to the framework. Here Demand class plays an essential part among the tiers.

*Using the General Intensional Programming System (GIPSY) for Evaluation of Higher-Order Intensional Logic (HOIL)* [18]

GIPSY helps as test bed for HOIL-based languages and also cyber forensic case analysis with event reconstruction. The GIPSY undertaking is a continuous exertion going for giving an adaptable stage to the examination on the intentional programming model as acknowledged by the most recent versions of the Lucid programming language, multidimensional context-aware language whose semantics is dependent on possible words semantics. GIPSY gives an incorporated schema to accumulating projects composed in theoretically all variants of Lucid, and even any language of intentional nature that could be deciphered into a "generic Lucid".

Eductive Model of Computation:

The idea of eduction could be depicted as “tagged-token demand-driven dataflow” computing (whereupon Lucid impacted a popular media stage and language called pure data). The focal idea to this model of execution is the thought of era, propagation, and utilization of requests and their ensuing values. Lucid programs are declarative programs where each identifier is characterized as a HOIL interpretation using other identifiers and an underlying algebra [20].

Intentional Logic and Programming:

Intentional programming could be utilized to take care of generally differentiated issues, which might be communicated utilizing enhanced languages of intensional nature. The GIPSY undertaking goes for the formation of a programming environment incorporating compiler era for all kinds of Lucid, a generic run-time system empowering the execution of programs composed in all kinds of Lucid. The objective is to give an adaptable stage to the examination on programming languages of intensional nature, keeping in mind the end goal is applicability of intensional programming to take care of paramount issues.

HOIL (Higher – Order Intensional Logic):
HOIL joins functional programming and intensional logics.

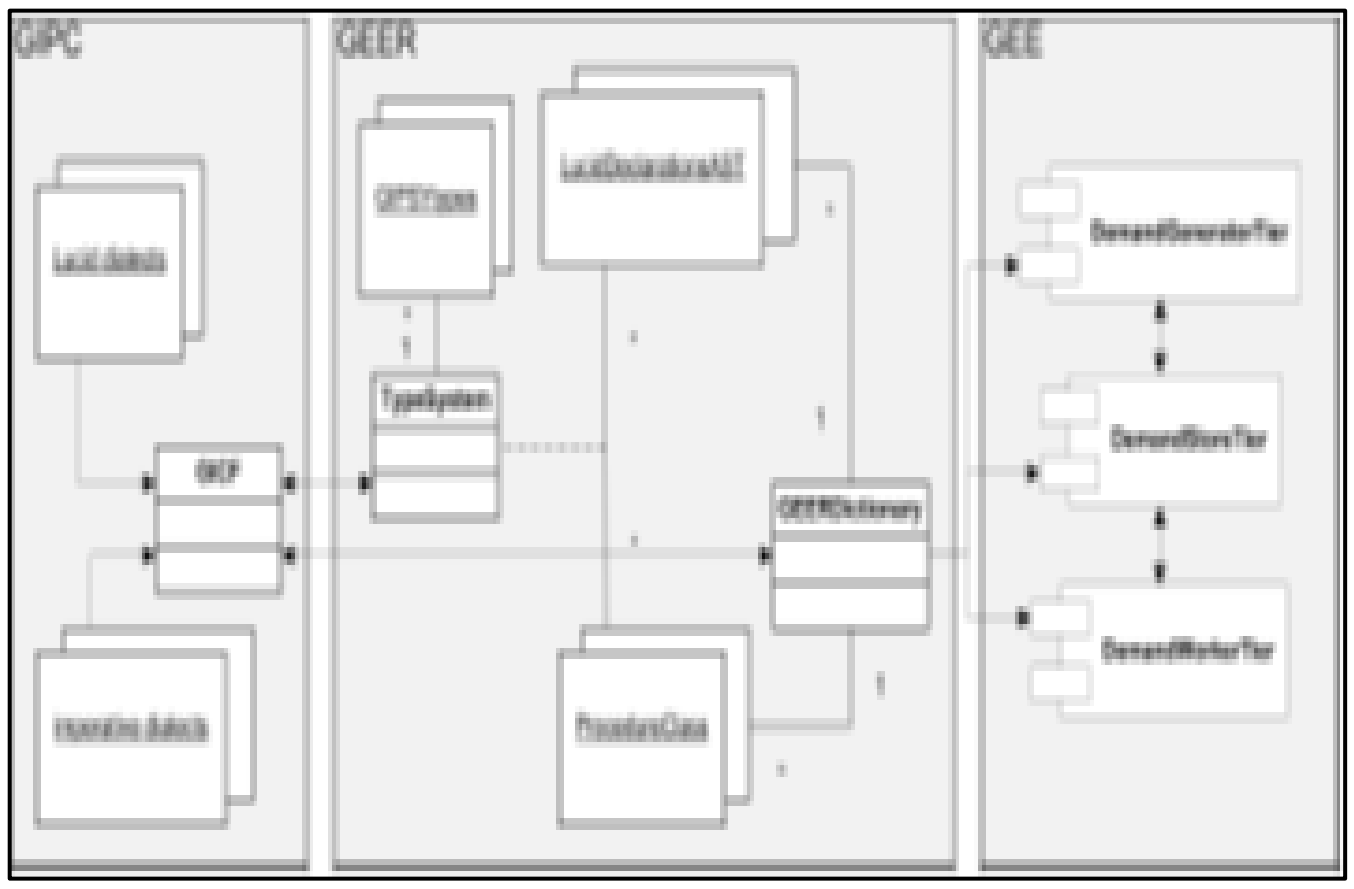

Figure 18: GIPSY’s GIPC-to-GEE GEER Flow Overview in Relation to the GIPSY Type System.

General Eduction Engine (GEE):

The outline architecture engineering embraced is a dispersed multi-tier architecture design, where every level can have any number of examples. The architecture planning bears resemblance with a peer-to-peer.

GIPSY Tier:

The architecture engineering embraced for this new development of the GIPSY is a multi-tier architecture where the execution of GIPSY programs is isolated in three different tasks assigned to separate levels.
GIPSY Node:

Uniquely, a GIPSY node is a computer that has registered for the facilitating of one or more GIPSY tier. GIPSY nodes are enrolled through a GIPSY Manager occurrence.

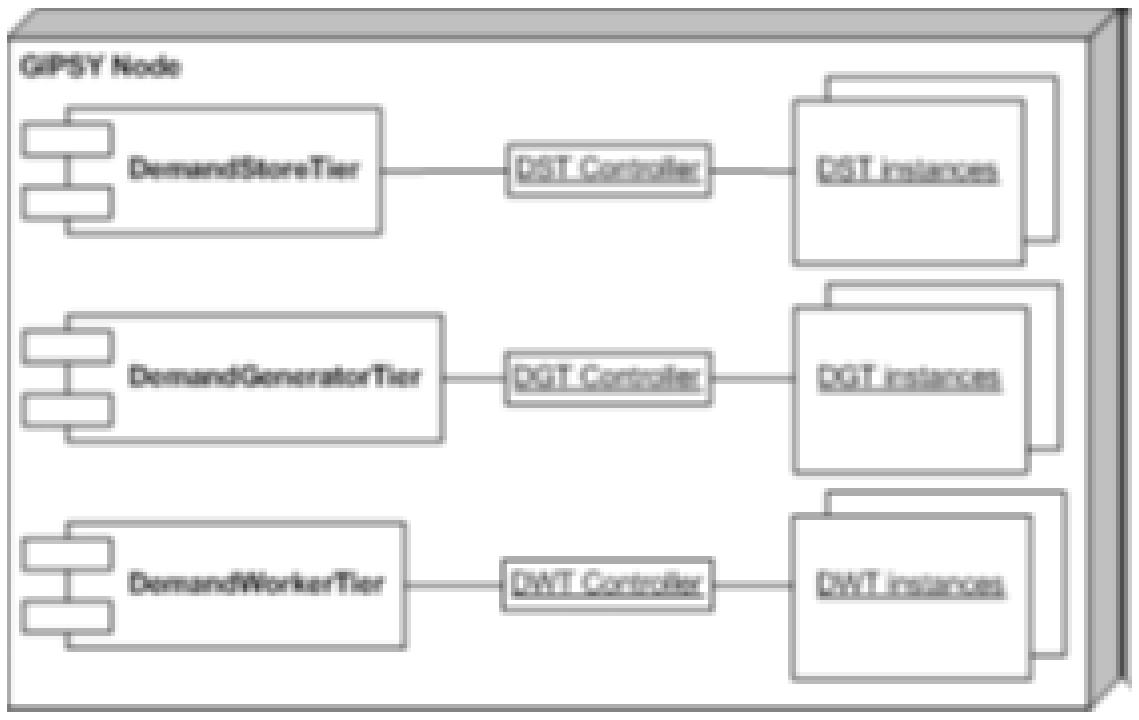

Figure 19: GIPSY Node

A GIPSY Instance is a situated of interconnected GIPSY Tiers sent on GIPSY Nodes executing GIPSY programs by offering their separate GEER occurrences. The Demand Worker Tier (DWT) processes procedural demands.

Context-Oriented Reasoning:

As said prior, the reasoning part of GIPSY is a distinction of a Lucid dialect as opposed to the building design, and the analysis has main concentration on the reasoning point. The architecture modeling is general enough to go past reasoning – in the essence it is an assessment of intensional logic expression.

In future and continuous work inside the context of GIPSY is a complete formalization of its hybrid intensional-basic sort framework, the update of the syntax and semantics of the Forensic Lucid dialect, and the multi-tier upgrade of the evaluation engine (GEE) including backing for OO intensional languages [20].

Finally, it's about modular intensional programming exploration stage, GIPSY, for reasoning undertakings of HOIL expressions. The concept of context as a first-class value is central in the programming paradigms GIPSY is built to explore, such as a family of the Lucid programming languages. At the time of this composition GIPSY has helps for compilation of GIPL, Indexical Lucid, Lucx, JLucid, and Objective Lucid and the execution of if the previous two with the other being finished. The DMS for distributed transport of the request has usage in JINI, plain RMI, and JMS.

*Distributed Eductive Execution of Hybrid Intensional Programs* [19]

The theory specifies the process of investigation of the intensional programming based on a framework known as LUCID, which checks with the declarative programming as the base in order to evaluate them in a multidimensional space. Besides in most of the cases the space for the intensional programming would be infinite and this process of execution is known as the Education, where each demand is propagated and stored as the identifier context pair. GIPSY provides a platform for the intensional programming languages which in turn is useful for LUCID programming.

The main concept of the Eductive model is the notion to propagation, generation and consumption of demands and their values. Lucid programs are kind of declarative programs where each identifier is identified as an expression with other identifiers and expressions. There is an initial demand for the identifiers and later based on the demands other demands are populated and are in turn computed out to the initial demand and represent dimensional abstract concepts.

Lucid being the descendant of ISWIM, it is being defined of data types and algebra. Due to their semantics, programs can be executed in parallel. However, in the years that immediately followed, the syntax and semantics of the Lucid language was modified, and the architecture of the GLU compiler and run-time system proved to be impossible to adapt to the latest changes in the language.

GIPSY is an evolving system having similar capacities as like GLU but it is more flexible. This system can cope with the range of lucid programs and exhibits qualities like language independence, scalability, flexibility of execution on so on.

The new system which has been developed for the fast evolution of the Lucid family is taken up by the well versed architecture of the GIPSY known as the General Intensional Programming Compiler (GIPC).

Architectural design:

The architectural design is a multi-tier architecture where many layers have their dependent work assigned, resembled as instances. Nodes/instances can be added any time to the system where a specific node does the required demand as per the request. Basically GIPSY tier, Node and Instance are the three different types of processing factors which take the demand into consideration and categorize accordingly as per the intensional demand and the procedural demand. Each node has demand generator, demand store, demand worker.

In essence, the paper presents a flexible and scalable infrastructure for eductive evaluation of hybrid intentional programming and focused on achieving goals like language independence, scalability, flexibility, opacity and observability.

*B. Summary*

A short analysis of the previous research studies introduces the basic concept of the system architecture of the study cases. Even though the studies included into the present document does not cover in totality the domain model and all requirements of our two background studies, certain of them were captured in the following section.

Certain quality requirements (and nonfunctional) were common to both studies, such as:

- Distributed computation, that means distributed architecture of the nodes have components of the systems deployed. This architectural pattern actually introduced the need for many others, identified further down. The backbone of a distributed system commonly used is the messaging system (such as Java Messaging System, JMS). Although, the systems supported also Web services and RMI calls, whereas advantages and disadvantages of different architecture would make the case of a separate study.

- Enterprise class monitoring, such as real time monitoring, profiling and awareness of the system about the status of its underlying subcomponents. That includes not only the current status, but also trending and historical profiling.
- Fault tolerant system requirement, including self-healing and self-optimization. This is one of the most critical factors contributing to the successful implementation of a real system. In reality, the so called "continuous available" system, which has 100% availability is extremely expensive to implement and achieve, although as the world emerges and data exchange spreads across the globe, there is a continuous increased need for these types of systems.
- Such distributed system architecture introduced new security requirements into the design of the systems. The attacks are more frequent lately and the cost associated with the security breaches represents an important factor, mainly due to lawsuits, privacy, data theft, etc. (many examples can be provided, with losses of magnitude of B$). The attackers are becoming more experts and dispose of more and more power in deploying their attacks.

- Scalability, parallel processing and loosely coupled systems.

Software Engineering (SE) as a discipline became finally a de-facto discipline in the design of the software systems. That could have led to the need for formal specifications, although this is difficult to attain. There have been many progresses in the SE in the latest 20-30 years, which is reflected in the complexity of the systems nowadays and in their intrinsic quality.

Another trend observed in the IT, given the amount of the data collected, that increases exponentially, which created the need for data mining, fuzzy logic and self-learning systems. MARF (and its distributed design DMARF) is one of the few systems capable of learning and analyzing fuzzy data and capable of taking decisions based upon what's been learned (there are also open source and commercial solutions, e.g. Tibco Spotfire for Pattern Matching Data Analysis, Tibco Master Data Management or Tibco Complex Event Processing, CEP, one of the leading providers according to Gartner quadrants).

GIPSY on the other hand, comes with a new approach in the programming languages, the intensional programming as whole new paradigm. Similar to DMARF, GIPSY had been designed in a distributed loosely coupled architecture approach and supports high availability and self-healing. GIPSY tries to implement an event driven and context oriented reasoning (which is related to the concept of CEP, described before). The hunger for data, data mining and rapid business decisions in the industry have been the engine of the investments into the research, creating the need to be ahead of the competition, that give a business advantage.

## IV. Requirements and Design Specifications

In this section, concepts, requirements, specifications and other aspects from the problem domain are depicted from the case studies analyzed in the first iteration of the document. The intent is to understand and problem domain and define the concepts, associations, and so on.

### *A. Personas, Actors, and Stakeholders*

<u>DMARF</u>

**Primary Persona:**

| **Name:** 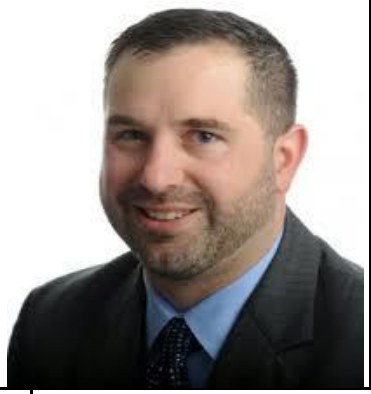 | James |
|---|---|
| **Age:** | 35 years old |
| **Occupation:** | James manages all aspects of Technology Transfer including: innovation relationships, invention disclosures, intellectual property, company collaborations, materials transfer, outsourcing contracts, and non-disclosure/confidentiality |
| **Background:** | James has a B.S. in chemistry from Centre College, a J.D. from Tulane University Law School and an LLM in Intellectual Property and Media Policy from the University of Illinois at Urbana-Champaign. He is also trained in Science Policy and Leadership from the American Association for the Advancement of Science. |
| **Goal:** | James would want to use the DMARF system to analyze the quality of the documents he creates and to be able to detect anomalies within their content. Since he is mostly busy with his business he does not have time to invest in setting |
| | up, maintenance, troubleshooting, etc. and he wants to have an autonomous system that heals itself and does not require much intervention. He wants the system available on demand, and cannot afford to lose too much time waiting for the processing to occur.<br>James knows a lot about technology, but he does not have enough time to delve into the details of the system. He would like to actually have someone else maintaining the system, performing administration and setup. |

**Actors:**

| | |
|---|---|
| **Admin:** | The role of the administrator is to maintain the system in good functional conditions and troubleshoot and correct any issues the system might run into. The admin would like to have simple monitoring and notification capabilities for the critical errors and wants to be notified to take action only when the system cannot heal itself. The admin is involved with the monitoring console, starting and stopping subcomponents of the system. |

**Stakeholders**:

| | |
|---|---|
| **Developer:** | The developer is involved with the implementation development, evolution and maintenance of DMARF system. The developer is responsible for the implementation of the system as design by the architect. |
| **Architect:** | The architect is responsible with the macro design of the system. He creates the blue print of the system. |

<u>GIPSY</u>

**Primary Persona:**

| | | |
|---|---|---|
| **Name:** | 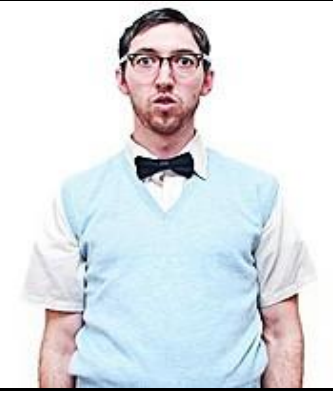 | Egbert |
| **Age:** | 28 years old | |
| **Occupation:** | Computer programmer | |
| **Background:** | Egbert has a B.S. in Computer Science and attending currently the classes towards Master in Software Engineering | |
| **Goal:** | Egbert is a nerd. He loves computers and he's an expert in computer programming in Java. He uses daily Java at work to develop middleware systems. Recently, he discovered the new intensional programming paradigm and he thinks this matches perfectly the requirements of the new system he just started working on. This particular system has specific software requirements that are difficult to be implemented by using the usual object oriented programming design. The system he's just started working on integrates the systems of the emergency services from hospitals with small carriers (such as hauling companies, taxies, limousines, etc.) for the remote areas, where the emergencies services have difficulties to reach due to the road and weather conditions. Egbert believes, given the high level of abstraction of the GIPSY programming language (intensional programming), he could use GIPSY as the framework for its' new system. His new system has to be deployed in large remotely distributed locations, where the maintenance personnel are not available; therefore the system must be capable of healing itself as much as possible. Similarly, the individual components or modules of the system might fail, disconnect, and so on, yet, they can be dynamically replaced by the system by coordinating deployment of the job/worker task towards a new node of the system.<br>In essence, Egbert wants to be able to act as the user of the new system he works on. He would create a set of applications developed in intensional programming paradigm as a library and share the library and GIPSY's framework to the emergency services. The functions made available in the library would satisfy the business needs of the remote emergency services. | |

**Actors**:

| | |
|---|---|
| **Admin:** | The role of the administrator is to maintain the system in good functional conditions and troubleshoot and correct any issues the system might run into. The admin would like to have simple monitoring and notification capabilities for the critical errors and wants to be |
| | notified to take action only when the system cannot heal itself. The admin interacts with the monitoring console of the system and support the primary actor accomplishing his tasks. |

**Stakeholders**:

| | |
|---|---|
| **Developer:** | The developer is involved with the implementation development, evolution and maintenance of GIPSY system. The developer is responsible for the implementation of the system as design by the architect. |
| **Architect:** | The architect is responsible with the macro design of the system. He creates the high level design of the system. |

## *B. Use Cases*

This section of the study introduces a set of fully dressed use cases. The intent is to identify concepts, entities and associations, which are expressed in bolded characters, as candidate for the conceptual classes. The verbs are expressed underlined, as candidates for the activity diagrams.

DMARF

DMARF is the autonomic distributed system version of MARF implemented as distributed system architecture. That means the underlying components of the system are deployed on distinct nodes in a clustered environment, which provides high-availability and better scalability to the system implementation. Due to the fact that MARF was designed in a modular approach, facilitated the distributed architecture implementation.

Assumptions:

- The system is deployed in a distributed architecture, therefore managed by an administrator;
- The nodes are managed and monitored solely by the system's internal monitoring system. No other external system is involved;
- The use case depicts features within the system boundaries;
- The management and monitoring by the admin are part of distinct use cases.

| **Use case UC1:** | ProcessDocument |
|---|---|
| **Scope:** | DMARF system |
| **Level:** | User level |
| **Priority:** | High |
| **Primary Actor:** | Client |
| **Summary:** | The use case describes the normal flow of processing a document through the system. |
| **Stakeholders and Interests:** | • Client: Wants accurate and fast processing of the document, low rate of errors;<br>• Admin: Wants high-availability, monitoring and alerting, easy recovery |
| **Preconditions** | -All subcomponents of the system are available and interconnected;<br>-System is ready for use;<br>-Document for processing available according to specifications;<br>- Client has access to the system interface and report result; |
| **Post conditions** | •Document is processed successfully;<br>• Results report returned by the system; |
| **Basic flow:** | 1. Client presents the **document** for processing to the system interface;<br>2. System load the **document**;<br>3. System performs processing of the **document**;<br>4. System returns **results set**;<br>5. Flow ends. |
| **Extensions:** | *Alternative Flow 1:* Unable to load document<br><br>2a) System tries to load the **document** but it fails;<br>3a) System returns report with **exception** unable to load the document;<br>4a) Flow ends.<br><br>*Alternative Flow 2:* Unable to process document<br><br>3b) System tries to process the **document** but it fails;<br>4b) System returns report with **exception** unable to process the document;<br>5b) Flow ends. |
| **Special requirements:** | System shall be continuously available as the requests from the client are received in a sporadic manner.<br>The interconnected components (distributed subcomponents of the system) shall expose interfaces and enforce policies for strict access to such interfaces, to avoid data tampering. Furthermore, the communication between such subcomponents must be encrypted for confidentiality and privacy.<br>Although subcomponents of the system might fail, disconnect, overload, etc., the system shall be capable of identifying such issues and selfheal.<br>There shall be no single point of failure (hence, the underlying components of the system shall have no single point of failure). |
| **Frequency of Occurrence:** | On demand. |
| **Technology and Data Variations List:** | The distributed architecture of the system could be accomplished using various transports (such as Web Services, Messaging Brokers, RMI, etc.). This aspect is hidden from the user perspective.<br>The interface made available to the client has the same specifications regardless the client context.<br>The message payload (document for processing) must conform to the initial requirements of the interface. |
| **Miscellaneous** | Client interface shall support various operation systems. The front end and back end of the system might be deployed on the strategic operating system platform. Is the client interface required to support mobile, since this had an exponential increase in the number of clients?<br>What if the so called “thin-client” interconnects with the front-end via large geographically dispersed components, what is the impact of high network latency? |

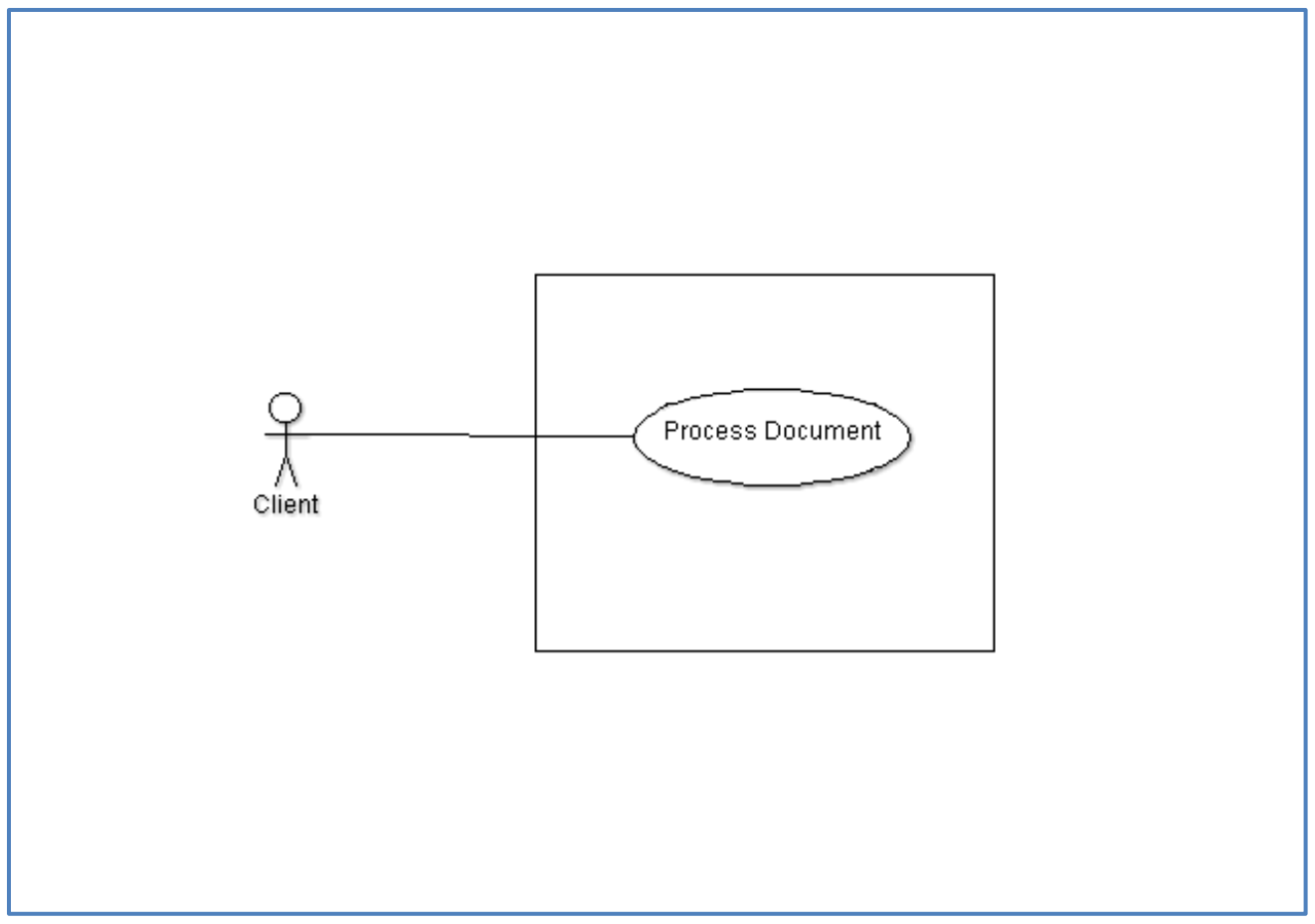

Figure 20 - DMARF Process Document Use Case diagram

GIPSY

GIPSY system provides a software platform for long term investigation of the intensional programming and consists of a flexible compiler and scalable runtime. The compiler translates the flavor of the intensional program into source language-independent resources and the runtime executes the generated code in a distributed manner [21].

Assumptions**:**

- The administrator of the system employs tasks in regards the configuration and setup of the GIPSY system;
- monitoring is performed by the internal components of the system;
- The use case depicts features within the system boundaries.

| | |
|---|---|
| **Use case UC1:** | ExecuteApplication |
| **Scope** | GIPSY system |
| **Level:** | User level |
| **Priority:** | High |
| **Primary Actor:** | Client |
| **Summary:** | The use case describes the normal flow of compiling and executing an application of intensional programming paradigm (e.g. Lucid) on the GIPSY system. |
| **Stakeholders and Interests:** | - Client: Wants to be able to compile and execute the source code;<br>- Admin: Wants high-availability, monitoring and alerting, self-recovery |
| | and self-healing; |
| **Preconditions:** | - All subcomponents of the system are available and interconnected;<br>- System is ready for use;<br>- Source code of the intensional application has been developed according to specifications and available for execution;<br>- Client has access to the system interface, compiler, and runtime components. |
| **Post conditions:** | • The intensional source code was compiled successfully;<br>• The translated application of the original source code was executed successfully;<br>• There were no exceptions encountered |
| **Basic flow:** | 1. Client presents the **source code** to the system for processing through the system interface;<br>2. System load the **source code**;<br>3. System performs compilation of the **source code**;<br>4. System returns **compilation results**;<br>5. System executes the new **compiled source code**;<br>6. System returns execution results;<br>7. Flow ends. |
| **Extensions:** | *Alternative Flow 1:* Unable to load the source code<br><br>2a) System tries to load the **source code** but it fails;<br>3a) System returns **exception** unable to load the source code;<br>4a) Flow ends.<br><br>*Alternative Flow 2:* Unable to compile the source code<br><br>3b) System tries to compile the **source code** but it fails;<br>4b) System returns report with **exception** unable to compile the source code;<br>5b) Flow ends.<br><br>*Alternative Flow 3:* Unable to execute the new compiled source code |
| | 5c) System tries to execute the new **compiled source code** but it fails;<br>6c) System returns **execution exception;**<br>7c) Flow ends. |
| **Special requirements:** | The interconnected components (distributed subcomponents of the system) shall expose interfaces and enforce policies for strict access to such interfaces, to avoid data tampering. Furthermore, the communication between such subcomponents must be encrypted for confidentiality and privacy.<br>Although subcomponents of the system might fail, disconnect, overload, etc., the system shall be capable of identifying such issues and selfheal.<br>There shall be no single point of failure (hence, the underlying components of the system shall have no single point of failure). |
| **Frequency of Occurrence:** | On demand. |
| **Technology and Data Variations** | The compiler and the execution engine of the system are deployed in a distributed architecture. This architectural aspect is hidden from the user perspective. The interface made available to the client has the same specifications regardless the client context. |
| **Miscellaneous:** | Client interface shall support various operation systems. Internal components of the system and their interconnection are hidden to the user. Similarly, the compiler, execution engine (manager and workers) is hidden behind the system interface. |

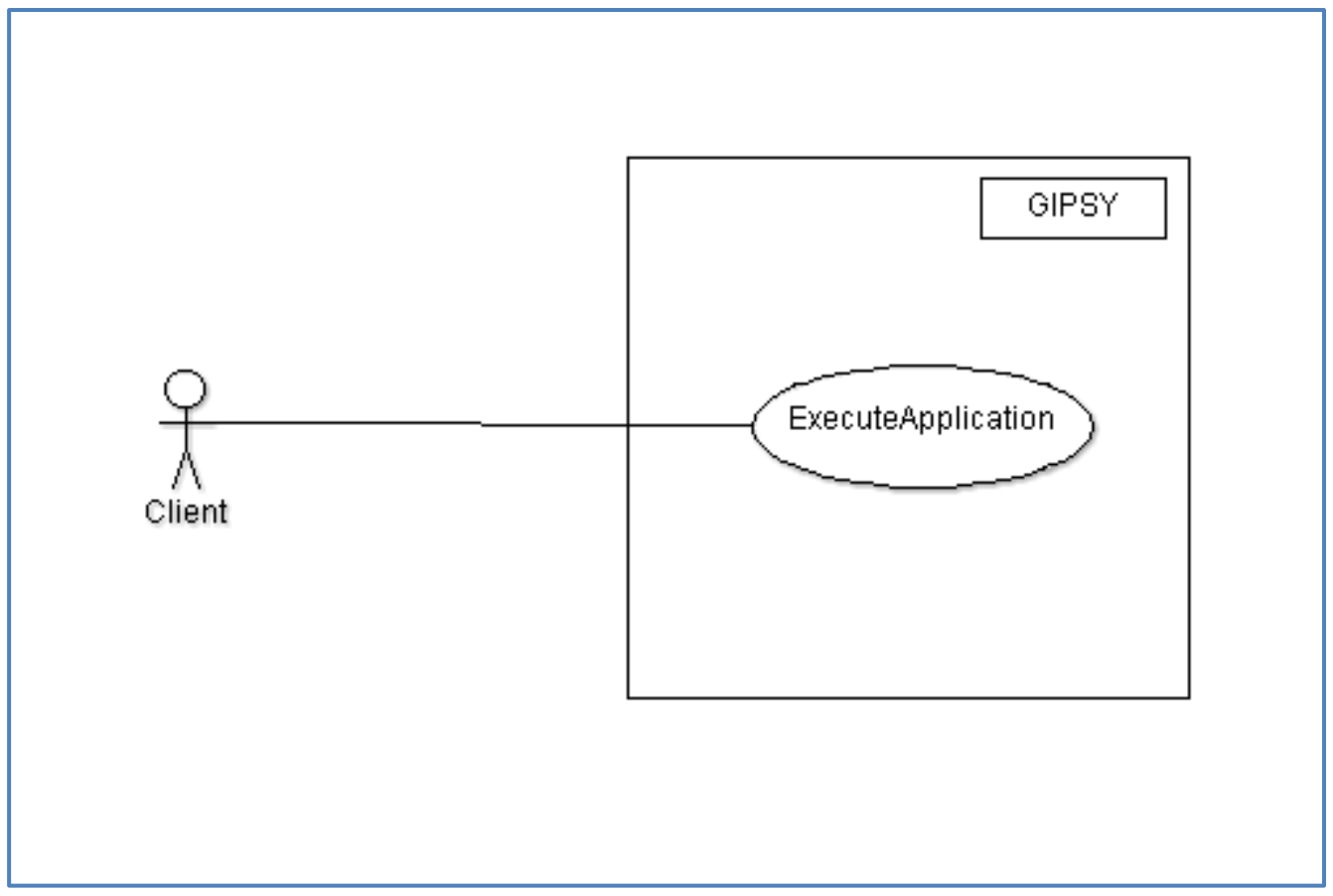

Figure 21 - GIPSY Execute Application Use Case diagram

## C. *Domain Model UML Diagrams*

DMARF

The Client presents **Document** in the form of a sample audio/text/image file to the DMARF System, where the Sample Loading Services (Loader) loads the **document** to the DMARF System and converts it for further processing using the Processor. The Processing Service accepts the incoming audio file sample from the Loader and does the required processing using the filters (FFT or CFE Filters). The features, which are extracted after the preprocessing file, will be segregated into training and classification by an algorithm. After the file extraction, the respective processed file would be compared with the existing document in training for data validation using Classification component and stores the result in **ReportResults** section.

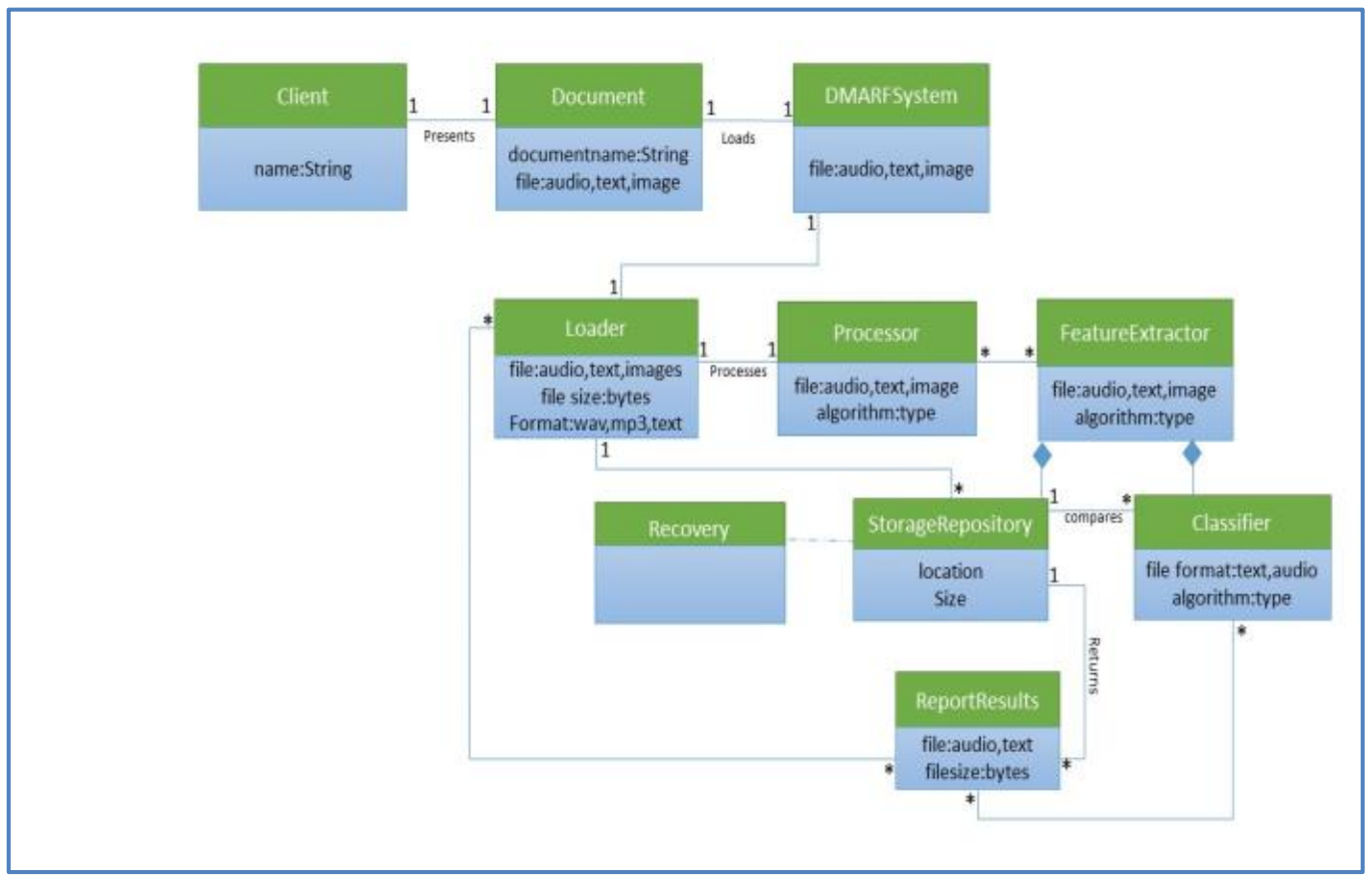

Figure 22 - Domain model for DMARF

GIPSY

The GIPSY run-time system is a distributed multi-tier and demand driven framework. As per the domain model depicts, the GIPSY network over an underlying physical network of computers, which has the run-time system that supports the demand driven computation of programs. The user presents the **source code** to the system and loads accordingly into the *GIPSYInstance* where the code is categorized into two parts: Lucid and Sequential. The categorized code will be compiled into C language. In GIPSYTier, the **compiled source code** is stored for processing and the *ConfigurationInfo* tracks the name, port number, IP address, service name of the request. The GIPSYNetwork is a TA (transport agent), which will help to carry the request, if the request is coming from the Remote connection. When the GIPSYNode accepts the request then the node will be registered and the corresponding request status will be in pending status, then the request will be sent to Demand Worker Tier for execution. The NodeConnection is responsible for instantiating and terminating the request, besides the GIPSYManager Tier keeps track of the request status. Once the request is completed where the corresponding status of the request will be in computed status and will be stored in GIPSYTier and returns to the ResultSet, where the client can view the results accordingly.

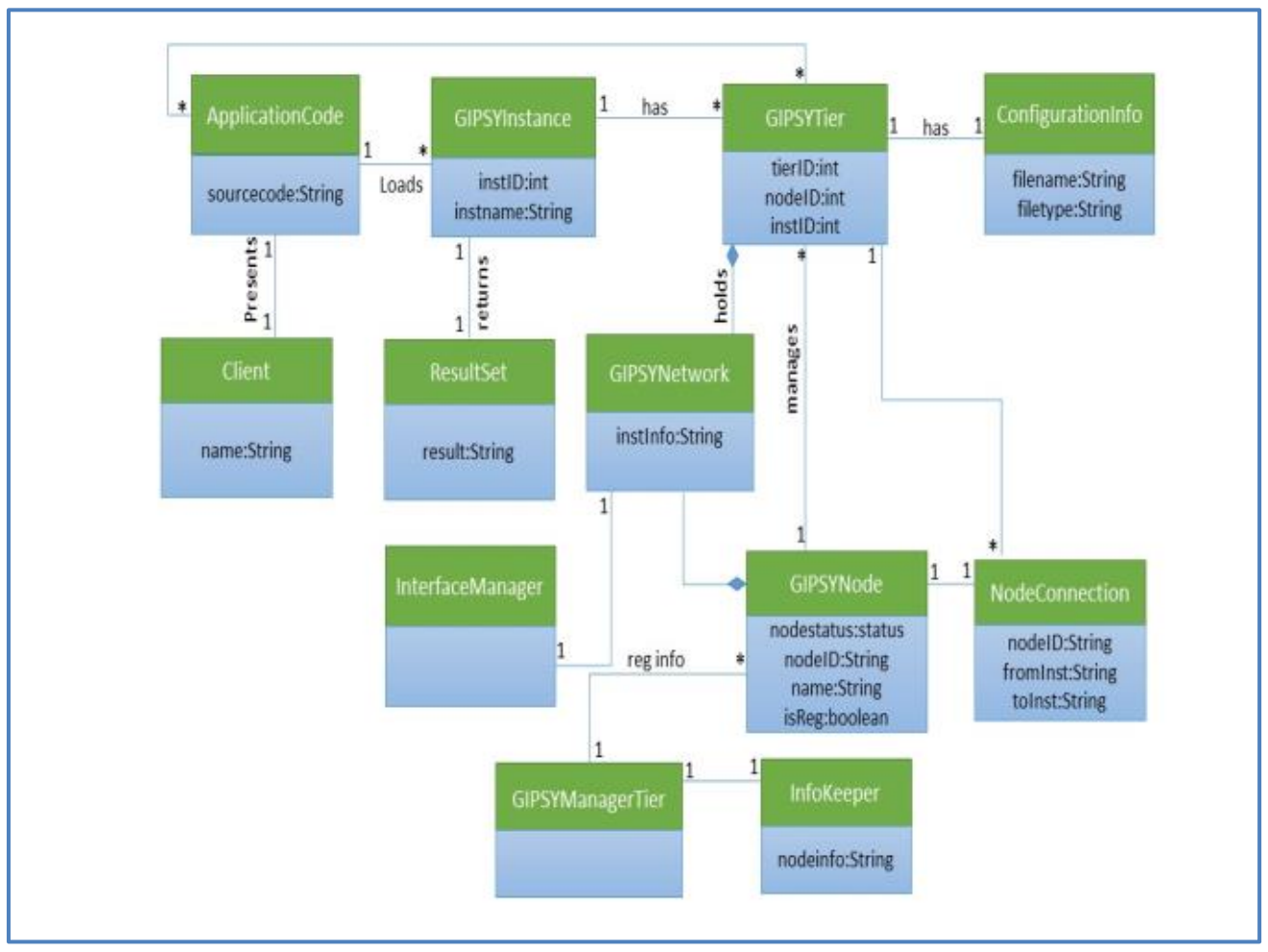

Figure 23 - Domain model for GIPSY

### D. *Fused DMARF-Over-GIPSY Run-time Architecture (DoGRTA)*

GIPSY system provides the framework for distributed architecture, having implemented transport agent components that handle the communication between different tiers. The agent is independent of the protocol implementation of the communication. By default, GIPSY supports the following protocols for the integration of its internal components: RMI, JINI and JMS.

There are four main components in GIPSY to handle the distributed aspect:

- Demand Generator (DG)
- Demand Store (DS)
- Demand Worker (DW)
- General Manager (GM)

The demand generators send demands for processing via the implementation of the transport agent, which abstracts the communication and provides a single interface across multiple underlying messaging protocols.

MARF can be expressed as being composed of the following modules:

- loader;
- preprocessor;
- feature extractor;
- classifier that is:
    = training processor;
    = classifier.

The system is implemented as a pipeline: messages are loaded into loader, and then sent to preprocessor, and so on, and then finally the result set is sent back to the client. Each individual stage across the pipeline can be integrated by calling out the GIPSY's components. MARF messages are sent to DG, then the DG sends them to the DS, then they go to the DW, where they are processed and finally the result set is returned back to the MARF.

For simplicity, let's consider the example of *Java Messaging System (JMS)* transport, where each module of MARF is interconnected with the next through message queues.

JMS insures the message delivery (and unique reliable delivery, of course depending on the persistence of the message). Commonly, the JMS messages are delivered via a *Message Driven Bean* (MDB), which has the method *onMessage* that is called whenever a message is received and the message broker initiate message delivery.

Similarly, MARF would rely on GIPSY's components to deliver and receive messages in their intermediate states along the pipeline of MARF's architecture. The DW would receive the message from previous stage (as sent by the DG), only once (although depending of the JMS delivery setup), perform processing, then return the result set the DG, which in turn returns the message to MARF. Next, MARF continues with the next step in its pipeline, again invoking DG, which sends message to DS, then DW and so on.

Furthermore, JMS supports multiple listeners (or clients) on one queue and message are delivered uniquely to one client (if set as exclusive) or distributed in round-robin manner. Hence, multiple clients (which is in fact the DW that implements the functionality of MARF subcomponents) are listening to one queue but only one gets the message. If a client disconnects, JMS automatically delivers the message to the next client. In essence, clients can be distributed across multiple nodes.

Another approach would be the scenario where MARF implements a call-back handler on each stage of the pipeline. Whenever the response set has finished processing by the DW and sent back to the DG, the DG would call back the MARF subcomponents and return the results set.

The following diagram describes high level how MARF could rely on GIPSY to implement a distributed architectural design:

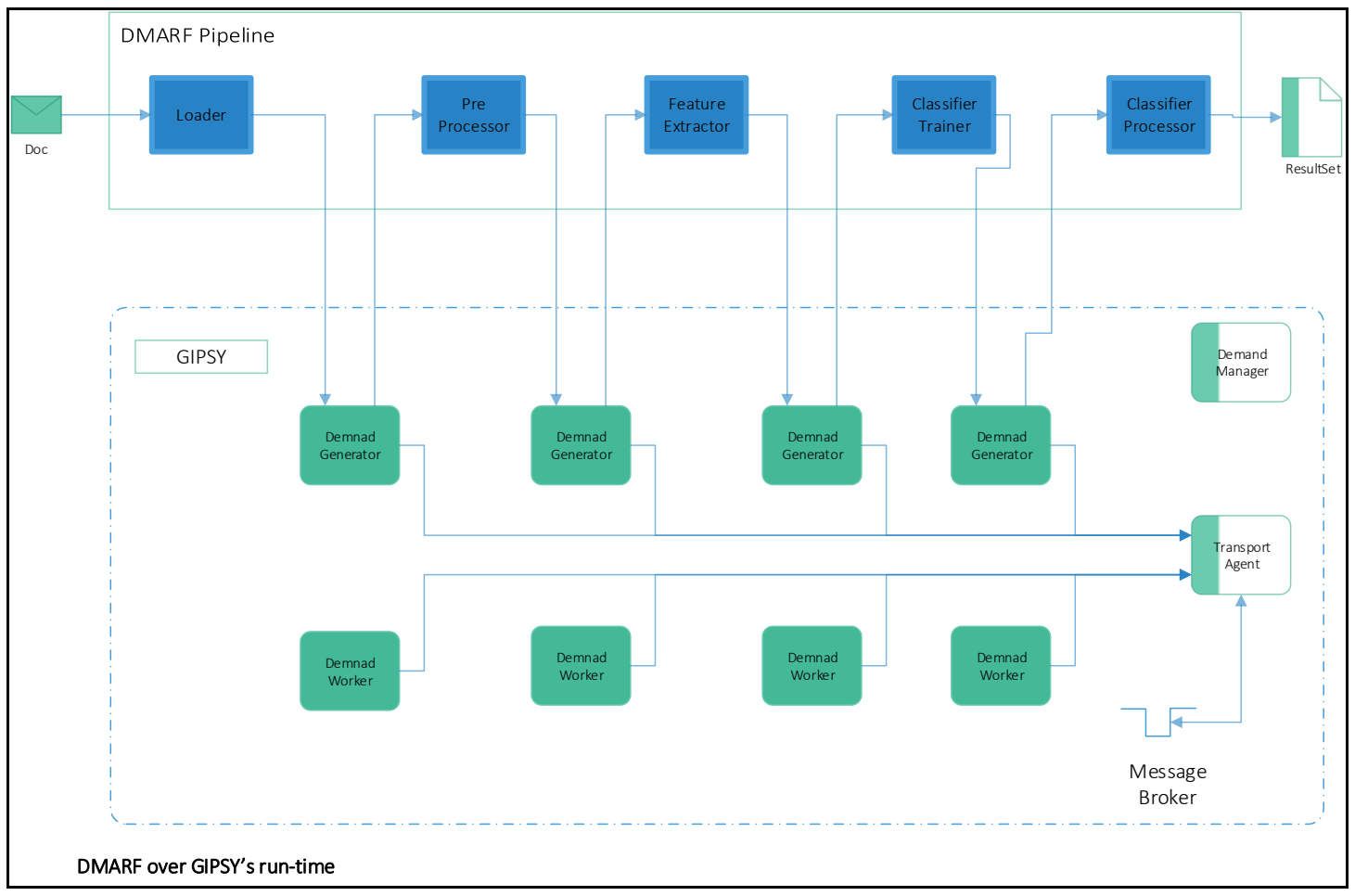

Figure 24 - Fused Architecture of DMARF over GIPSY
(See also the annex for detailed diagram)

The TA is the interface (the contract) and *Message Broker (MB)* would compose the backbone of the system, the message bus.

As shown on the diagram, the subcomponents of the pipeline of the MARF system are associated to a DG from the GIPSY system. Alternatively, the same DG and DW can be used. The architecture can be further extended so that MARF components are using the TA interface - in that manner we ensure that the DG is high available too.

JMS [22] offers a great solution for distributive send messages back and forth and many features are already implemented within the standard message brokers. Yet, JMS is no bullet-proof, and it can fail as well - the solution, in order to eliminate the single point of failure, is adding a secondary JMS instance in the connection factory of the GIPSY TA. The TA would automatically detect if one broker does not respond and fail over to the second one. In such manner, the failover happens dynamically, and no messages are lost.

Similarly, using Web Services as the underlying transport, the DG would invoke the DW. However, the delivery and fault tolerance must be managed at the web service level, on both client and server. The fault tolerance, in such case, would be achieved adding a secondary web service server with a load-balancer in front of both, which will detect whenever a web service goes down. In such case, the WS must implement the reliable messaging protocol (RM), where the TA is called synchronously by the demand generator or asynchronously and a given callback handler is exposed to the TA.

The advantage of WS over JMS for instance is that the WS if called synchronously, the entire process flow can be executed in sync. In contrast, the JMS based implementation, once the message is sent to the broker; the producer of the message does not have control over the response, and does not know if the processing happened or if it was successful.

### *E. Actual Architecture UML Diagrams*

The domain model of both systems, DMARF and GIPSY is quite complex and the scope of the study is to introduce the main concepts of them. Only the interesting classes to our study are depicted in the following diagrams.

Few tools for reverse engineering were briefly evaluated and selected one is used across all study for the consistency. The tools evaluated were:

- Dynamic Interactive Views for Reverse Engineering (DIVER)[4];
- ObjectAid UML Explorer[5];
- Enterprise Architect – Sparx Systems[6];
- ArgoUML[7];
- Visual Paradigm[8];
- ALTOVA UModel[9].
- JDeodorant (see appendix 3 for detailed description).

The selected tools were finally Altova UModel part of Mission Kit 2014, which provides a 30 days trial, McCabe IQ and JDeodorant.

<u>DMARF</u>

DMARF system is composed of the following main modules:

- Loader;
- Preprocessor;
- Feature extractor;
- Classifier.

Loader module contains a set of classes providing features for the initial step of the flow, the loading of the document. The loader provides a common interface and, as specified by the *AudioSampleLoader* interface, and then implemented on multiple classes:

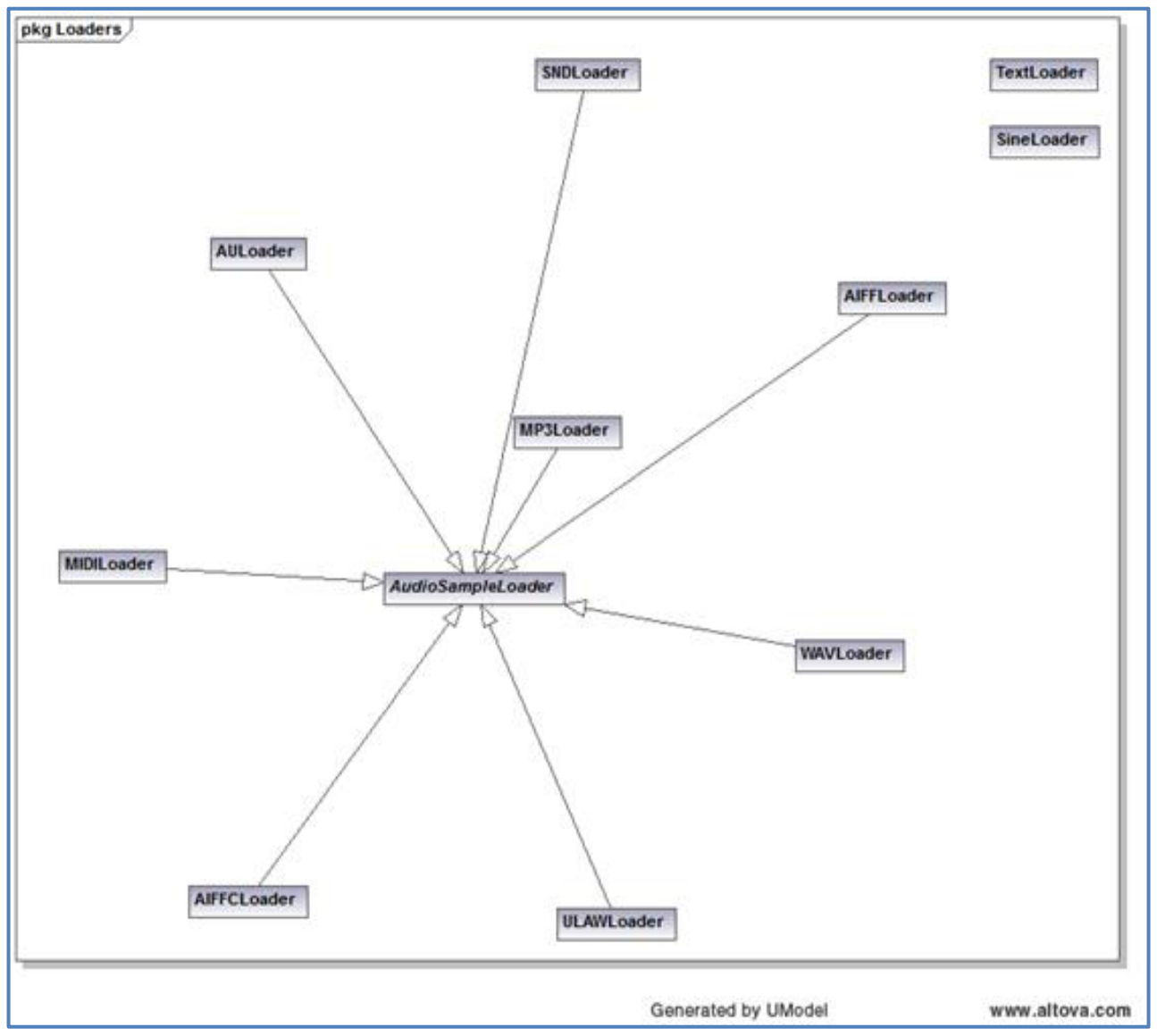

Figure 25 – Loader UML class diagram

Next component in the pipeline is the preprocessor that performs specific operations on the normalized object.

4 University of Victoria, the CHISEL group
5 ObjectAid, Atlanta, GA, USA, http://www.objectaid.com
6 UML Modeling and Lifecycle Tool Suite, Sparx Systems, http://www.sparxsystems.com
7 Tigris ArgoUML, CollabNet, http://argouml.tigris.org/
7
8 Visual Paradigm Code Engineering, Visual Paradigm, Hong Kong
9 Altova UModel, Altova Inc., part of Altova MissionKit 2014, Austria

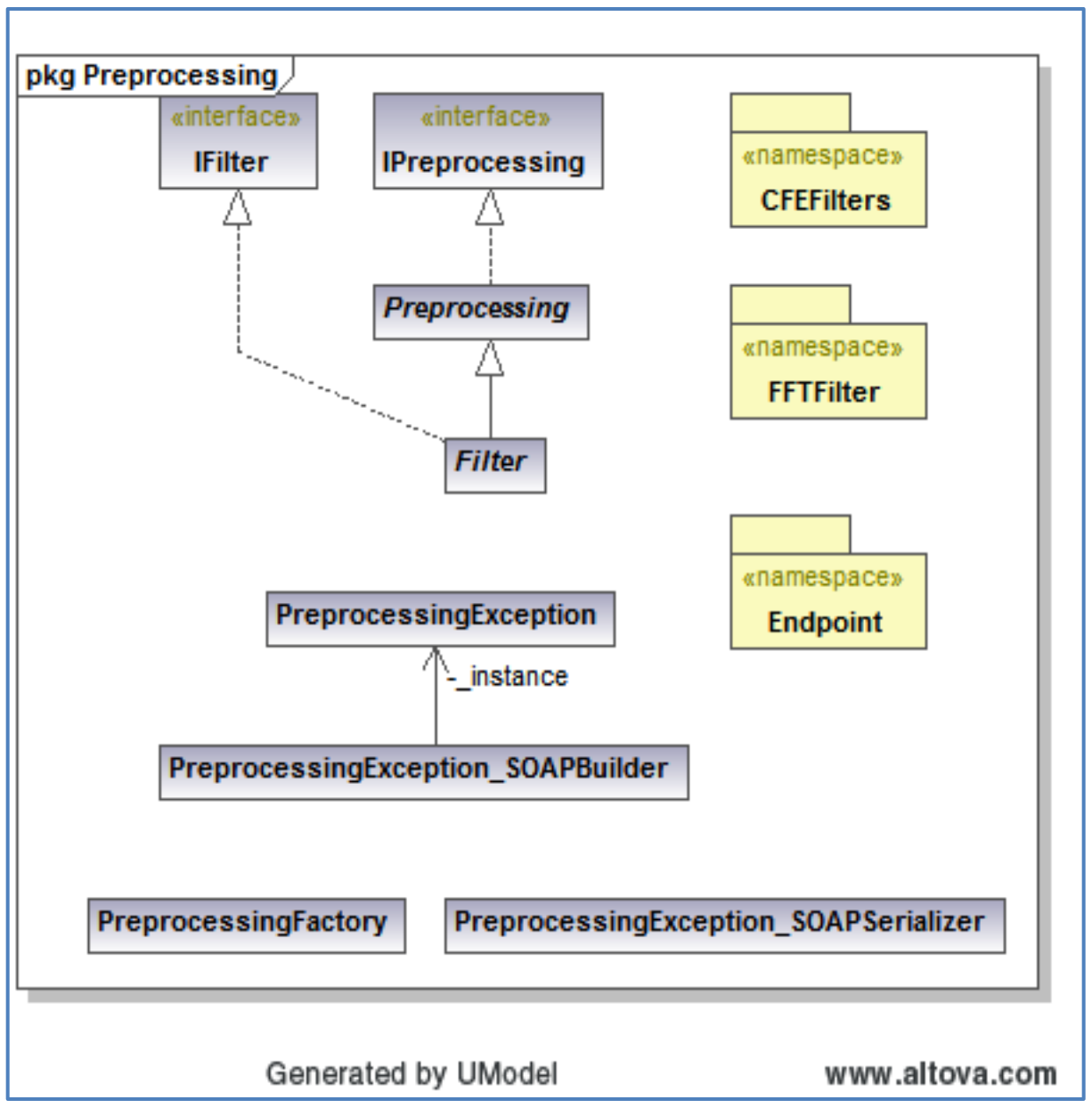

Figure 26 - Preprocessor UML class diagram

Next, in the pipeline is the feature extractor. The feature extractor makes use of few design patterns, such as the factory and aggregate patterns:

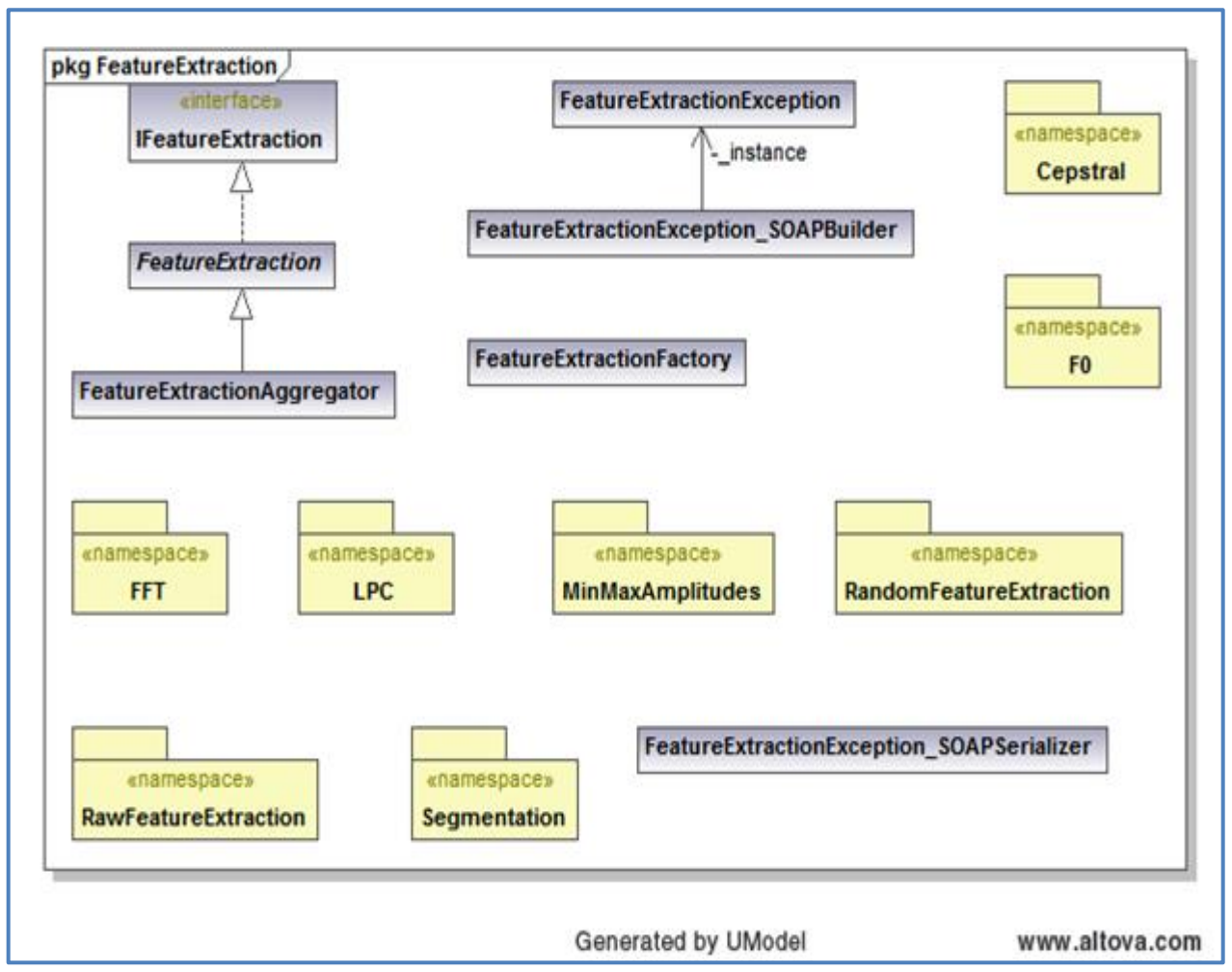

Figure 27 - Feature extraction UML class diagram

Finally, the last component in the pipeline of DMARF system is the natural language processor (NLP) parser:

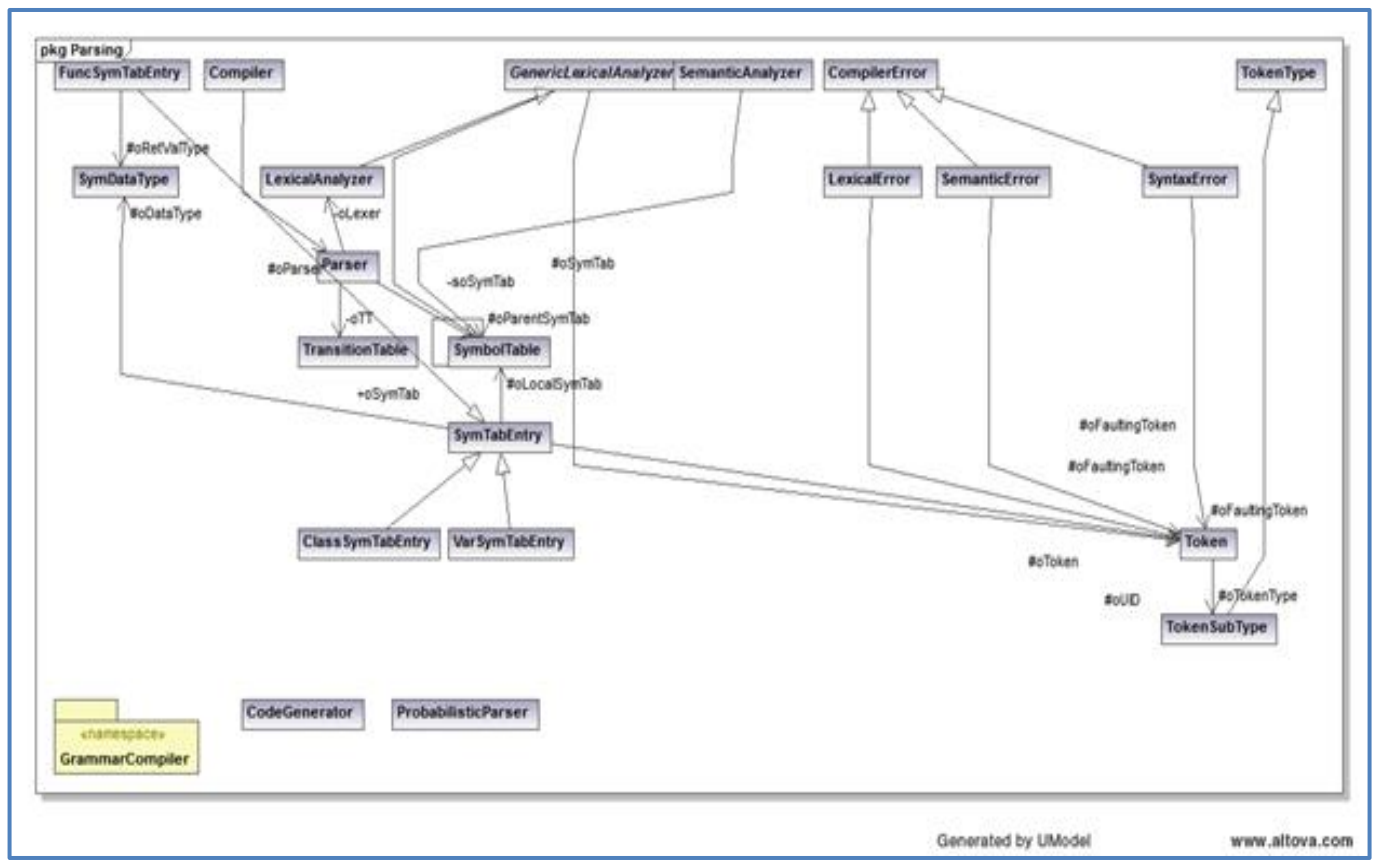

Figure 28 - Natural Language Parser (NLP) UML class diagram

## GIPSY

GIPSY system is composed of the following main packages:

- General Eduction Engine (GEE);
- General Intensional Programming Compiler (GIPC)
- Runtime Interactive Programming Environment (RIPE)

Once again, only certain classes are described on the study. Few components are described and the UML class diagram is shown. Yet, the system is not presented by following the dataflow.

The runtime engine, the general eduction engine (GEE) has been implemented under GEE package is composed of the following and contains the *Executor* class along with few other modules.

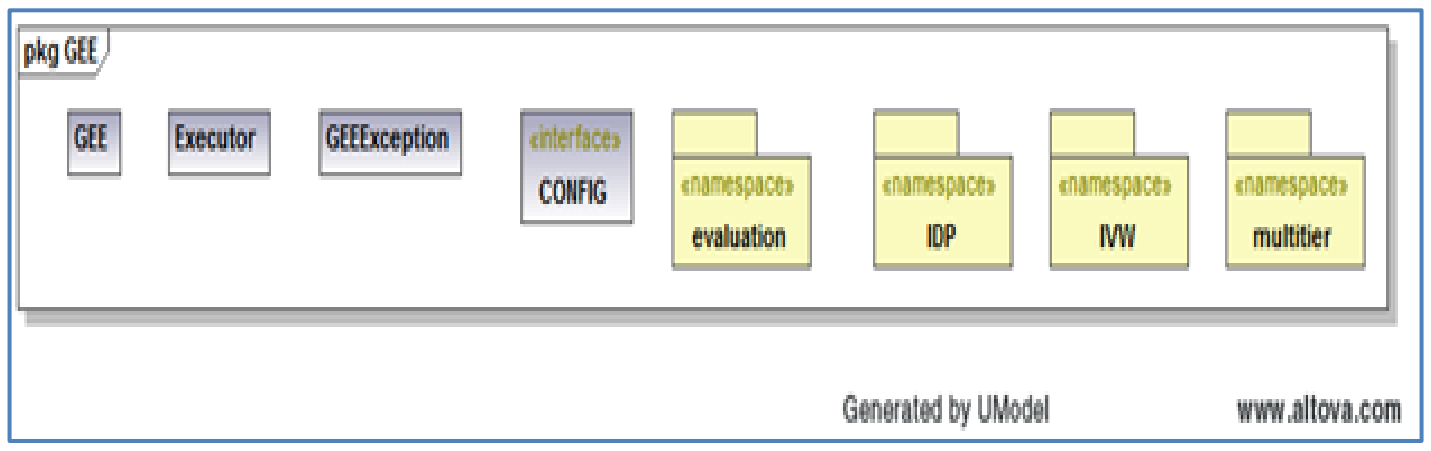

Figure 29 - General Eduction Engine (GEE) UML class diagram

Part of the runtime engine, there is the Intensional Demand Propagator (IDP) that manages the demands and the interface with the transport agent (TA). It includes the Demand Worker and Demand Dispatcher class implementations:

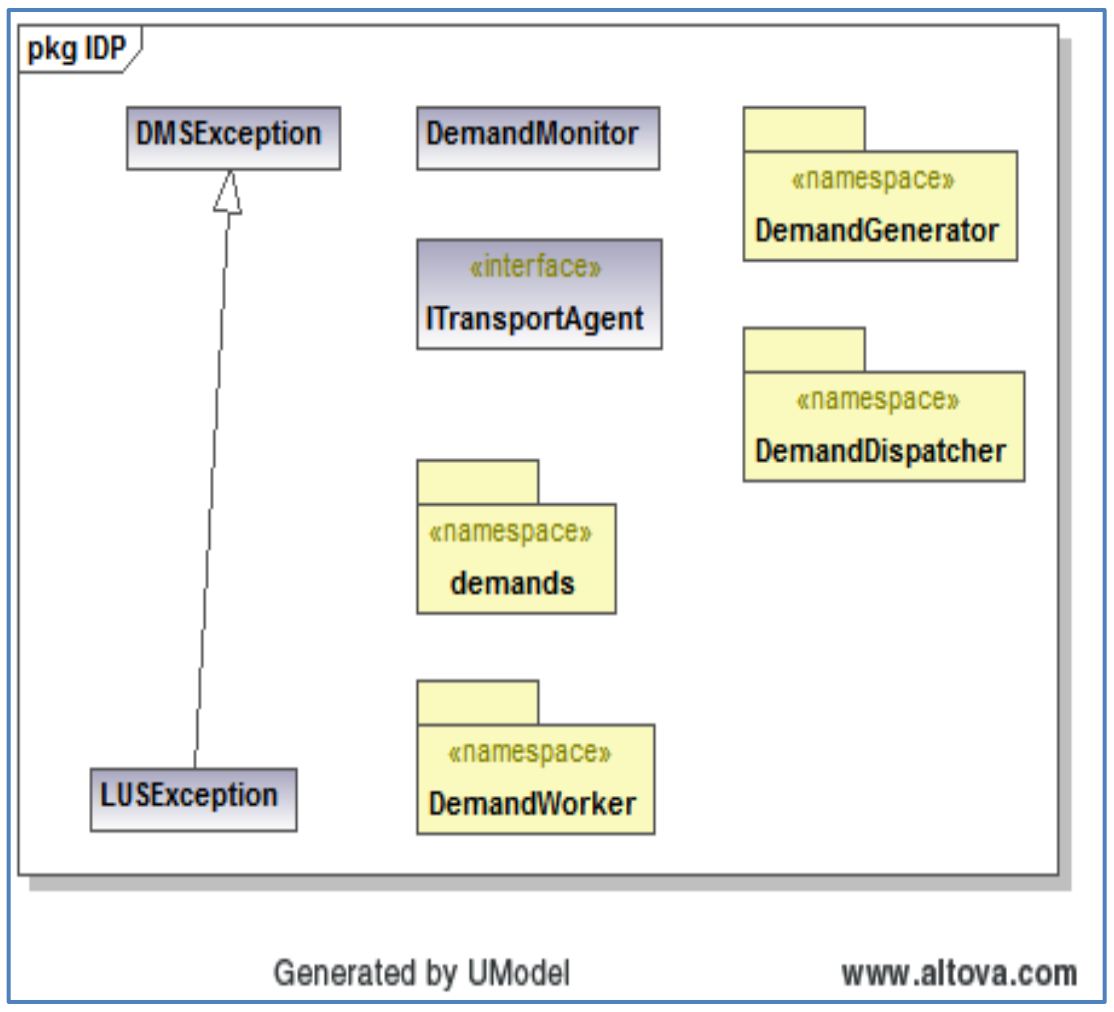

Figure 30 - Demand Propagator UML class diagram

Demand Dispatcher is defined by the *IDemandDispacher* interface, and set as an abstract class:

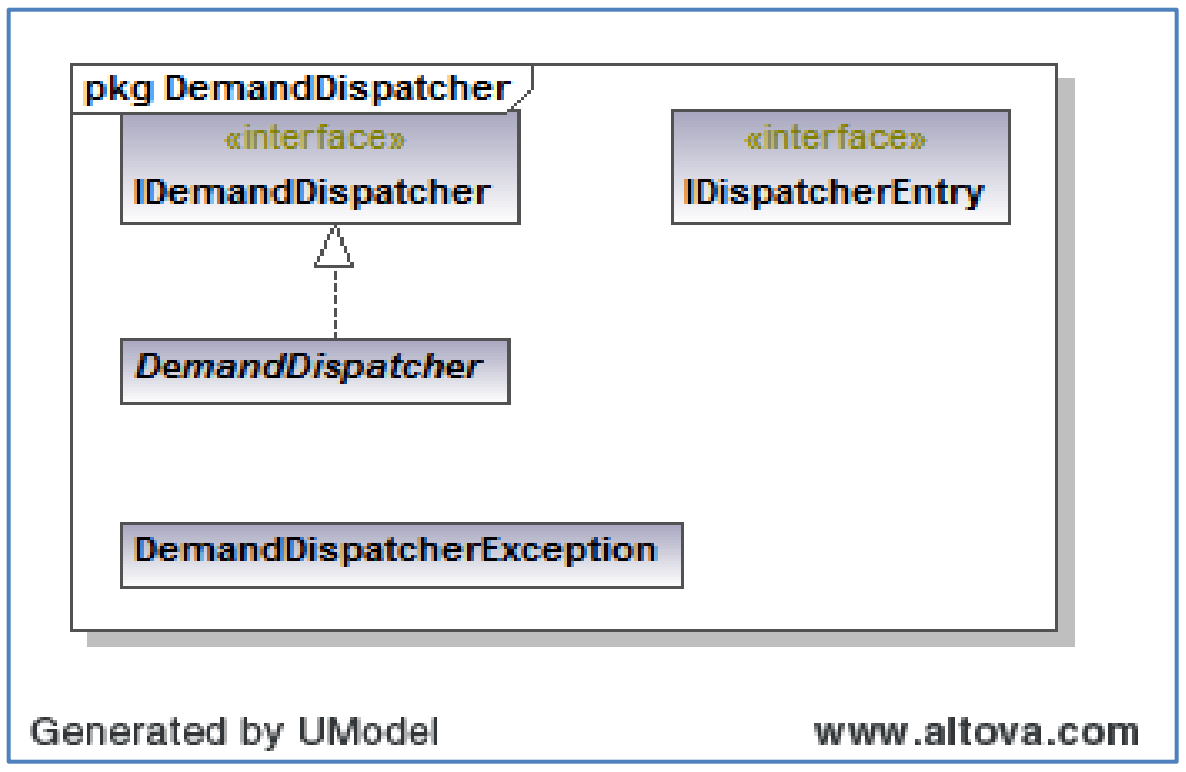

Figure 31 - Demand Dispatcher UML class diagram

Demand Generator actually constructs the demand:

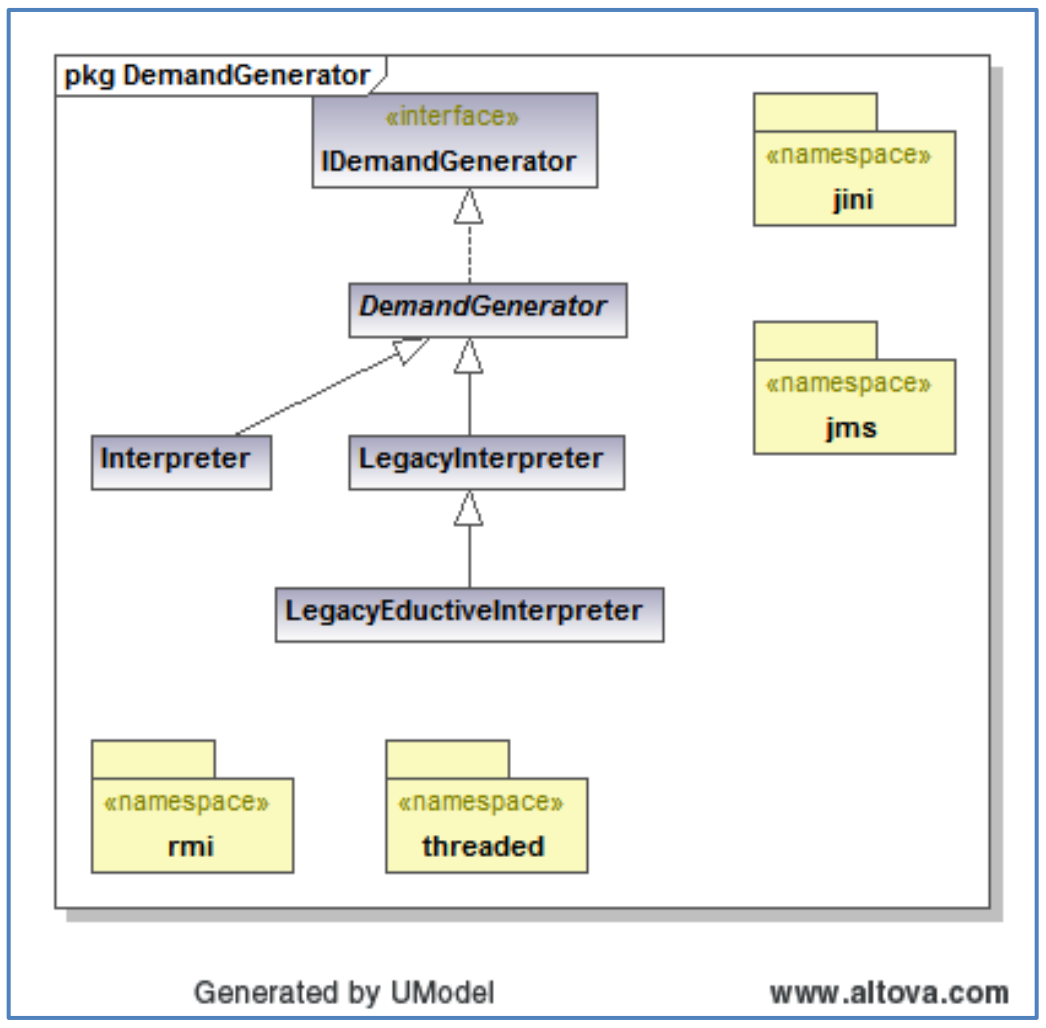

Figure 32 - Demand Generator UML class diagram

Finally, the Demand Worker (DW) component performs the crunching of the data:

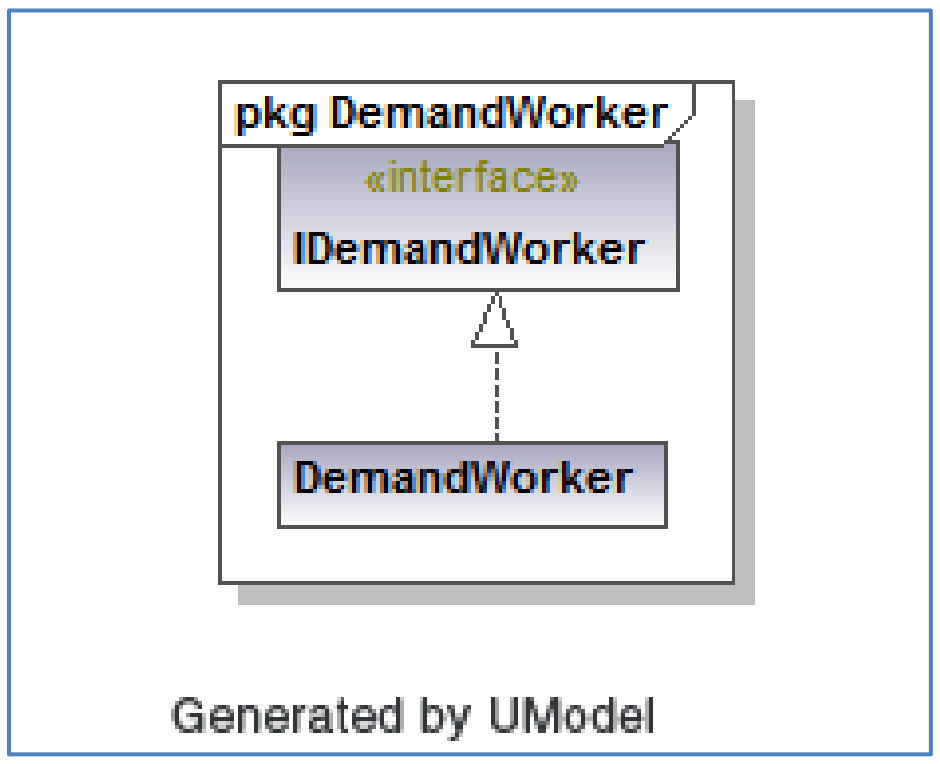

Figure 33 - Demand Worker (DW) UML class diagram

Comparing this UML diagrams that were extracted from the implementation using the reverse engineering approach against the domain model, several discrepancies and gaps were identified. The domain model describes the features at higher level, and it can be mapped in certain cases to modules in the implementation. Individual subcomponents of DMARF for instance, in implementation they were implemented using decorator pattern, which provides a common interface and then easily new responsibilities and functions can be added dynamically or statically. While in the domain model, the conceptual class is represented by one single object, the implementation actually provides a common interface then individual classes implement the interface and provide distinct functionality. There is a gap between the conceptual models, as certain transport related characteristics are not shown on the domain model, although they were inherent to the system.

In GIPSY domain model, the Transport Agent has been depicted from *GIPSYNetwork*, but in the implementation, this is mapped to several modules, classes and interfaces. They abstract in fact the transport manner between the Demand Generator and Demand Worker. The domain model described in the previous section, in fact maps to modules in the implementation, and that's the reason of higher abstraction approach used in the domain model. The implementation specifics are hidden in multiple concrete classes (design classes) and in fact the interface specifies the interface between the components.

For example the *DemandGenerator* class, implements the interface *IDemandGenerator* , all the demand generator must adhere to it. This class hierarchy implements the interpretor function in the GIPSY system:

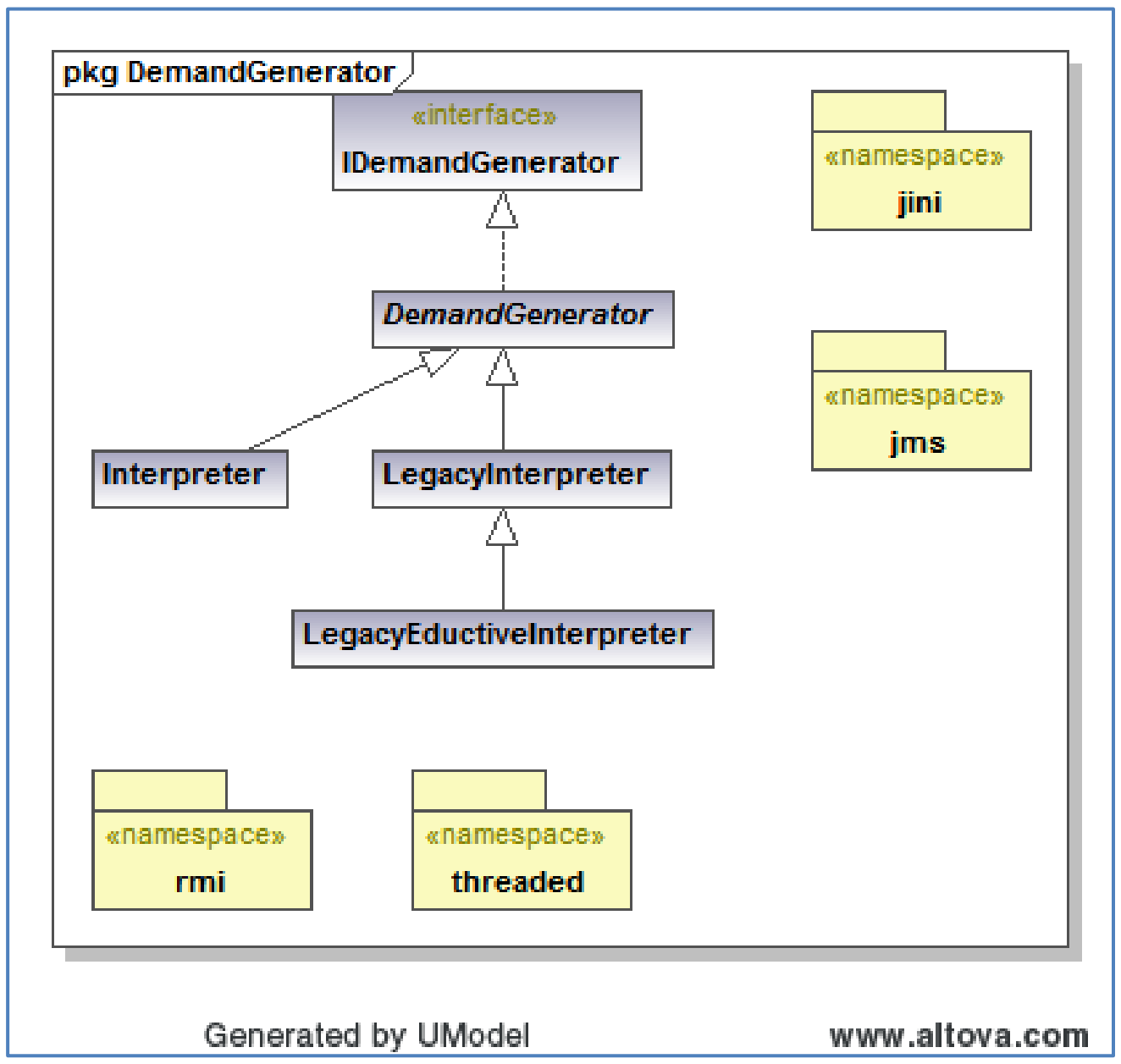

Figure 34 - *DemandGenerator* UML class diagram

This is the class hierarchy:

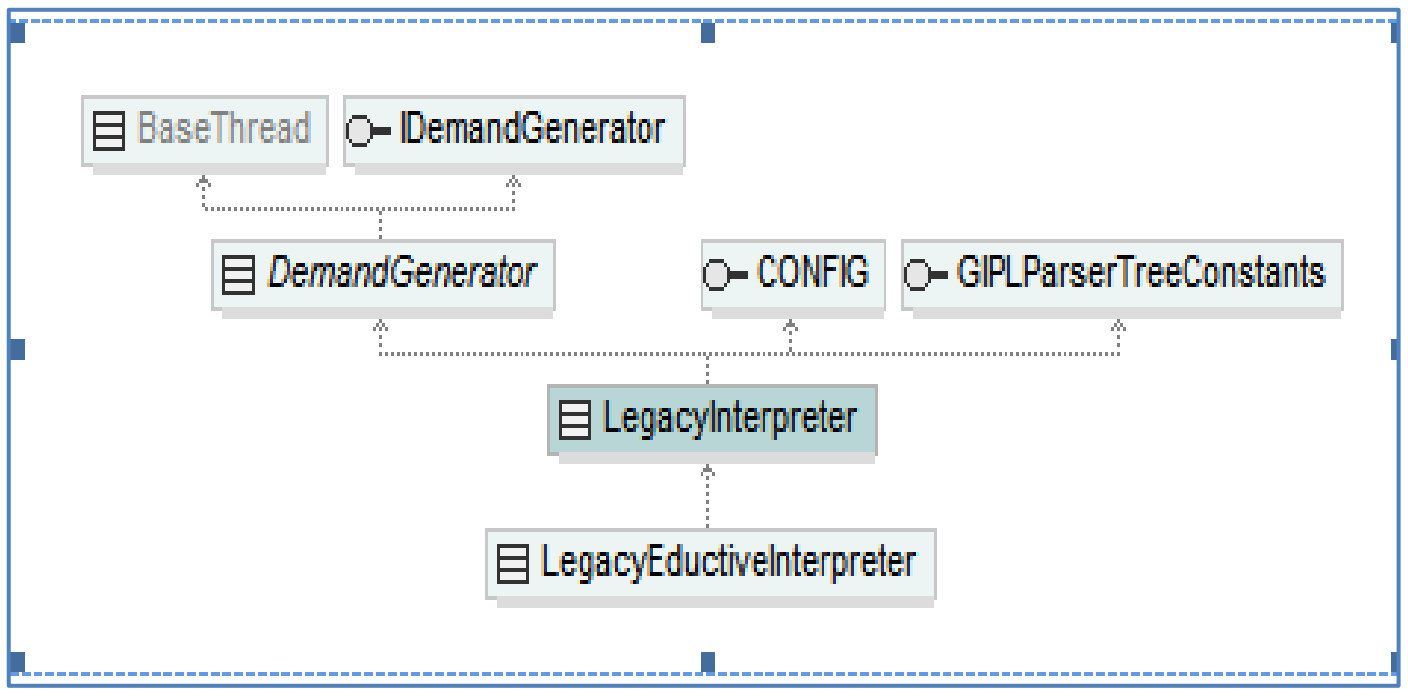

Figure 35 - *DemandGenerator* class hierarchy diagram

```
public interface IDemandGenerator
{
        void setDemandDispatcher(IDemandDispatcher
poDispatcher);
        void setGEER(GIPSYProgram poGEER);
        String generateDemand(int piID, int[] paiContext);
        GIPSYType generateDemandAndWait(IDemand poDemand);
       void generateDemand(IDemand poDemand);
        IDemand getComputedDemand(DemandSignature
poSiganture);
        IDemand getComputedDemand(DemandSignature
poDemandSignature, GEERSignature poGEERSignature);
        IDemand getDemand();
       ……
}
```

Class *DemandGenerator* implements the *IDemandGenerator* and it is abstract, being implemented by several concrete classes:

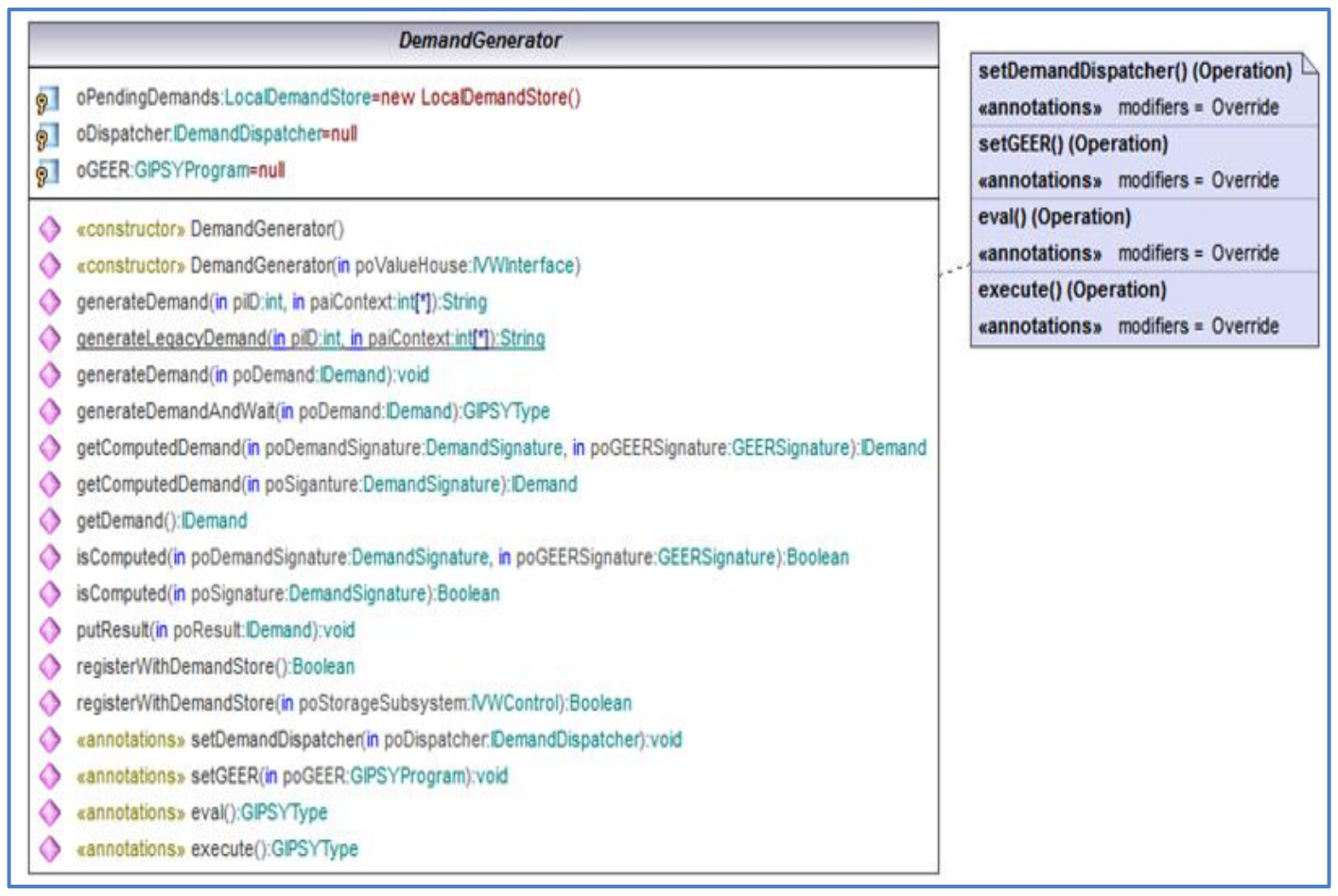

Figure 36 - *DemandGenerator* operations

Class *LegacyInterpreter* extends the previous abstract class:

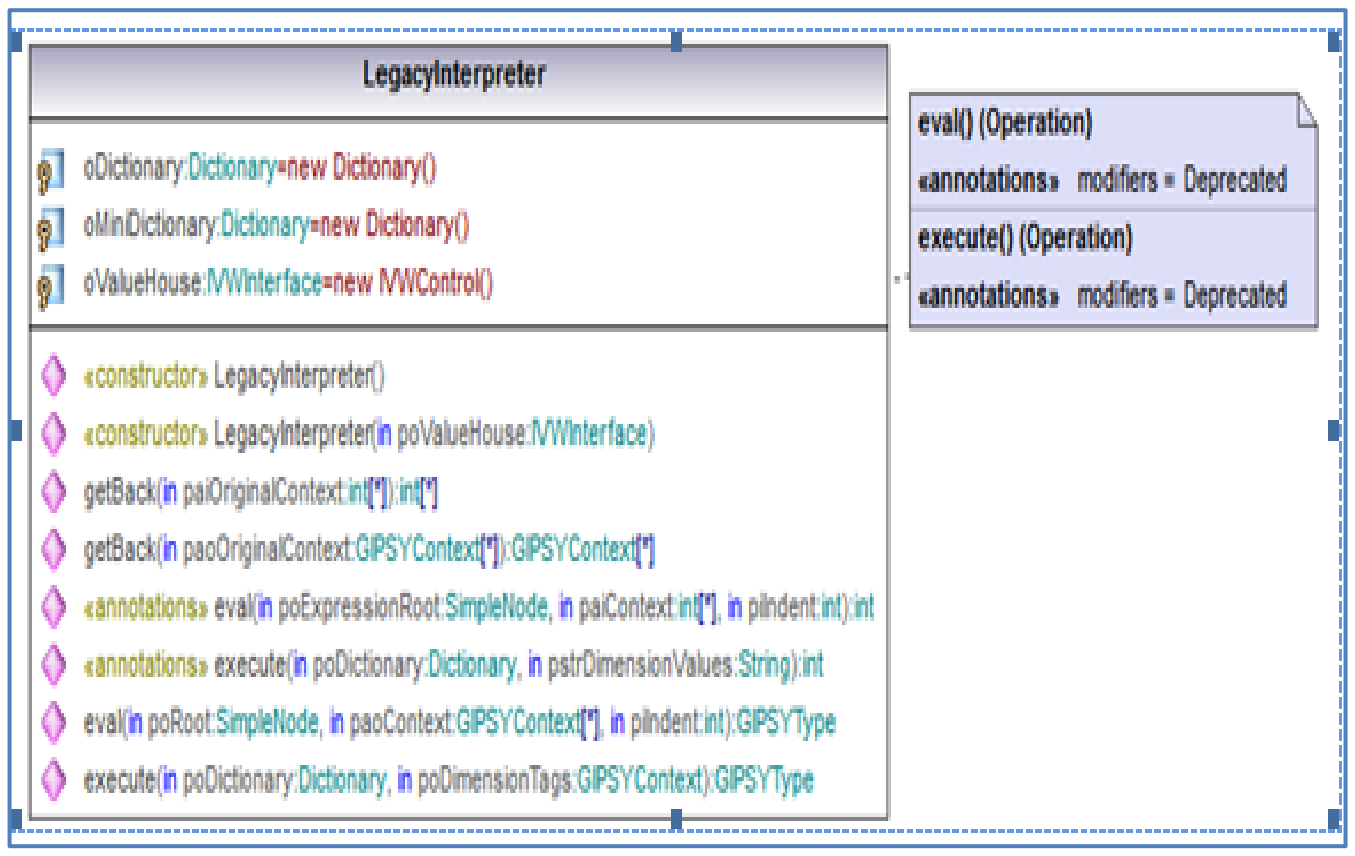

Figure 37 - *LegacyInterpreter* operations

This class is further extended by subclass *LegacyEductiveInterpreter*:

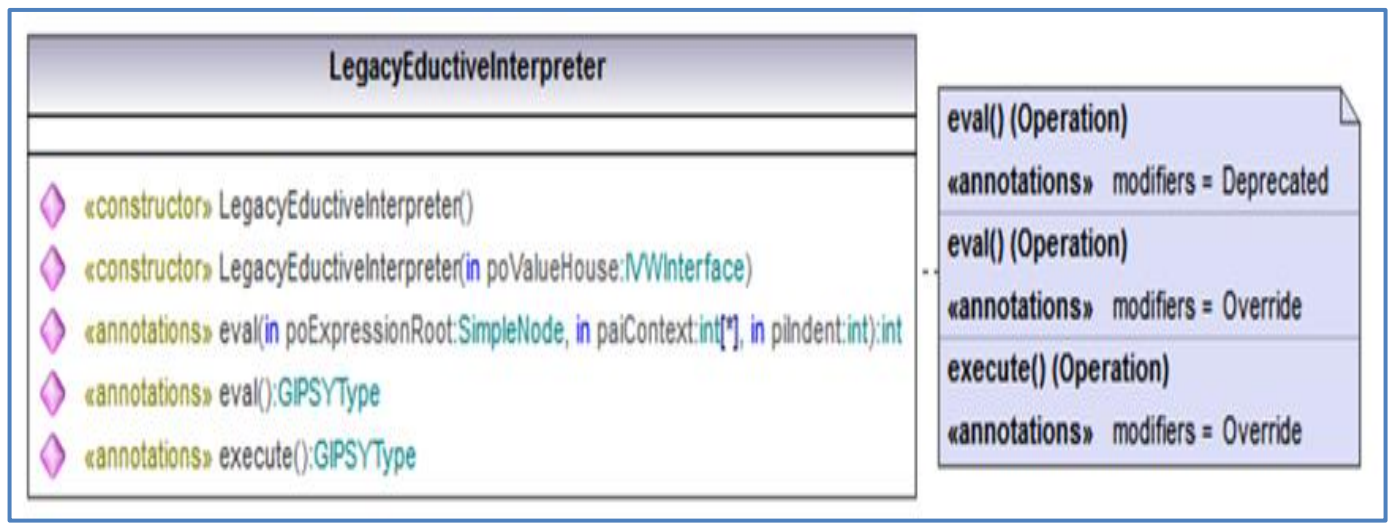

Figure 38 - *LegacyEductiveInterpreter* operations

On DMARF system, for example the feature extractor is composed of certain interfaces and classes as described in the following diagram:

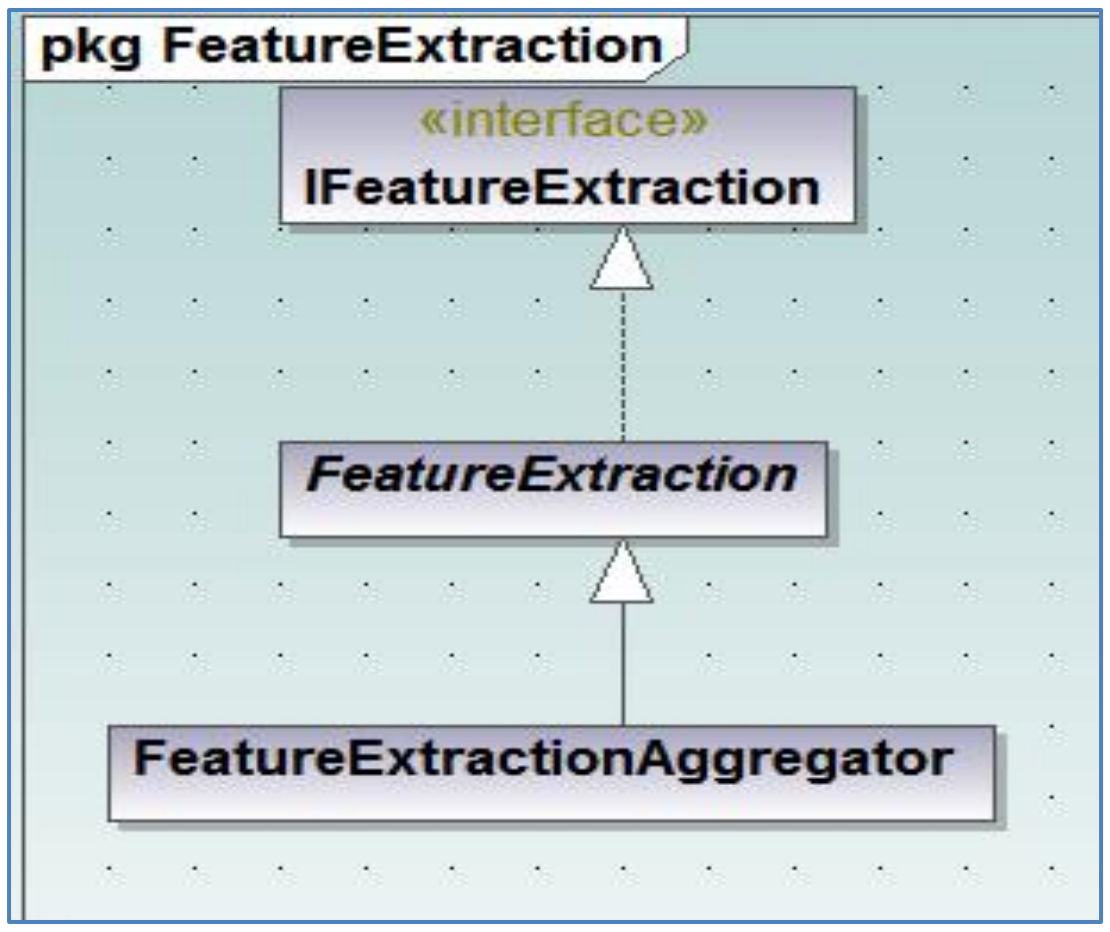

Figure 39 - *FeatureExtraction* class hierarchy and interface

The interface *IFeatureExtraction* defines the specification of the feature extractor. The interface is then implemented in the abstract class *FeatureExtraction* and finally in the concrete class *FeatureExtractionAggregator*, that implements the Aggregator design pattern. The aggregator clones the incoming preprocessed sample for each feature extractor and runs each module in a separate thread. At the end, the results of each thread are collected in the same order as specified and returned as a concatenated feature vector.

The abstract class *FeatureExtraction* provides implementation of certain features:

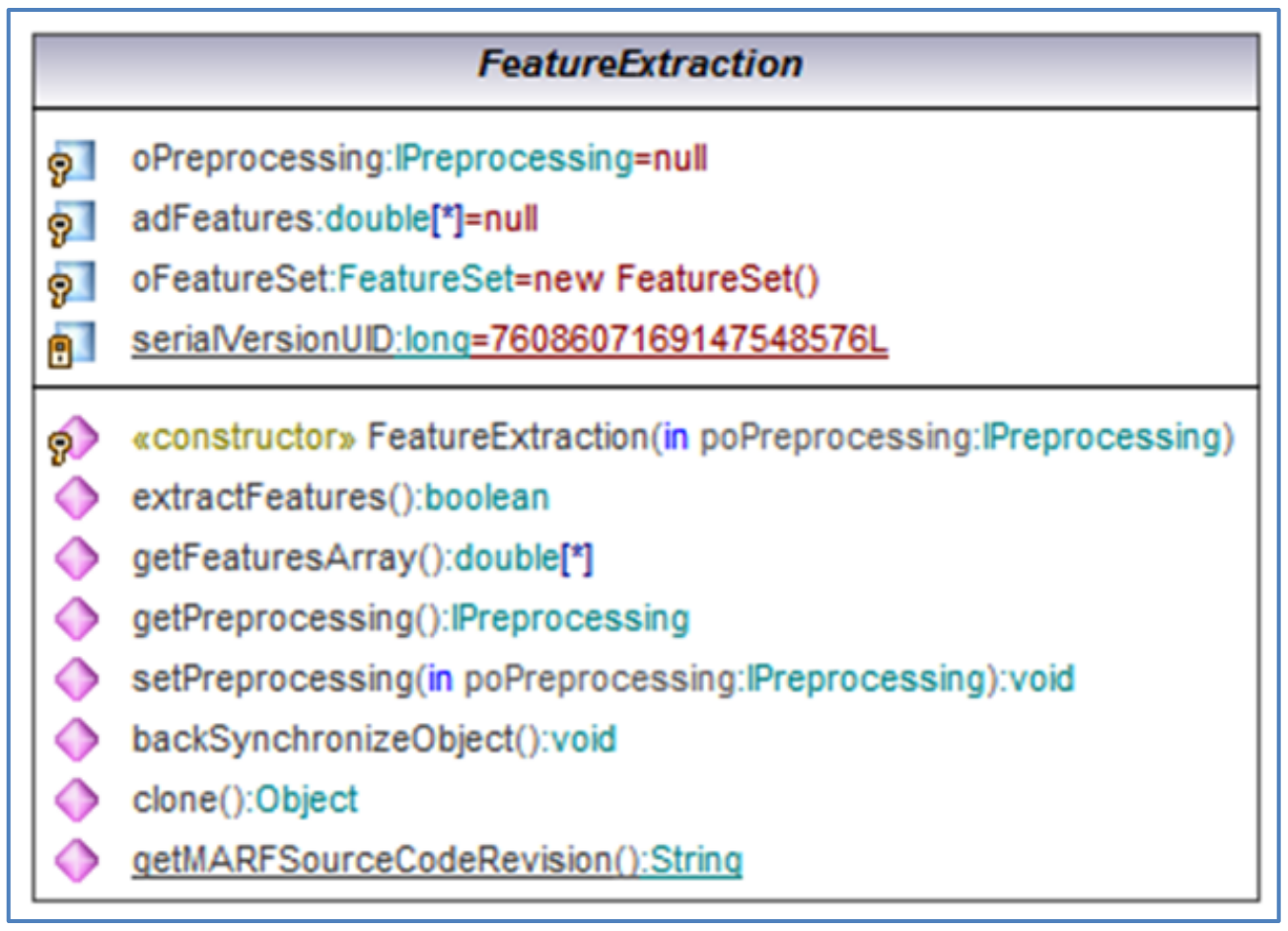

Figure 40 - *FeatureExtraction* class methods and attributes

Finally, the aggregator *FeatureExtractionAggregator* class which is a concrete object:

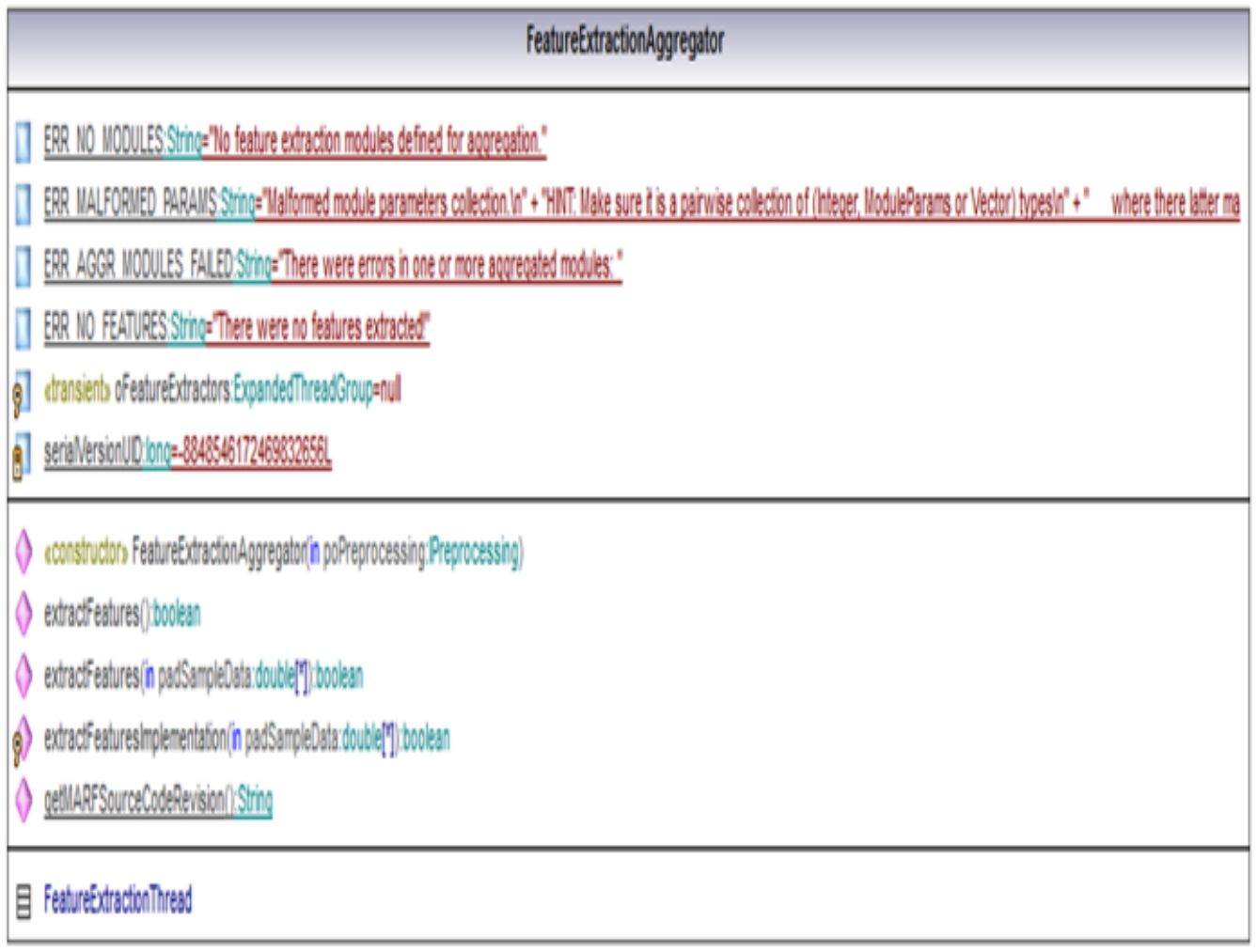

Figure 41 - *FeatureExtractionAggregator* methods and attributes

Following section, based on the actual design described herein, some refactoring is suggested

## V. METHODOLOGY

### A. Refactoring

#### a. Identification of "code smells" and system level refactoring

DMARF

In the DMARF source code, between the classes identified by McCabe tool as being problematic, the class *NeuralNetwork* was found with high complexity on the method *NeuralNetwork.createLinks(org.w3c.dom.Node):*

| Modules | Cyclomatic Complexity | Essential Complexity | Design Complexity | Lines |
|---|---|---|---|---|
| NeuralNetwork.createLinks(org.w3c.dom.Node) | 28 | 18 | 28 | 13 |

The method creates input and output links for a Document Object Model (DOM) XML node.

First of, the method uses hard coded values, and the DOM object seems to be parsed by a series of conditions:

```
if (strName.equals("layer")) { .... }
```

Second, the method is quite complex and it spawns over 131 lines of code with a cyclomatic complexity of 28 (McCabe recommends under 10 for maintainability). It is unknown at this point the exact purpose of the DOM object, but probably by using Java for XML binding (JAXB) and Java for XML Parser (JAXP), an XML schema could be defined then the DOM object can be actually converted to an Plain Old Java Object (POJO), which will be much easier to validate, marshal and unmarshal (serialize and deserialize) the object representation. In addition, the schema can be defined beforehand and the XML document validated as soon as it gets loaded, against the XSD, which would eliminate a semnificative number of lines handling exceptions. Performing a search on the usage of the class, it seems to be called only by this method:

```
public final void initialize(final String pstrFilename, final boolean pbValidateDTD)
{
    ……
    Debug.debug("Parsing XML file...");
     Document oDocument = oBuilder.parse(new File(pstrFilename));

    // Add input layer
    this.oLayers.add(this.oInputs);

    // Build NNet structure
    Debug.debug("Making the NNet structure...");
    buildNetwork(oDocument);

    // Add output layer
    this.oLayers.add(this.oOutputs);

    // Fix inputs/outputs
    Debug.debug("Setting the inputs and outputs for each Neuron...");
    this.iCurrenLayer = 0;
    createLinks(oDocument);
    …….
}
```

JAXB provides extremely good support for binding XML to Java and reverse. Command line tools can easily generate XML Schema Definitions (XSD) from POJO and similarly, from POJO to XSD. If the DOM node does not change the structure, JAXB could be a better alternative. Here is an example of DOM vs JAXB parser[10], for a random class *PersonList.class* (the comparison is only shown for information purpose of the size of the program, there are others factors to take into account, such as memory usage, performance, etc)*:*

JAXB parser

```
public void testUnMarshallUsingJAXB() throws Exception
{
        JAXBContext jc = JAXBContext.newInstance(PersonList.class);
         Unmarshaller unmarshaller = jc.createUnmarshaller()
        PersonList obj = (PersonList)unmarshaller.unmarshal(new
File(filename));
}
```

DOM Parser

```
public void testParsingWithDom() throws Exception
{
        DocumentBuilderFactory domFactory =
DocumentBuilderFactory.newInstance();
    DocumentBuilder builder = domFactory.newDocumentBuilder();
         Document doc = builder.parse(filename);
       List personsAsList = new ArrayList();
       NodeList persons = doc.getElementsByTagName("person");
              for (int i = 0; i <persons.getLength(); i++)
               {
              Element person = (Element)persons.item(i);
               NodeList children = (NodeList)person.getChildNodes();
              Person newperson = new Person();
              for (int j = 0; j < children.getLength(); j++){
              Node child = children.item(j);
               if (child.getNodeName().equalsIgnoreCase("id")) {
                          newperson.setId(child.getNodeValue());
                 } else if (child.getNodeName().equalsIgnoreCase("name")) {
             newperson.setName(child.getNodeValue());
          }
       }
       personsAsList.add(newperson);
    }
}
```

And the complexity of the code increases with the number of arguments of the test class, while on JAXB the size is constant, regardless the complexity of the class to parse.

## GIPSY

One of the bad smells identified by McCabe IQ as risky module is within *GEE.java* class, on the method *startServices()*, that has the high cyclomatic complexity of 20 and the essential complexity of 9 as measured by the tool. The McCabe risk criteria are evaluated as following:

| | |
|---|---|
| High: | Essential Complexity > 4 |
| Medium: | Cyclomatic Complexity > 10 |

10 Staveley, Alex, Dublin Tech, blogspot.ca, JAXB, SAX, DOM comparative performance, 2011

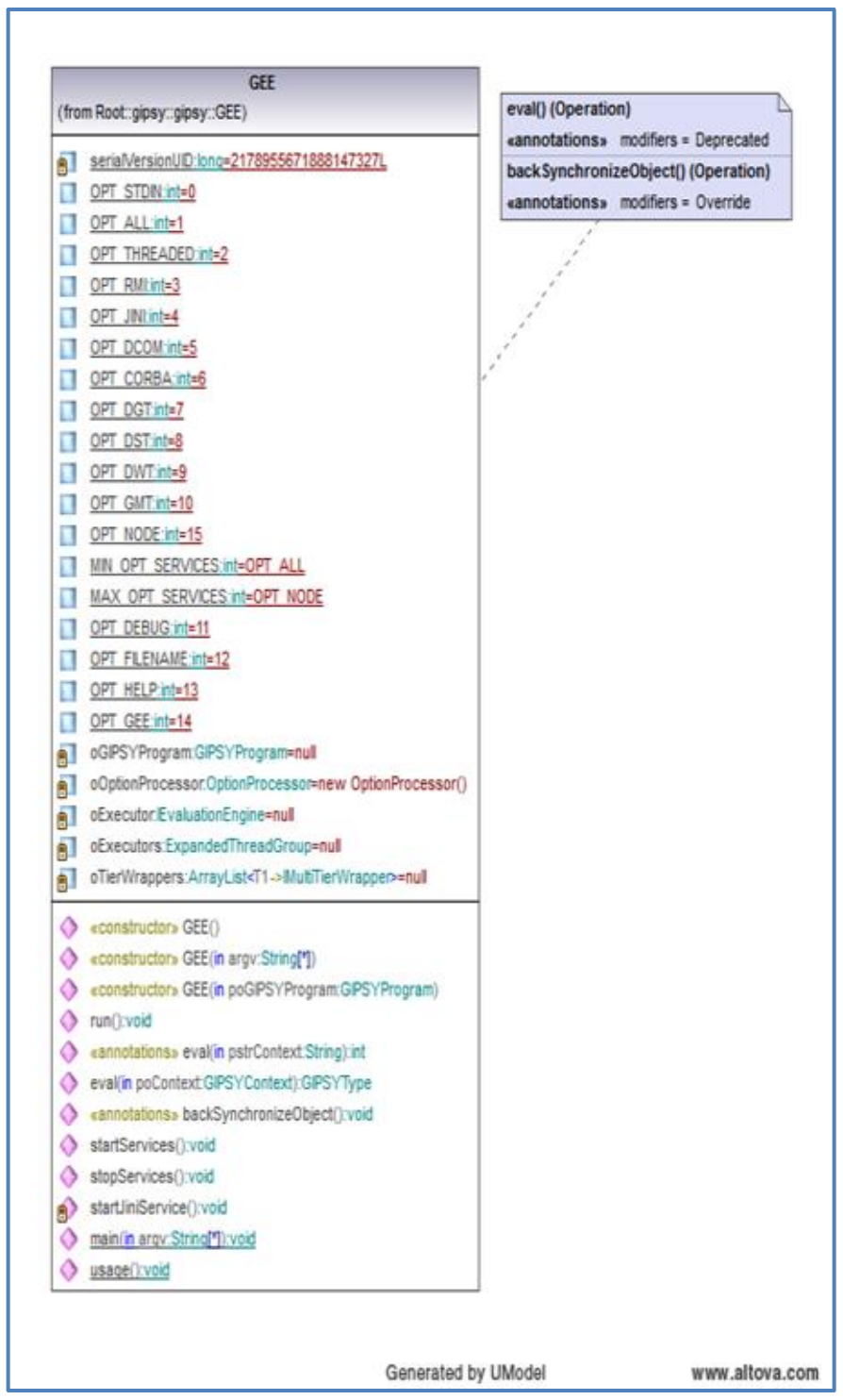

Figure 42 - Problematic class GEE.java UML class diagram

Another change in the architecture can be implemented for the Transport Agent using the Facade design pattern. Facade behaves as a door to a complex system and provides one single interface to the subsystem. The details of the transport can be all hidden behind this interface. Facade is not the only entry point to the sub-system but is a convenient point of communication to the subsystem and client can always have the direct access to the subsystem.

This method helps developing the subsystem independently without affecting the clients using them.

Facade pattern class diagram[11] will look like this:

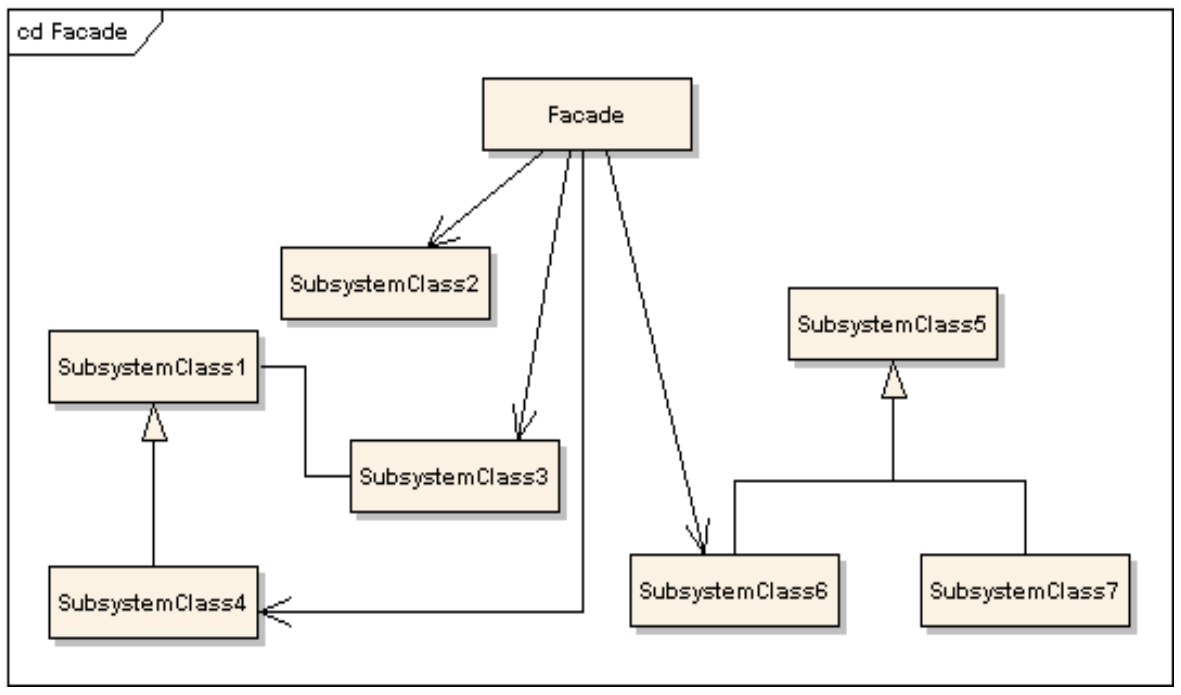

Figure 43 - Facade design pattern diagram

11 Design pattern in simple example

Several other few examples (some reported by FindBugs[12] and EasyPMD[13] tools):

FindBugs

EasyPMD

Few class level refactoring:

1. Test for floating point equality on *marf.Classification.Distance.Distance.classify(doubel[]):*

```
if(dCurrentDistance == dMinDistance) {
        Debug.debug("This distance had happened before!");
}
```

However, this is helper test for debugging, hence this is less important.

2. Dead store to local variables in *marf.Preprocessing.CFEFilters.CFEFilter*

```
double dUpperBound = (2 * Math.PI) - (2 * Math.PI / padSample.length);
```

3. Inefficient invocation of String constructor, on *marf.Stats.WordStats*

```
public WordStats(final WordStats poWordStats)
{
    super(poWordStats);
    this.strLexeme = new String(poWordStats.strLexeme);
}
```

12 Eclipse plugin, FindBugs in Java, Bill Pugh and David Hovemeyer, http://findbugs.sourceforge.net/
13 EasyPMD NetBeans plugin, Gianluca Costa, http://gianlucacosta.info/

*b. Specific refactoring to be implemented in PM4 – planned refactoring*

## Weighted Method per class (WMC)

WMC is the count of methods implemented in a class. It is recommended WMC does not exceed the value 14. Using the McCabe tool, we found that few of the class has high WMC value (>14) and one of them is *StorageManager* class which has the WMC value of 39. So by doing refactoring we can maintain the WMC value of the *StorageManager* class to threshold value i.e. 14. The following four methods are categorized under long method bad smells:

**public abstract class** StorageManager **implements** IStorageManager

```java
public synchronized void dumpBinary()
throws StorageException
{
    try
    {
        ObjectOutputStream oOOS = null;

        if(this.strFilename == null || "".equals(this.strFilename))
        {
            // Assume STDOUT if no filename specified.
            oOOS = new ObjectOutputStream(System.out);
        }
        else
        {
            FileOutputStream oFOS = new FileOutputStream(this.strFilename);
            oOOS = new ObjectOutputStream(oFOS);
        }
```

```java
public synchronized void dumpGzipBinary()
throws StorageException
{
    try
    {
        GZIPOutputStream oGZOS = null;

        if(this.strFilename == null || "".equals(this.strFilename))
        {
            // Assume STDOUT if no filename specified.
            oGZOS = new GZIPOutputStream(System.out);
        }
        else
        {
            FileOutputStream  oFOS = new FileOutputStream(this.strFilename);
            oGZOS = new GZIPOutputStream(oFOS);
        }
```

```java
public synchronized void restoreBinary()
throws StorageException
{
    try
    {
        ObjectInputStream oOIS = null;

        if(this.strFilename == null || "".equals(this.strFilename))
        {
            // Assume STDOUT if no filename specified.
            oOIS = new ObjectInputStream(System.in);
        }
        else
        {
            FileInputStream oFIS = new FileInputStream(this.strFilename);
            oOIS = new ObjectInputStream(oFIS);
        }
```

```java
public synchronized void restoreGzipBinary()
throws StorageException
{
    try
    {
        GZIPInputStream oGZIS = null;

        if(this.strFilename == null || "".equals(this.strFilename))
        {
            // Assume STDOUT if no filename specified.
            oGZIS = new GZIPInputStream(System.in);
        }
        else
        {
            FileInputStream oFIS = new FileInputStream(this.strFilename);
            oGZIS = new GZIPInputStream(oFIS);
        }
```

The following 2 methods are under type checking bad smells:

```java
public synchronized void dump()
throws StorageException
{
    switch(this.iCurrentDumpMode)
    {
        case DUMP_GZIP_BINARY:
            dumpGzipBinary();
            break;

        case DUMP_CSV_TEXT:
            dumpCSV();
            break;

        case DUMP_BINARY:
            dumpBinary();
            break;

        case DUMP_XML:
            dumpXML();
            break;

        case DUMP_HTML:
            dumpHTML();
            break;

        case DUMP_SQL:
            dumpSQL();
            break;

        default:
            throw new StorageException("Unsupported dump mode: " + this.iCurrentDumpMode);
    }
```

```java
public synchronized void restore()
throws StorageException
{
    switch(this.iCurrentDumpMode)
    {
        case DUMP_GZIP_BINARY:
            restoreGzipBinary();
            break;

        case DUMP_CSV_TEXT:
            restoreCSV();
            break;

        case DUMP_BINARY:
            restoreBinary();
            break;

        case DUMP_XML:
            restoreXML();
            break;

        case DUMP_HTML:
            restoreHTML();
            break;

        case DUMP_SQL:
            restoreSQL();
            break;

        default:
            throw new StorageException("Unsupported dump mode: " + this.iCurrentDumpMode);
    }
```

Attributes:

| |
|---|
| *iCurrentDumpMode : int*<br>*strFilename:String*<br>*oObjectToSerialize: Serializable*<br>*bDumpOnNotFound:Boolean.* |

UML Class diagram of *StorageManager* class:

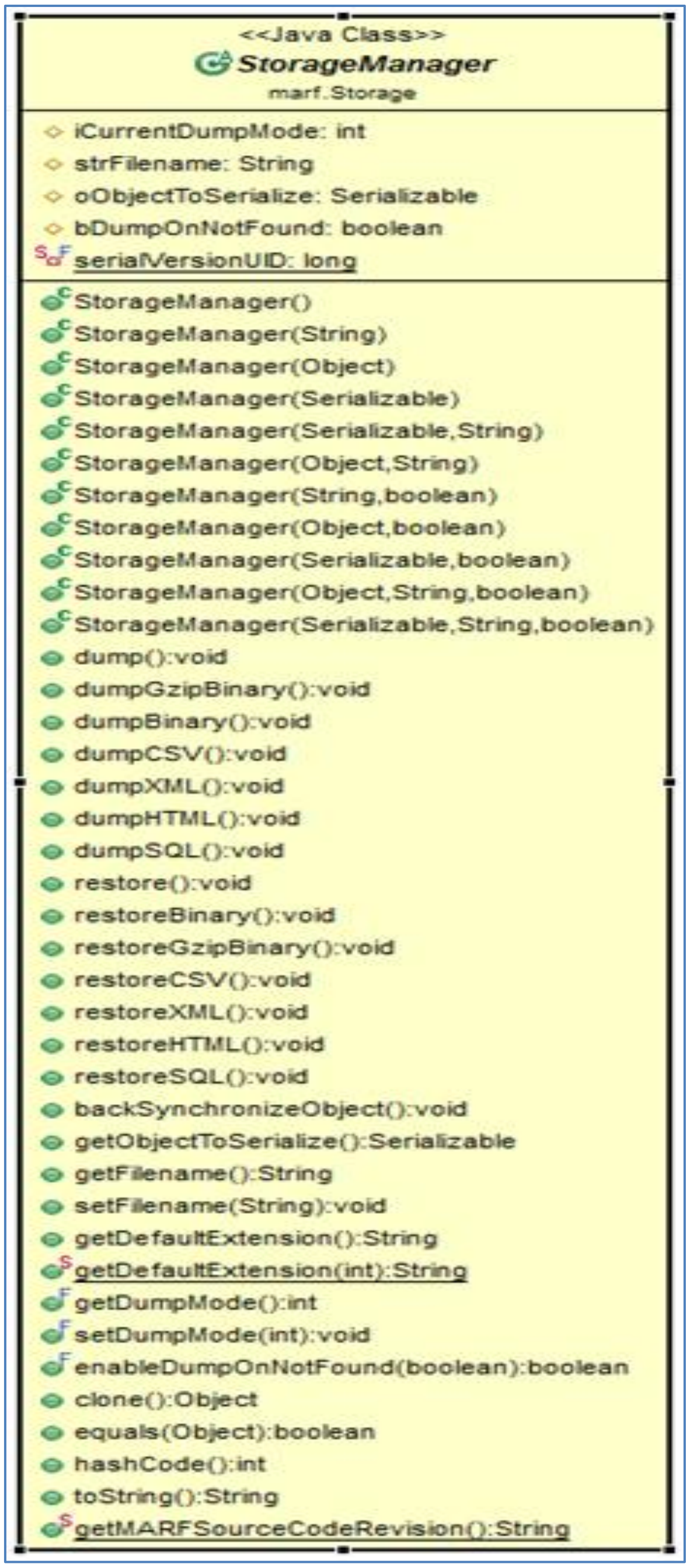

Figure 44 – *StorageManager* UML class diagram.

The unit test case for the StorageManager Class is found in the test.java class.

```
import marf.Storage.StorageManager;
…..
public class test
{
…..
public static String getMARFSourceCodeRevision()
{
return "$Revision: 1.59 $";
}
……
private static void revision()
```

```
{
// marf.Storage
…..
StorageManager.class.getName() + ": " +
StorageManager.getMARFSourceCodeRevision() + "\n" +
…..
  }
}
```

## GIPSY

Making use of JDeodorant tool we have identified the bad smells in *GIPSYGMTOperator* class. Upon running the metrics for the above class we got the results. WMC, total number of attributes, number of public attributes, total number of public methods, and number of base classes are out of bound values. By doing the refactoring we can maintain the threshold value for the above metrics thereby the problematic class can be resolved well.

*Public void updateStateProperties(Object obj)*

```
try
{

    if (obj instanceof String)
    {
        String stateName = (String) obj;
        GIPSYTier oGIPSYTier = GlobalInstance.getInstance()
                .getGIPSYTierByName(stateName);
        if (oGIPSYTier != null)
        {
            oActionsLog.showStateProperties(oGIPSYTier);
        }
    }
    else if (obj instanceof GIPSYTier)
    {
        GIPSYTier oGIPSYTier = (GIPSYTier) obj;
        if (oGIPSYTier != null)
        {
            oActionsLog.showStateProperties(oGIPSYTier);
        }
    }
    this.repaint();
```

*Public void startInstance()*

```
try
{
    // -- Start the GMT first.
    System.out.println("Starting the GMT...");
    oGIPSYEntityController.startGMTNode();
    System.out.println("GMT started successfully...");
    oGIPSYEntityController.allocateTier();

}
catch (Exception e)
{
    System.err
            .println("An error occured while trying to start instances.\n Error"
                    + e.getMessage());
}
```

Attributes in the class are as follows:

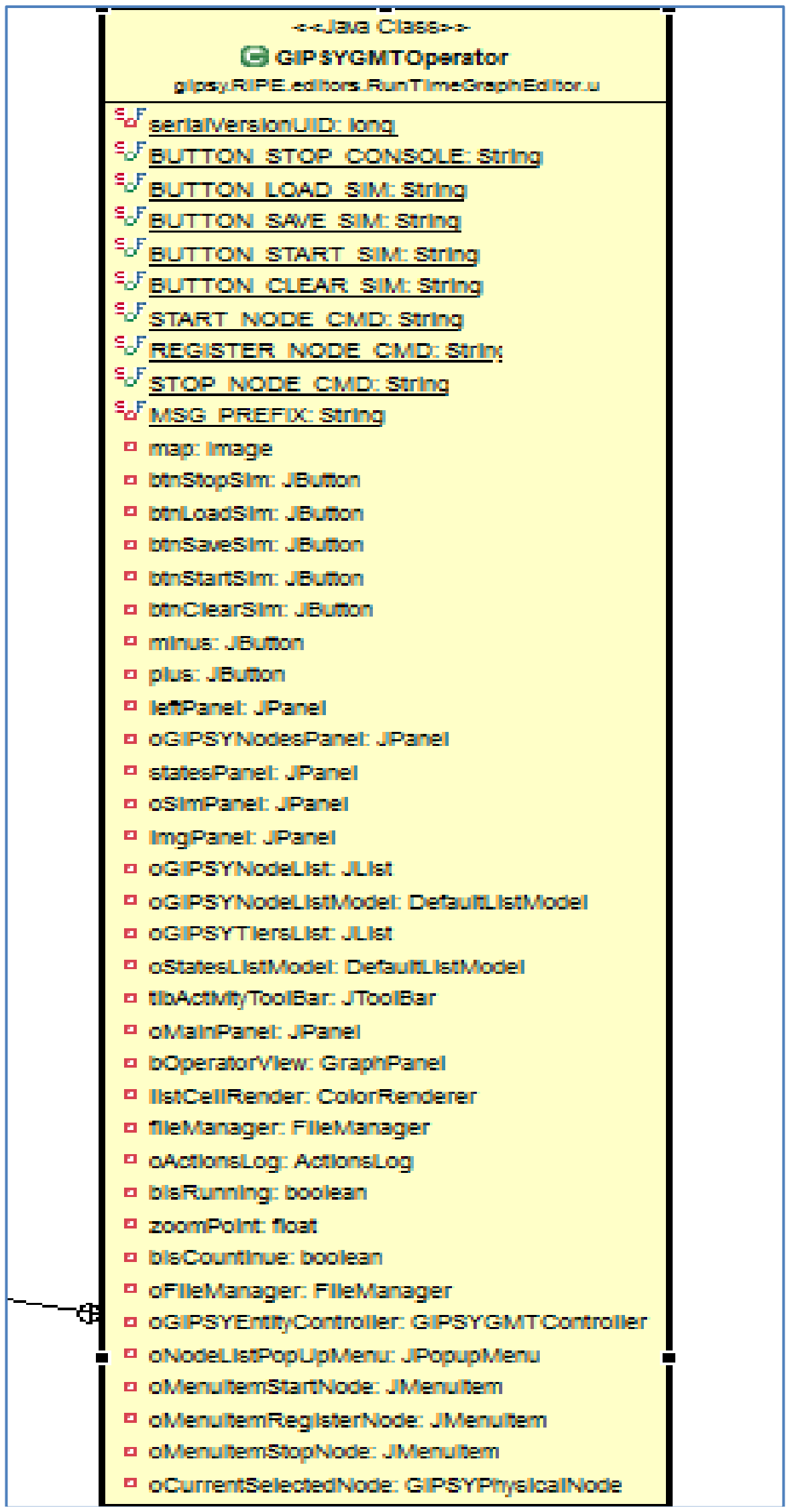

Figure 45 – *GIPSYGMTOperator* UML Class Diagram, attributes

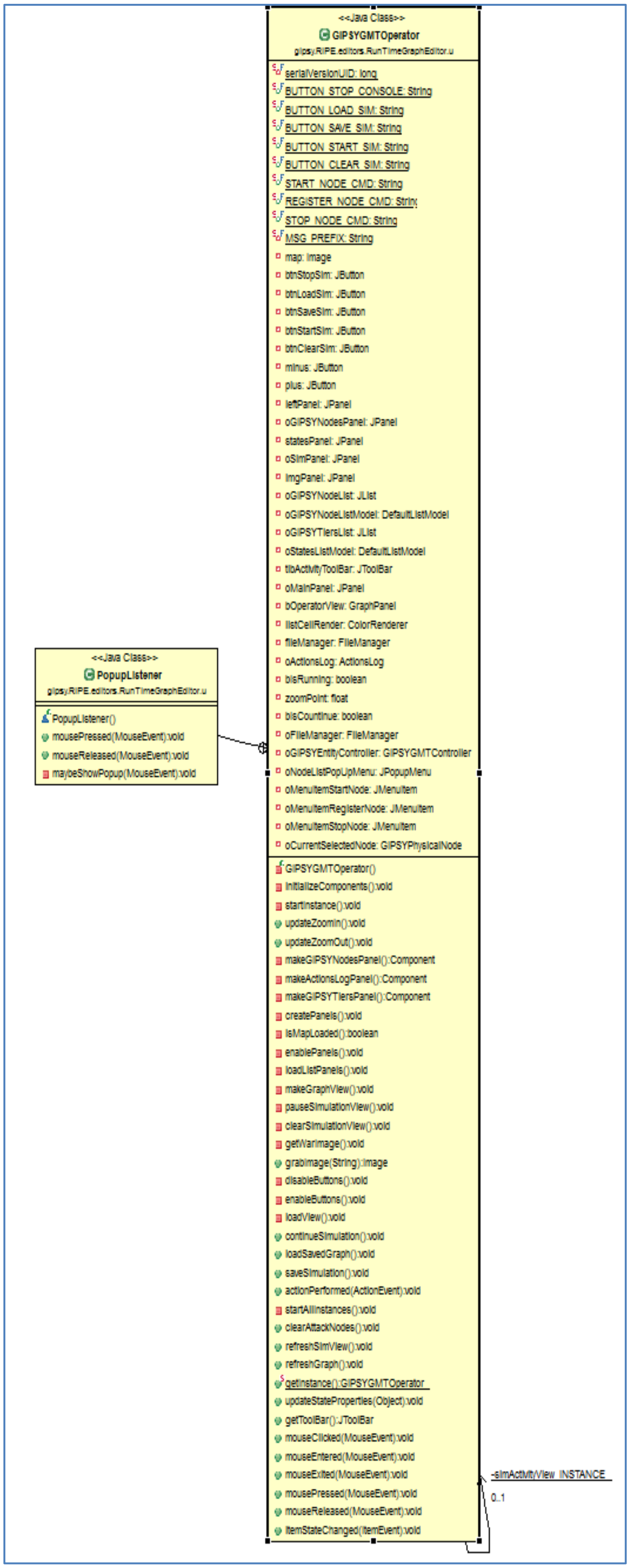

Figure 46 - *GIPSYGMTOperator* UML class diagram, methods

GIPSY Test case

The unit test case for the above GIPSY class is not found in the gipsy.tests. So, we are implementing the Junit test cases for the above class.

*B. Identification of the design patterns*

<u>DMARF</u>

<u>Singleton pattern</u>

The singleton pattern [31] ensures that only one single instance of a class is created. This is easily achieved hiding the constructor and providing a separate method that calls the constructor only after validation that the instance does not exist. The method must be synchronized to avoid multiple threads calling the instance concurrently, which might cause multiple instances of the same class being created. Usually, the singleton pattern is used as a placeholder for configurations, or static data, where the effort of getting the data is semnificative (the tradeoff is the memory consumption, since the instance remains in the memory for the application lifecycle, once instantiated). The instance of the class is retrieved by a static method. Example of how a class is defined as a singleton:

```
public class MySingleton{
	MySingleton instance=null;
	private MySingleton(){ ....}
	public static synchronized getInstance()
	{
		if(instance == null)
		{
				/* Invoke the constructor only if the instance is null
				instance = new MySingleton();
		}
		return instance;
	}
}
```

Java Enterprise Edition (JEE5 and above) provides the singleton pattern through annotations and the Java container manages the lifecycle of the instance of the class that is annotated as singleton, for example:

```
@Singleton
public class MySingleton{ .....}
```

Here is simple diagram of the pattern [23]:

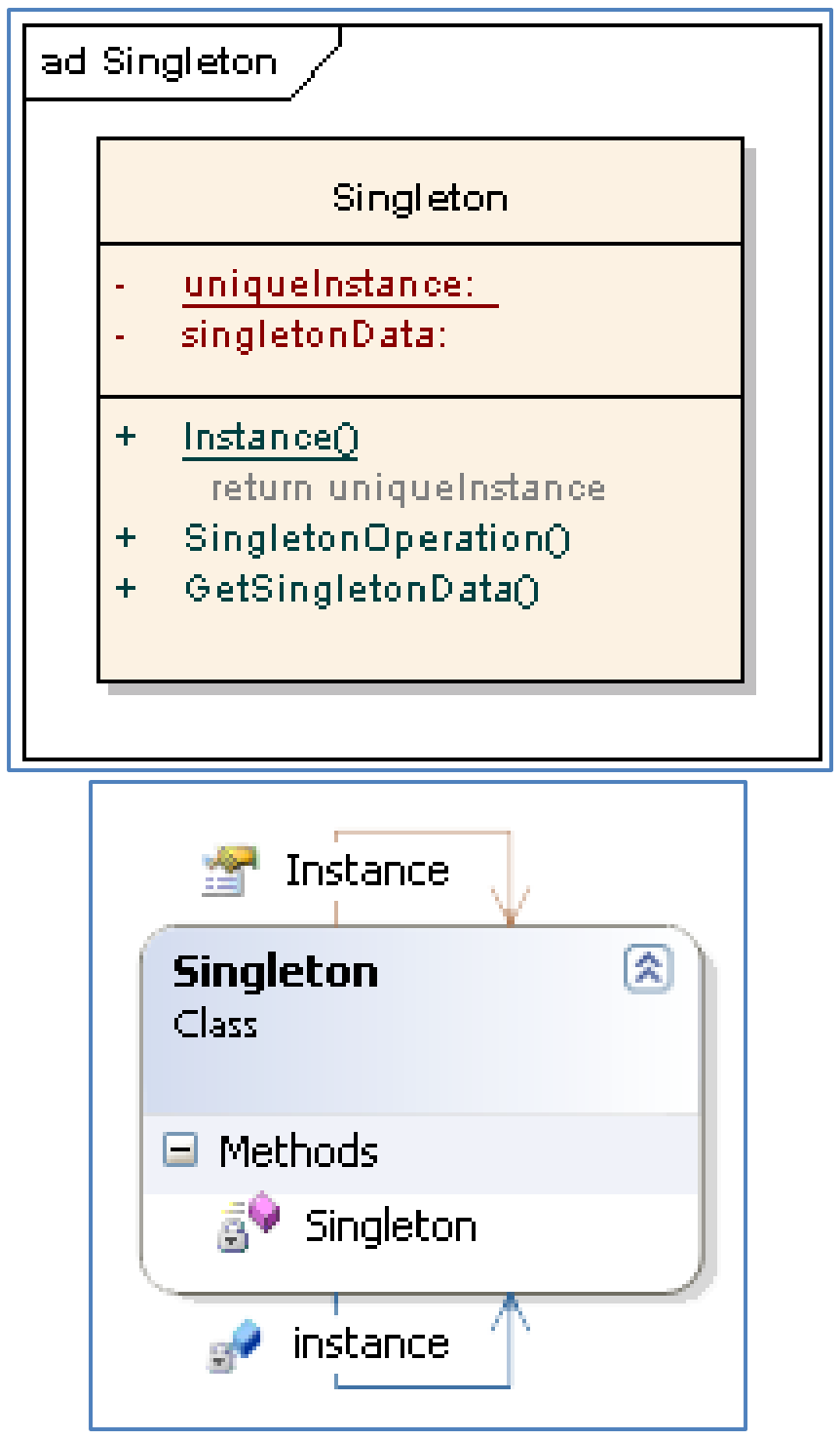

Figure 47 - Singleton Design pattern Diagram

Example of a singleton pattern in DMARF is class *marf.util.OptionFileLoader:*

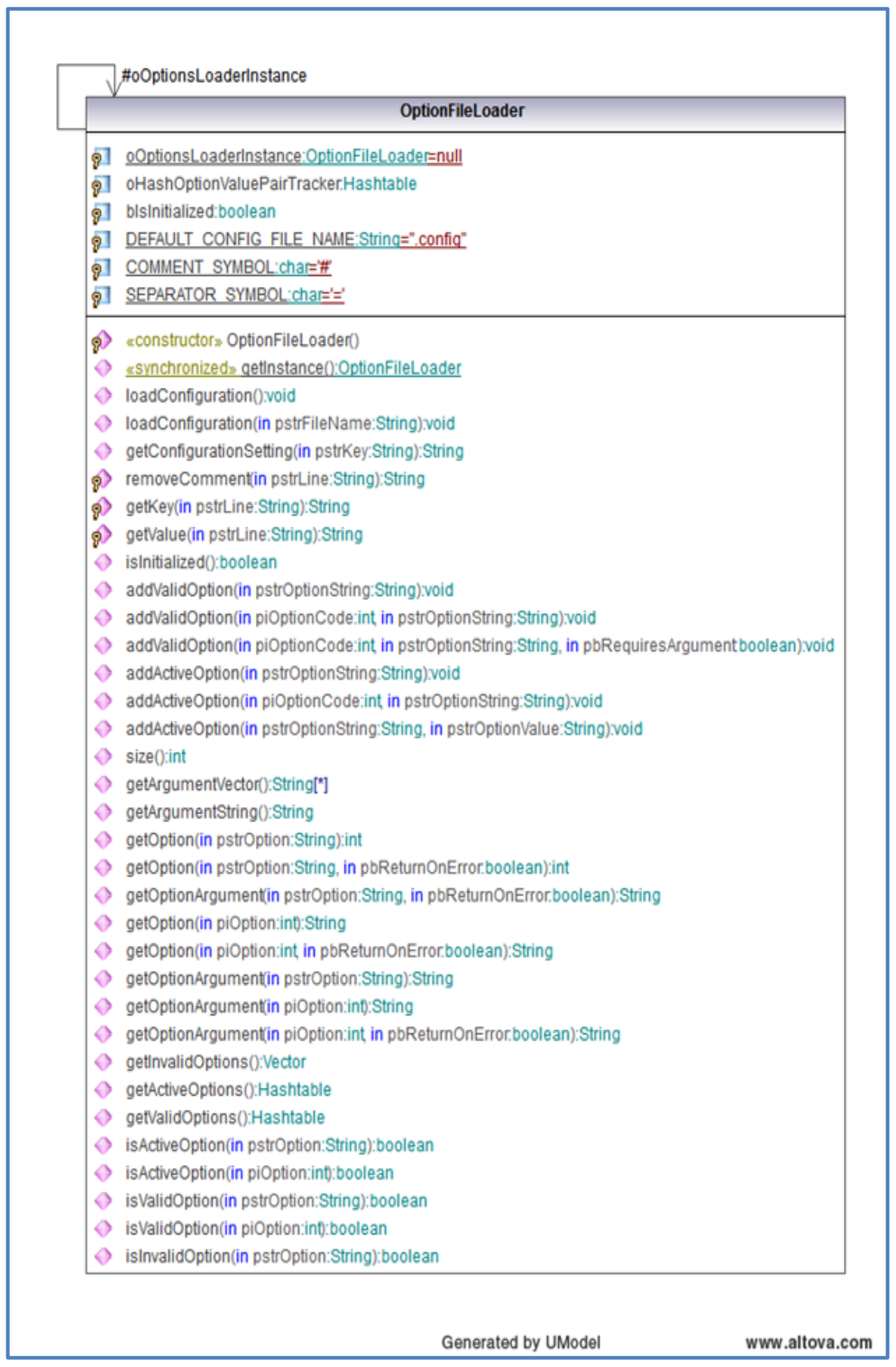

Figure 48 - *OptionFileLoader* Singleton pattern class diagram

Here is how the singleton class is used in DMARF:

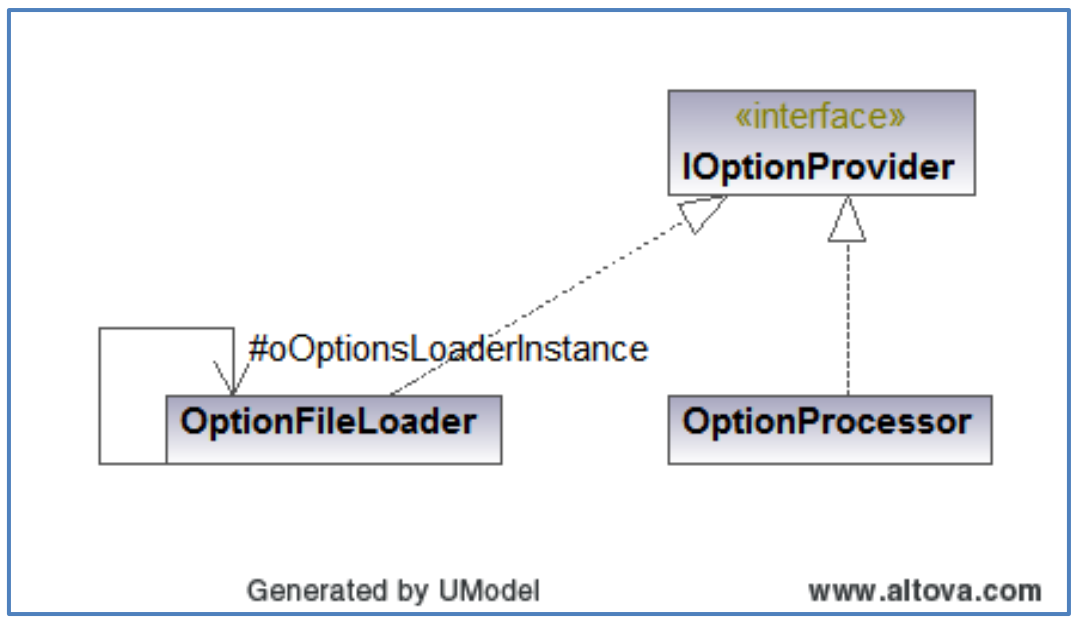

Figure 49 - OptionFileLoader Singleton pattern module

```
package marf.util;

public class OptionFileLoader implements IOptionProvider
{
    protected static OptionFileLoader oOptionsLoaderInstance = null;
              ………………….
    protected OptionFileLoader() { ….. }
    public static synchronized OptionFileLoader getInstance()
    {
              if(oOptionsLoaderInstance == null)
              {
                 oOptionsLoaderInstance = new OptionFileLoader();
              }

              return oOptionsLoaderInstance;
    }
…………..
}
```

## Factory pattern

Factory pattern [26], [[29] is the widely recognized pattern in modern programming industry in designing the systems. It helps the design to be more customizable i.e. one can add new products into the design easily. It may add up a bit of complexity to the design but it helps in reducing the effort in adding new components to the design any time later. The implementation of factory pattern is, it contains a super class that specifies all the generic and specific behavior of the product and the subclasses are being delegated the duty of creating the objects as per the requirement. The main motive behind the use of deferring the duty of creating the objects to the subclass instead of using new operator is considered to be harmful as it creates a reference for the new objects being created and it results in ending up with high coupling as there is new additions to the design at later point of time. Example of how the classes are defined in Factory Pattern:

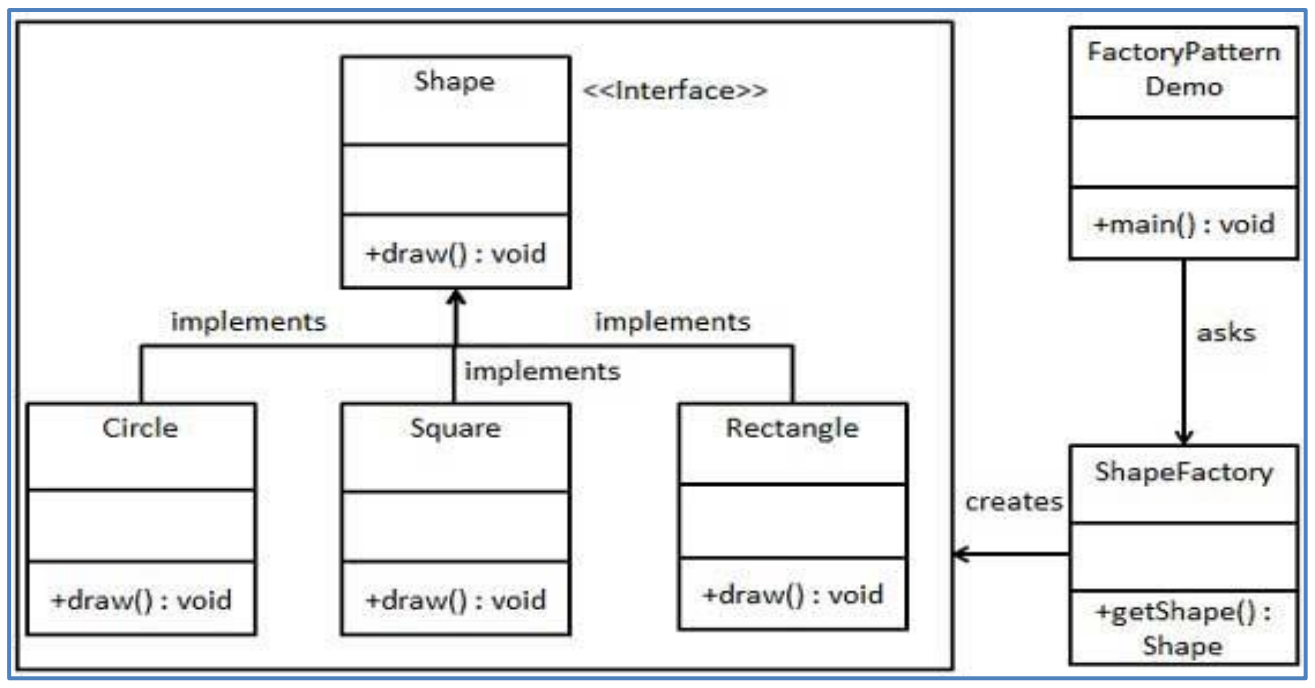

Figure 50 – Factory Pattern Structure

From the above figure it clearly explains how the factory patterns works. Here the ShapeFactory acts like a factory class which creates the shape Interface and the FactoryPatternDemo class will use the ShapeFactory to get a shape object.

```
public interface Shape {
   void draw();
}
```

```
public class ShapeFactory {

public Shape getShape(String shapeType)
{
    If (shapeType == null)
      {
        return null;
      }
      if(shapeType.equalsIgnoreCase("RECTANGLE"))
{
        return new Rectangle();
} else if(shapeType.equalsIgnoreCase("SQUARE")){
        return new Square();
}
      return null;
}
}
```

Example of a Factory Pattern in DMARF

```
public interface IMARFException
{
IMARFException create();
IMARFException create(String pstrMessage);
IMARFException create(Exception poException);
IMARFException create(String pstrMessage, Exception poException);
}public class MARFException extends Exception implements
IMARFException
{
….
….
public IMARFException create()
….
public IMARFException create(Exception poException)
….
public IMARFException create(String pstrMessage)
….
public IMARFException create(String pstrMessage, Exception
poException)
```

```
….
}
```

Class Diagram

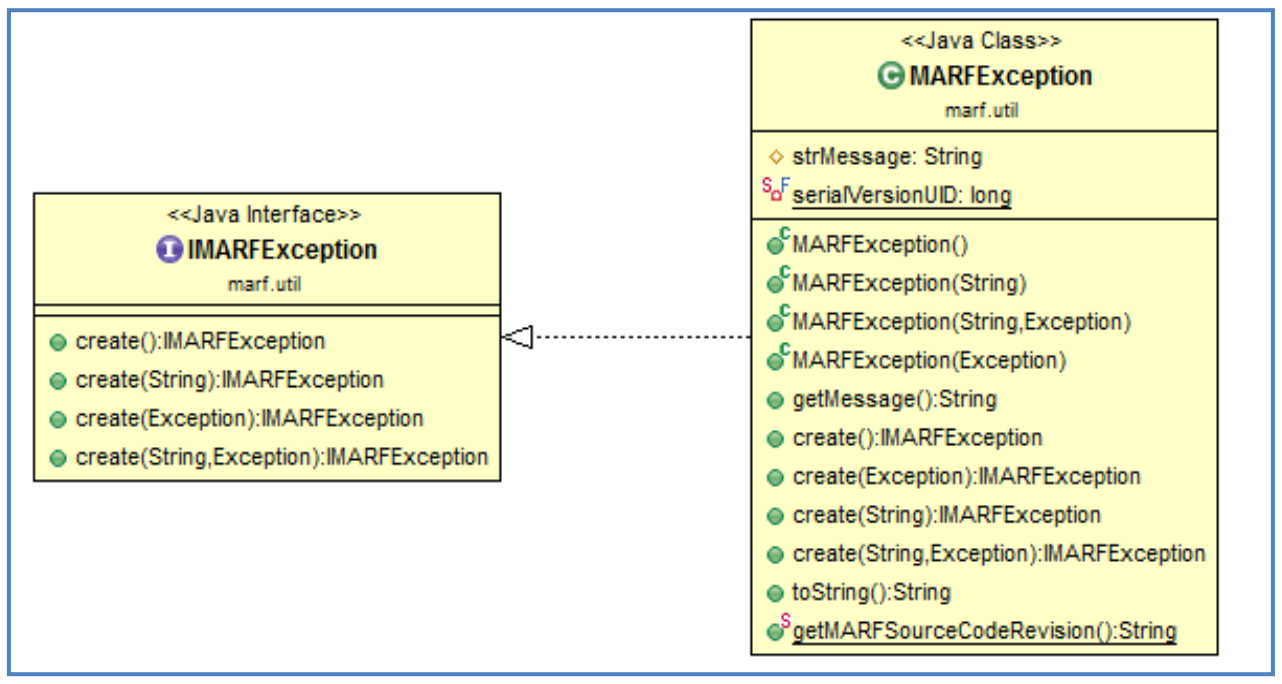

Figure-51 MARFException Factory Pattern Class Diagram

DMARF has implied the rules of factory pattern. In factory pattern, the emphasis on creating objects is deferred to the subclasses. The super class will have the details of the object behavior. But it won't take the responsibility to create the objects. Here the *MARFException* class contains the abstract methods that help the other subclasses to create the objects on their own based on the requirements.

## Prototype Pattern

The motivation for prototype pattern [28], [34] comes with the need of reusing the code for creating objects. If the cost of creating new object is considered to be expensive, then the best way to avoid it is by cloning the similar object. By using the similar objects code as a prototype, programmers can clone it and can create new objects. This turns out to be the efficient way to confine creation of object to subclass similar to the factory pattern. But the prototype pattern is different from it in one major way i.e. using prototype pattern one can create only one object. The implementation focuses on creating a prototype interface that creates a clone of the current object. It is considered that using new operator will increase the coupling between objects and thereby reduces the flexibility of the design.

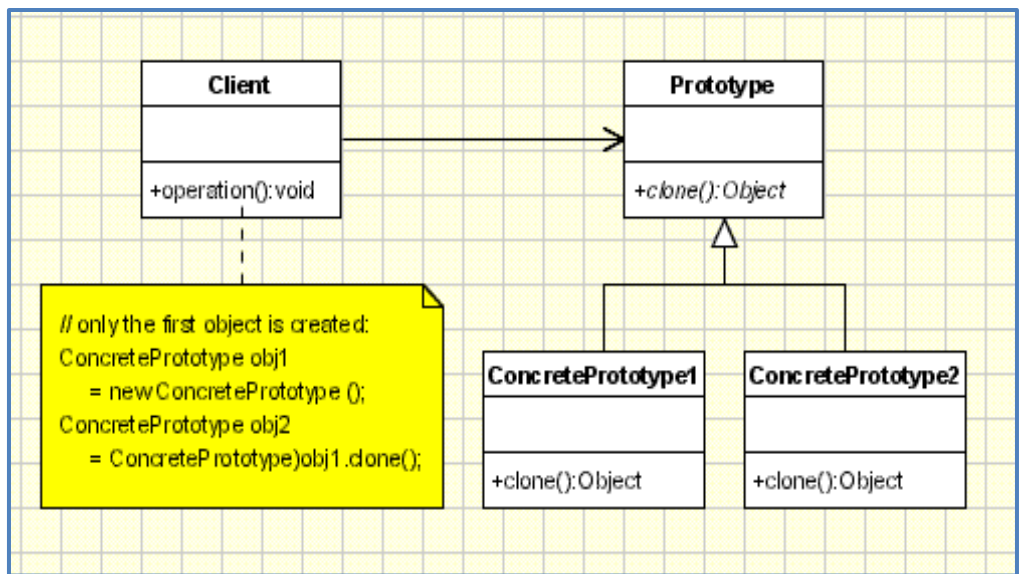

Figure 52- Prototype Pattern Structure

```
import java.io.FileInputStream;
…
public abstract class StorageManager implements IStorageManager

{
..
..
public StorageManager(Serializable poObjectToSerialize, String
pstrFilename)
public StorageManager(Object poObjectToSerialize, String pstrFilename)
..
..
public synchronized Object clone()
…
}
public abstract class Classification extends StorageManager implements
IClassification

{
…
…
private final void saveTrainingSet()throws StorageException
public static TrainingSet loadTrainingSet(int piDumpMode, String
pstrFilename)throws StorageException
…
…
public Object clone()
{
Classification oClone =(Classification)super.clone();
oClone.oResultSet = (ResultSet)this.oResultSet.clone();
oClone.oTrainingSet = (TrainingSet)this.oTrainingSet.clone();
oClone.oFeatureExtraction = this.oFeatureExtraction;
return oClone;
}
….
}
```

Class Diagram

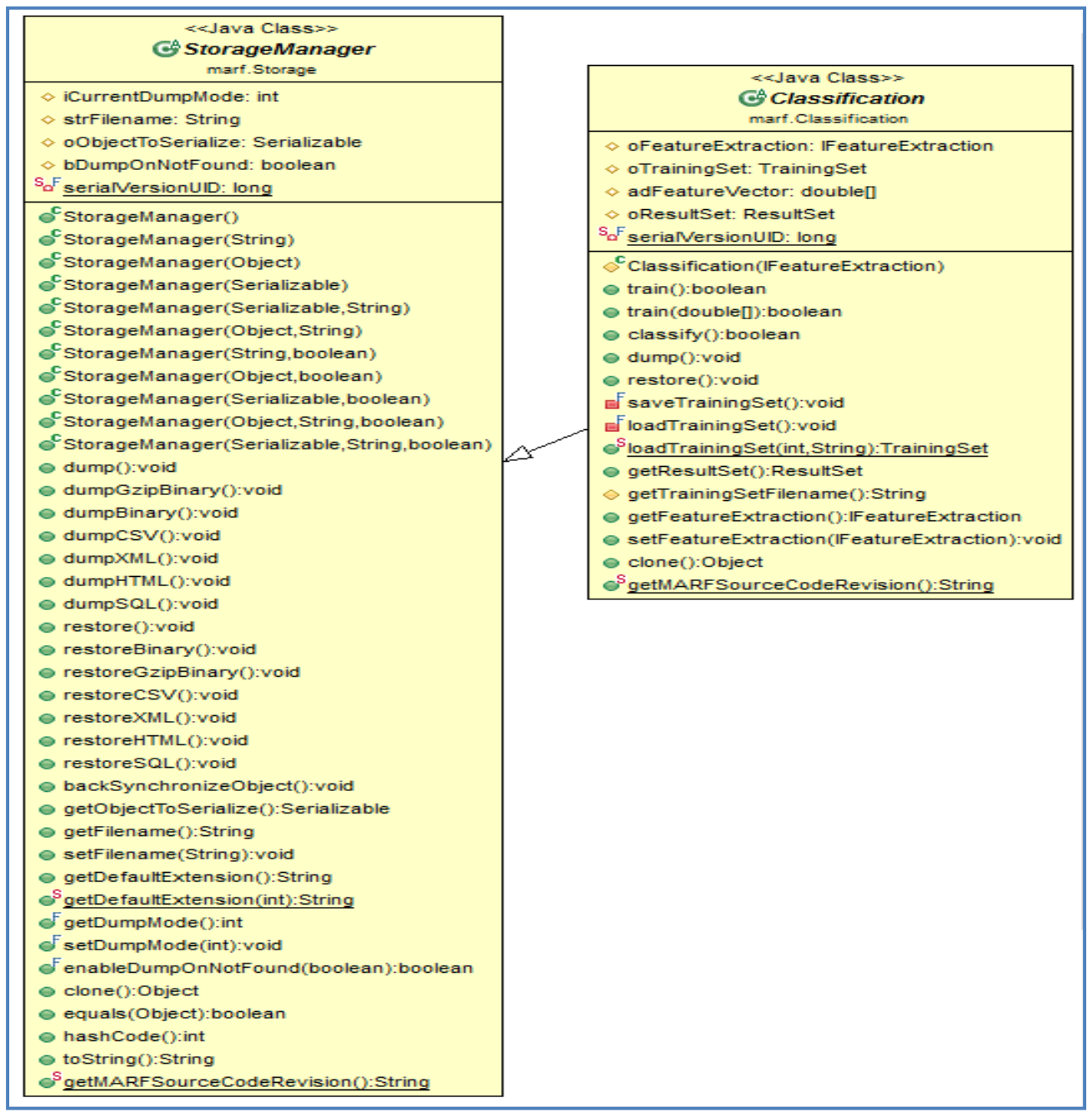

Figure 53- StorageManager Prototype Pattern Class Diagram

In MARF, the *StorageManager* class opts the prototype pattern to use the objects that are costly to create newly. It uses the similar class that has the methods and thereby results in cost effective usage of the objects. There are methods in the *StorageManager* class that instantiates some objects in the class Classification. In here, the concrete methods/classes are being used based on the prototype and are thereby customized accordingly.

## Strategy pattern

It defines the family of the algorithms and encapsulates each one, also makes them interchangeable. It lets the algorithm change independently from the clients that use it. Collects the abstraction in an interface and put implementation details in derived classes. [28]

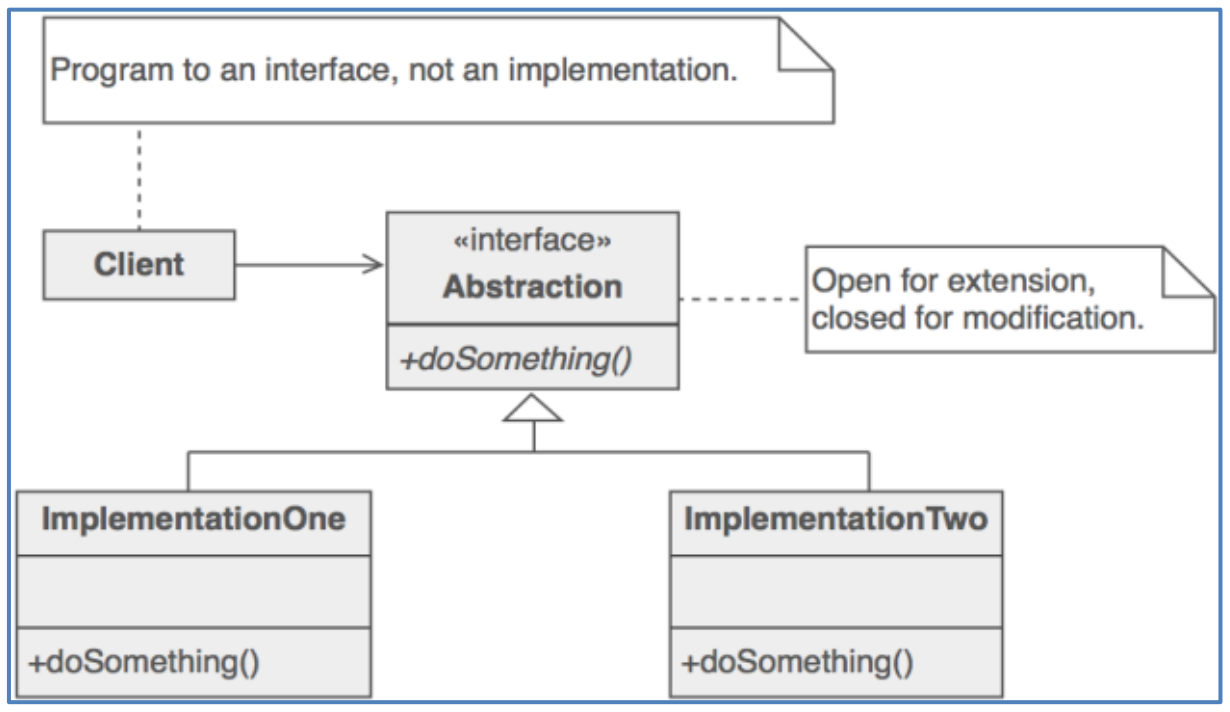

Figure 54 - Domain Diagram for Strategy Pattern

```
public interface IPreprocessing extends Cloneable

{
…..
boolean preprocess()throws PreprocessingException;

boolean removeNoise()throws PreprocessingException;

boolean removeSilence()throws PreprocessingException;

boolean normalize()throws PreprocessingException;
…
…
boolean cropAudio(double pdStartingFrequency, double
pdEndFrequency)throws PreprocessingException;

}

public abstract class FeatureExtraction extends StorageManager
implements IFeatureExtraction
{
….
protected FeatureExtraction(IPreprocessing poPreprocessing)

public boolean extractFeatures()throws FeatureExtractionException

public void setPreprocessing(IPreprocessing poPreprocessing)
….

}
```

Class Diagram

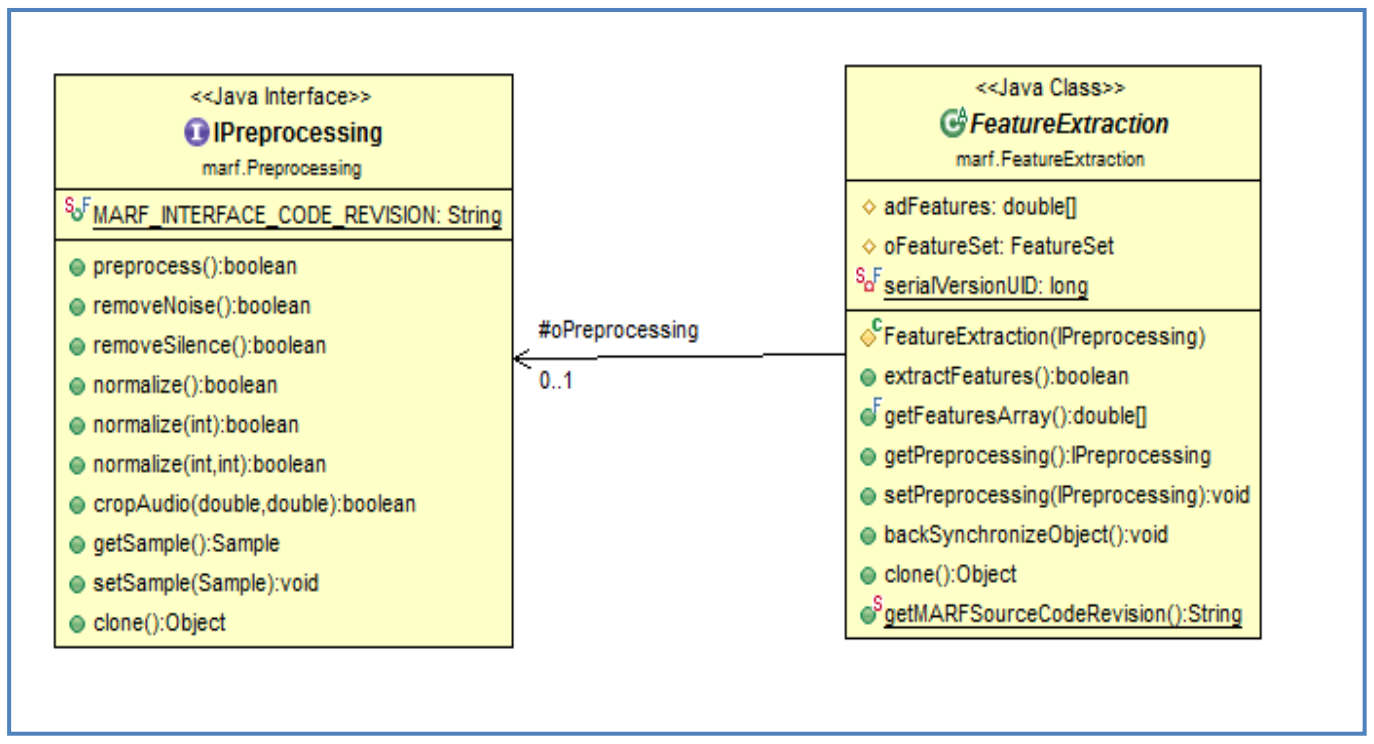

Figure – 55 IPreprocessing Strategy Pattern Class Diagram

This part of DMARF has been implicated with the strategy pattern. The feature extraction class acts as a super class and it contain the properties that most of the subclasses inherit from it. Thus if the super class wants to introduce a new feature, then it may not be needed by the subclass or the subclass is not supposed to inherit that new property. Hence *IPreoprocessing* interface has been brought in to do the job. It simply provides the desired method to the required subclass.

## GIPSY

### Decorator pattern

This pattern is used as extending functionality and provides a malleable alternative sub classing for that particular functionality. It adds an extra responsibility to respective object dynamically. Makes core object looks interesting by recursively wrapping it with respective to the client-specified. But adding a behavior to an individual objects in run-time the concept of inheritance remains nothing because it being static it applies to complete class but this pattern suggests client the ability to specify whatever features they desired.[25]

Example: Let's assume user interface toolkit and you need adding borders and scroll bars to windows. We can define inheritance concept.

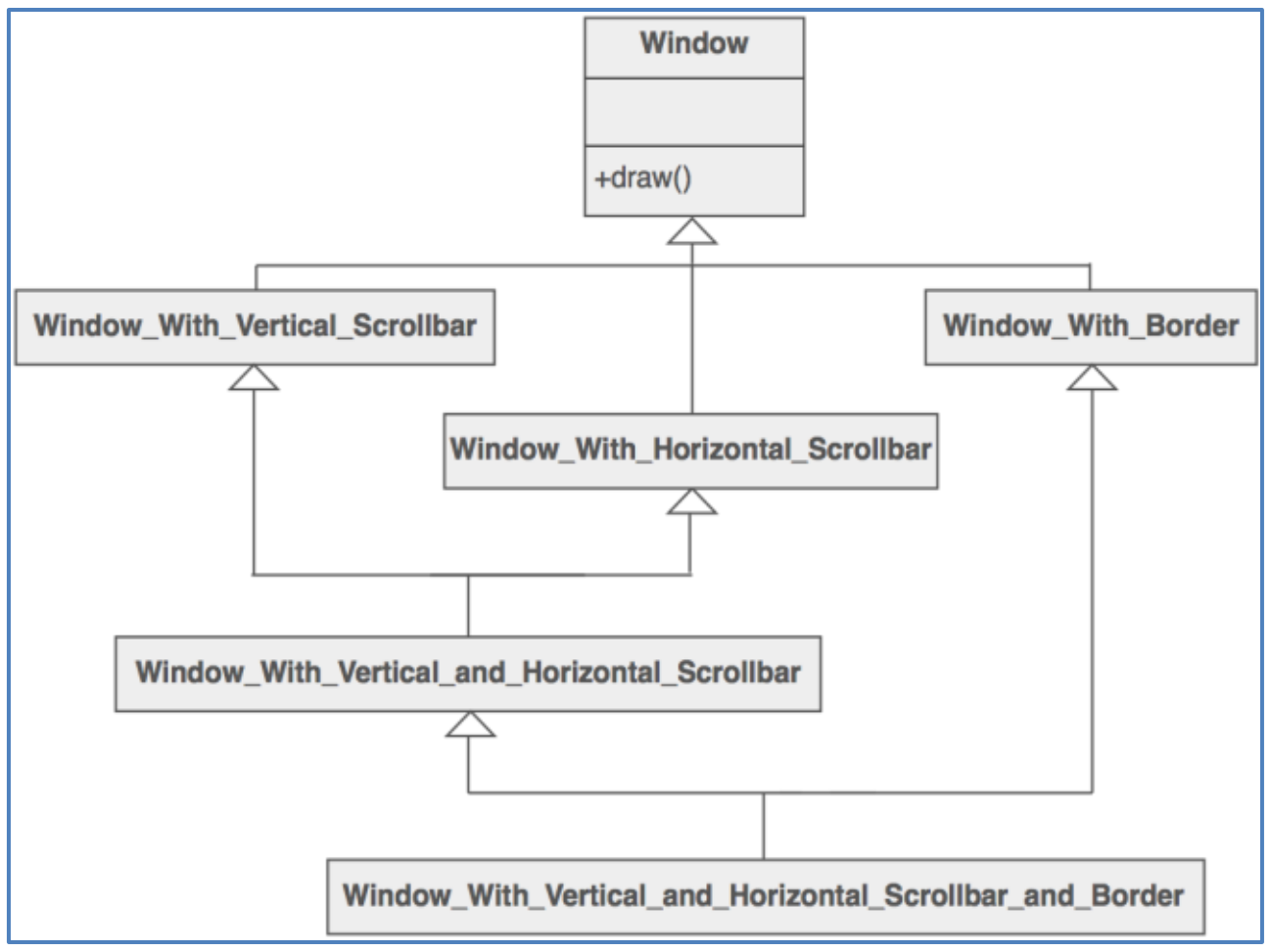

Figure 56 – Decorator pattern structure 1

```
Widget* aWidget = new BorderDecorator(
newHorizontalScrollBarDecorator(
newVerticalScrollBarDecorator(
new Window( 80, 24 ))));
aWidget->draw();
```

### Structural Decorator pattern

Request for extended functionality from client applies on *CoreFunctionality.doThis()*. This is client interest or may not interest in *OptionalOne.doThis()* and *OptionalTwo.doThis()*. These classes work for Decorator base class and this class always works to wrappee object.

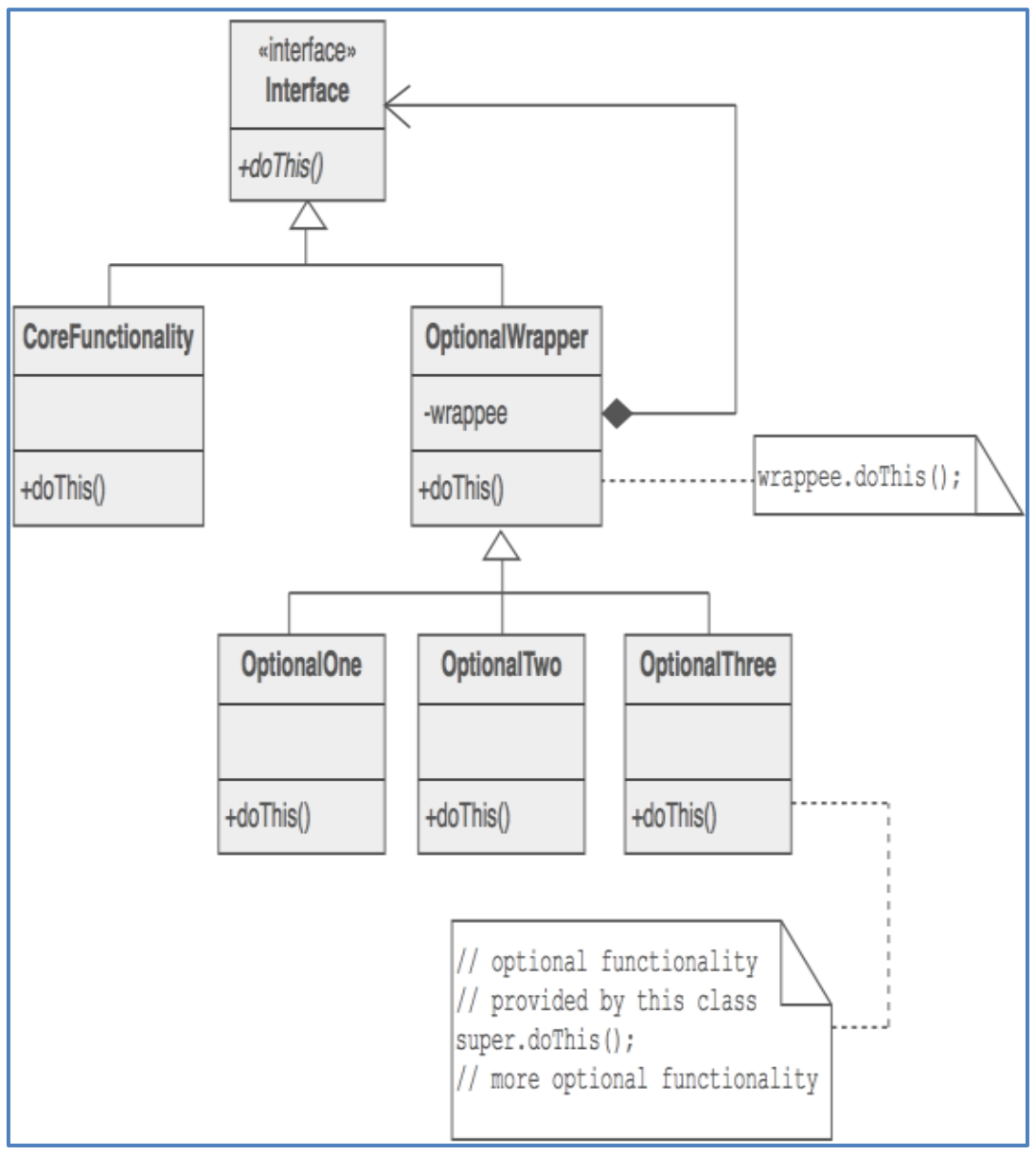

Figure 57 – Decorator pattern Structure 2

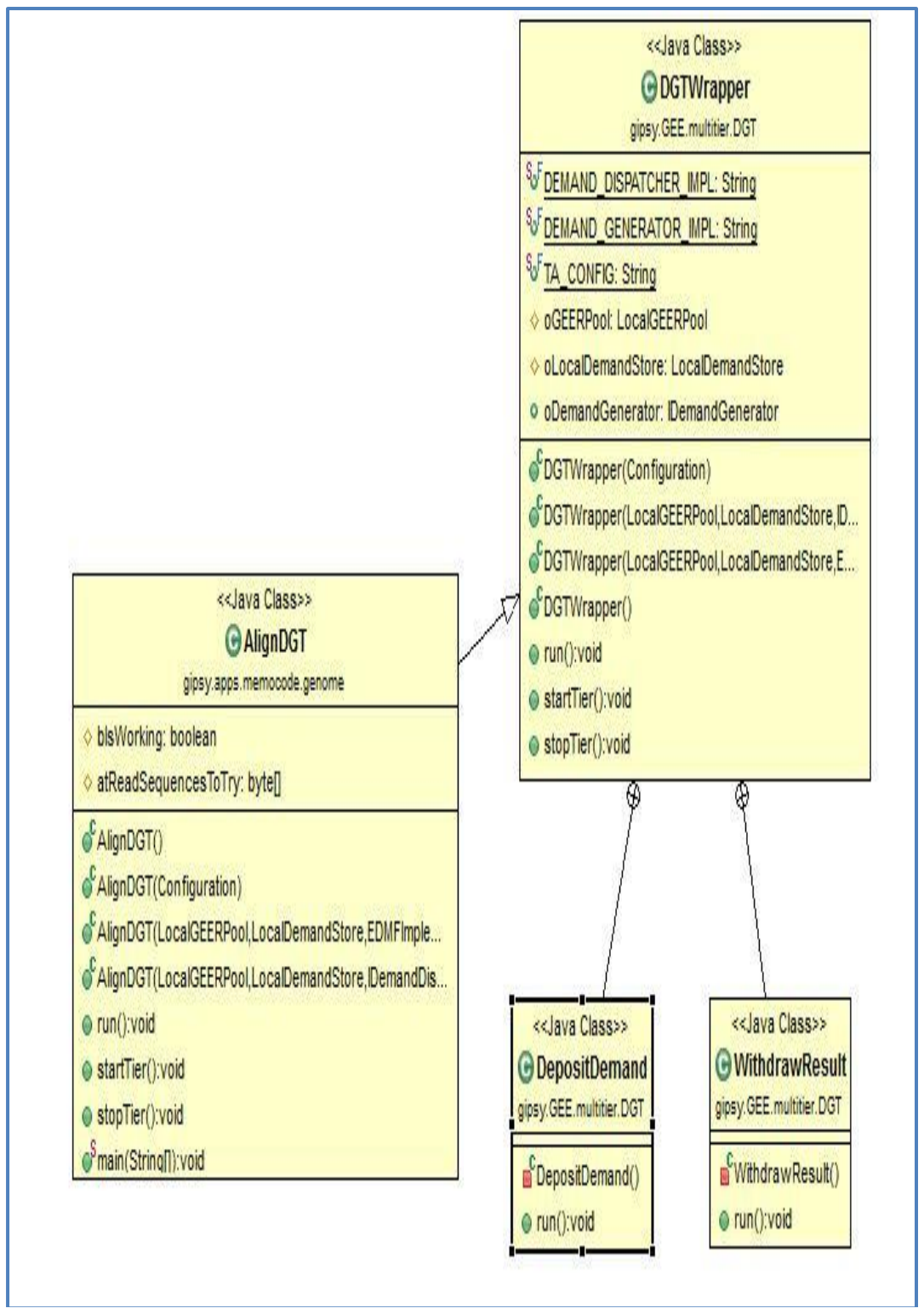

Figure 58 – AllignDGT class Diagram

```
Public class AlignDGT
extendsDGTWrapper
{
……………………………………
publicvoidstartTier()
throwsMultiTierException
{
try
{
// Create a TA instance using the configuration
Configuration oTAConfig =
(Configuration)this.oConfiguration.getObjectProperty(DGTWrapper.TA_C
ONFIG);
ITransportAgentoTA = TAFactory.getInstance().createTA(oTAConfig);

// Create a DemandDispatcher instance using the configuration
String strImplClassName =
this.oConfiguration.getProperty(DGTWrapper.DEMAND_DISPATCHER_
IMPL);
Class<?>oImplClass = Class.forName(strImplClassName);
Class<?>[] aoParamTypes = new Class[] {ITransportAgent.class};
Constructor<?>oImplConstructor =
oImplClass.getConstructor(aoParamTypes);
Object[] aoArgs = new Object[]{oTA};
this.oDemandDispatcher =
(DemandDispatcher)oImplConstructor.newInstance(aoArgs);
this.oDemandDispatcher.setTAExceptionHandler(this.oTAExceptionHandl
er);
newBaseThread(this).start();
}
catch(MultiTierExceptionoException)
{
throwoException;
}
catch(Exception oException)
{
thrownewMultiTierException(oException);
```

```
}
}
```

## Observer Pattern

This pattern defines one-to-many dependency between objects so that whenever an object changes its state all its dependents are notified. This pattern [32], [33] maintains all the list of all its dependents which are called observers.

Observer pattern is used to model the dependent functionality with observer hierarchy and subject abstraction for independent functionality. Observer pattern plays a key role in Model-View-Controller architectural pattern. In this pattern all the observers should register themselves with an object that is observable and then start the processing. This observer is responsible for extracting information from the subject. Implementation of this pattern requires explicit registration and explicit deregistration as the subject holds strong references to the observers.

Example class Diagram for an observer Pattern

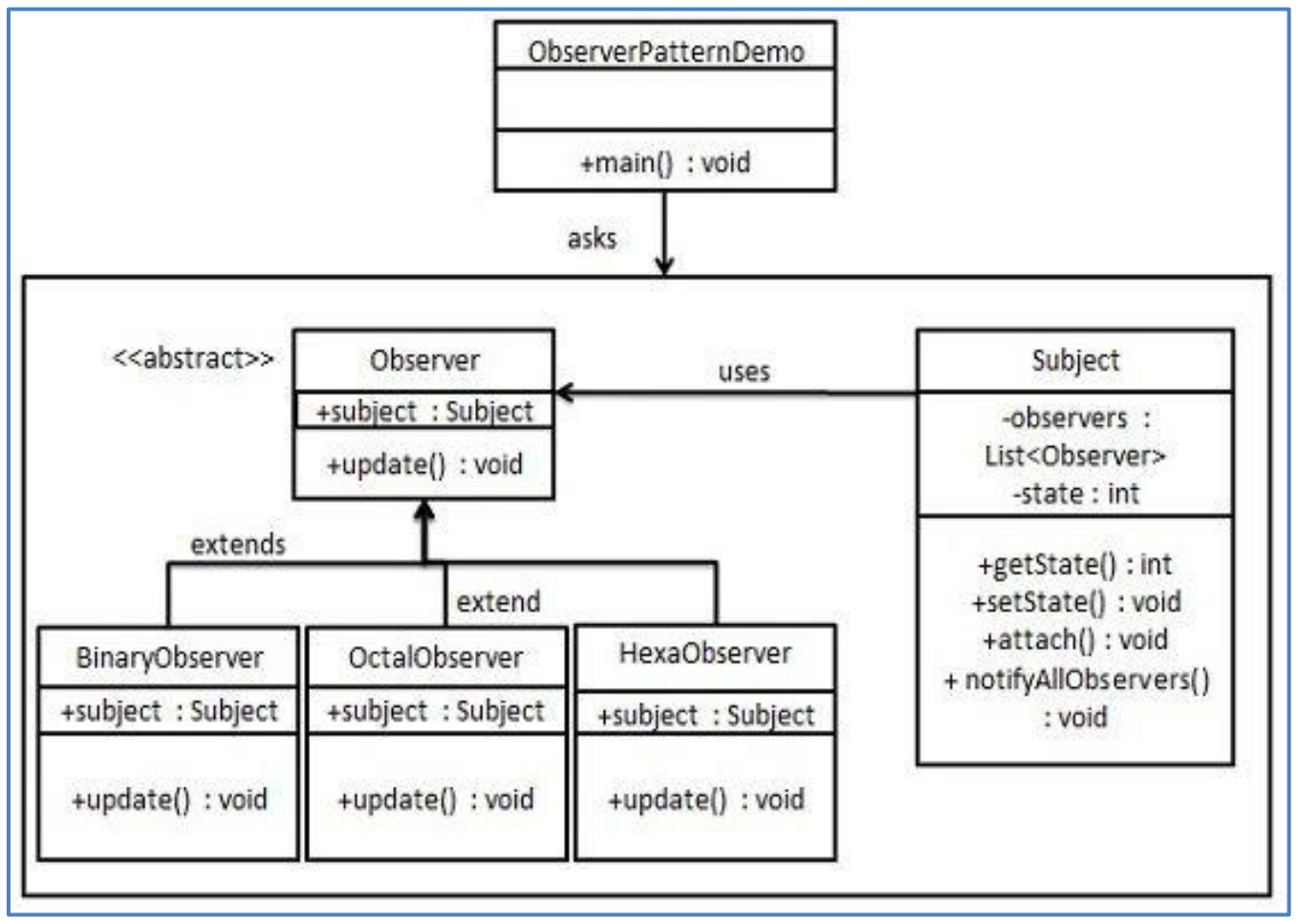

Figure 59 – Observer Pattern Structure

The figure describes observer pattern uses three actor classes. Subject, Observer and Client.

```
public abstract class Observer {
   protected Subject subject;
   public abstract void update();
}
```

```
public class Subject {
   private List<Observer> observers
      = new ArrayList<Observer>();
   private int state;

   public int getState() {
      return state;
   }
```

```
public void setState(int state) {
   this.state = state;
   notifyAllObservers();
}

public void attach(Observer observer){
   observers.add(observer);
}

public void notifyAllObservers(){
   for (Observer observer : observers) {
      observer.update();
   }
}
```

Example of an Observer Pattern in GIPSY

The Demand class implements the *IDemand* interface class and this interface has methods which are overridden in the Demand class.

```
public interface IDemand extends ISequentialThread, Cloneable

{

 public GIPSYContext getContext();

      ….

 public void setContext(GIPSYContextpoContext);

void setSignature(DemandSignature poSignatureID);
DemandSignature getSignature();
void setType(DemandType poType);
DemandType getType();
void addTimeLine(String pstrTierID);
void addTimeLine(TimeLine poTimeLine);
long getAccessNumber();
      ….
       ….

}
public abstract class Demand extends FreeVector<Object> implements
IDemand

{
.....

public Demand(String pstrName)

public DemandSignature getSignature()
….
….

}
```

Class Diagram:

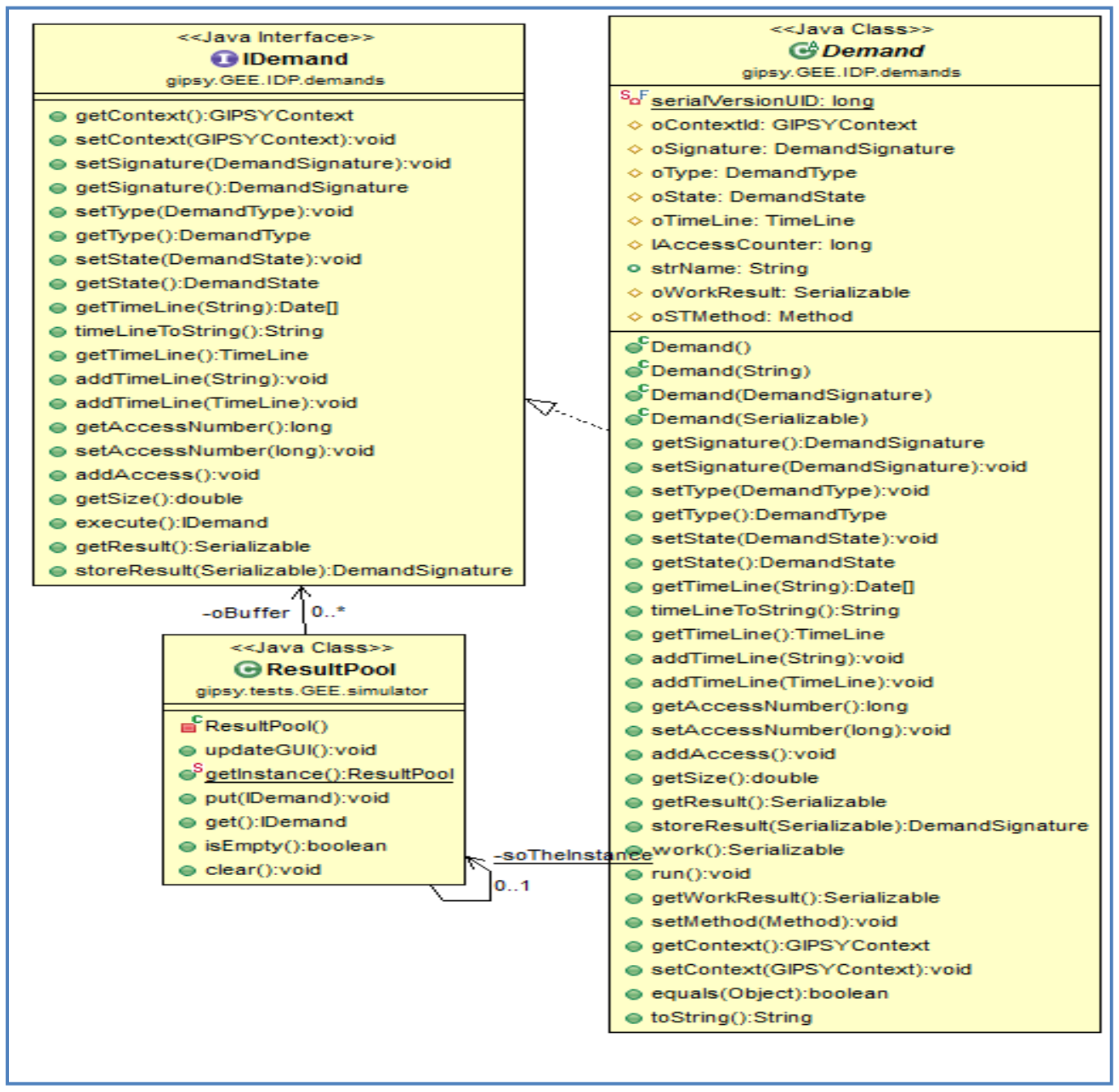

Figure 60 – *IDemand* Observer Pattern Class Diagram

*gipsy.GEE.IDP.demands* is the class that is acting as subject here and is sending the data to the observers via the *gipsy.GEE.IDP.demands* interface which acts as a listener interface. Using this interface the observers will receive the event notices of the above mentioned class. As shown in the above figure, the class *resultpool* is one such observer class that is interested in the data of the demands class.

<u>Singleton Pattern</u>

The singleton pattern [29], [31] is used for managing internal or external resources and provides a global point of access to an object. It creates only one instance of a class. It is simple design pattern its associate only single class that is answerable to itself, so to fix it that creates only one instance.

The range of this pattern is very expressive that involves a static member in this pattern class, a private constructor and a static public method that sends a reference to the static member.

The implementation of these patterns involves in accessing resource in shared mode, factories, and logger classes and configures classes, etc.

Example class diagram for Singleton pattern

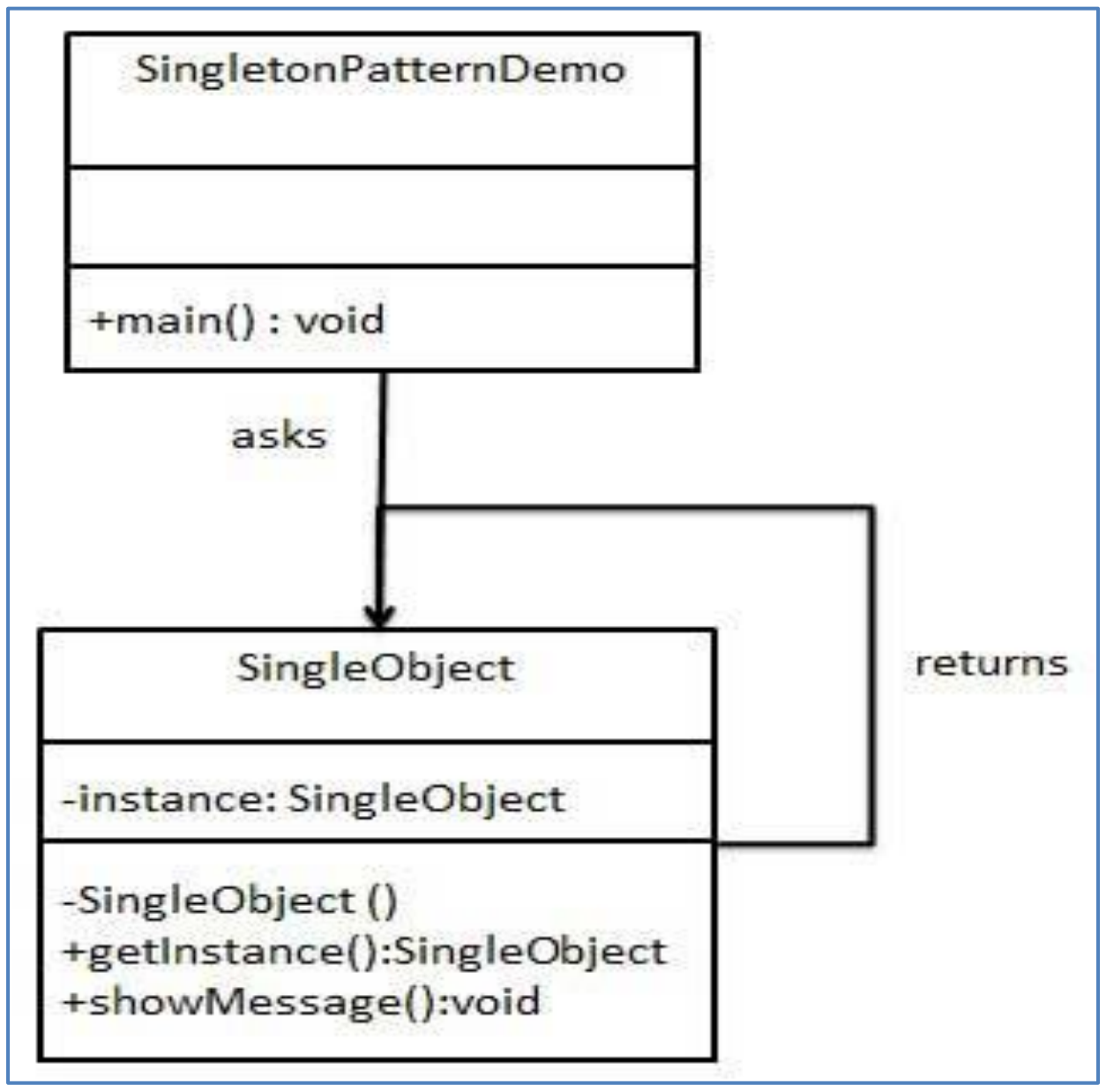

Figure 61 – Singleton Pattern Structure

Above figure shows how the single pattern creates an object. SingleObject class constructor should be declared as private and it has a static instance of itself. SingleObject class has its own static method to get its static instance.

```
public class SingleObject {

   //create an object of SingleObject
   private static SingleObject instance = new SingleObject();

   //make the constructor private so that this class cannot be
   //instantiated
   private SingleObject(){}

   //Get the only object available
   public static SingleObject getInstance(){
      return instance;
   }

   public void showMessage(){
      System.out.println("Hello World!");
   }
}
```

Example of Singleton Pattern in GIPSY

```
public class DemandClassPool
{
…
public static synchronized DemandClassPool getInstance()
{
if(null == soInstance)
{
soInstance = new DemandClassPool();
}
return soInstance;
```

```
}
…
…
public static synchronized DemandClassPool getInstance()
public synchronized void put(String pstrItem)
public synchronized String get()
…
…
}
```

Class diagram

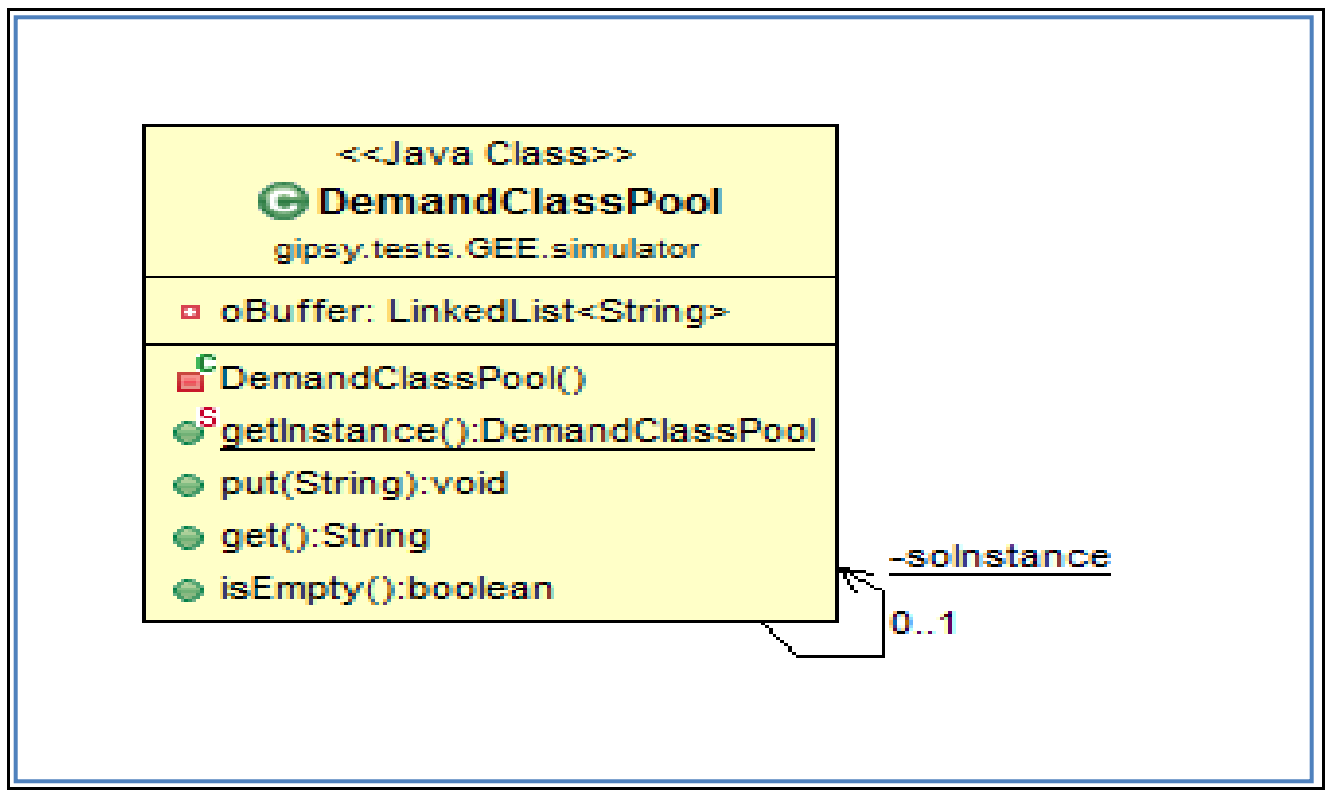

Figure 62 – *DemandClassPool* Singleton Pattern Class Diagram

As per the terms of singleton pattern, the MARFCATDWT class has static method and static object that are supposed to work for only one call at a time. Whenever a demand worker is given to work on a demand to process, it need to work on the demand alone, without diverting on the other demands being generated and are required to be processed. Hence the object being static is only assigned to the concerned class that will process the demand.

## Proxy Pattern

The proxy pattern [28], [29], [33] is a structural pattern that is used when one needs the ability to control the access to an object. In its implementation, it creates a proxy interface that executes the exact methods in the real object interface. Whenever the object need the original object, then it is guided to execute that object and thereby reduces the cost. By creating the proxy, it reduces the cost effect on using the real subject without actually calling it, unless it is actually needed.

Example class Diagram of proxy pattern

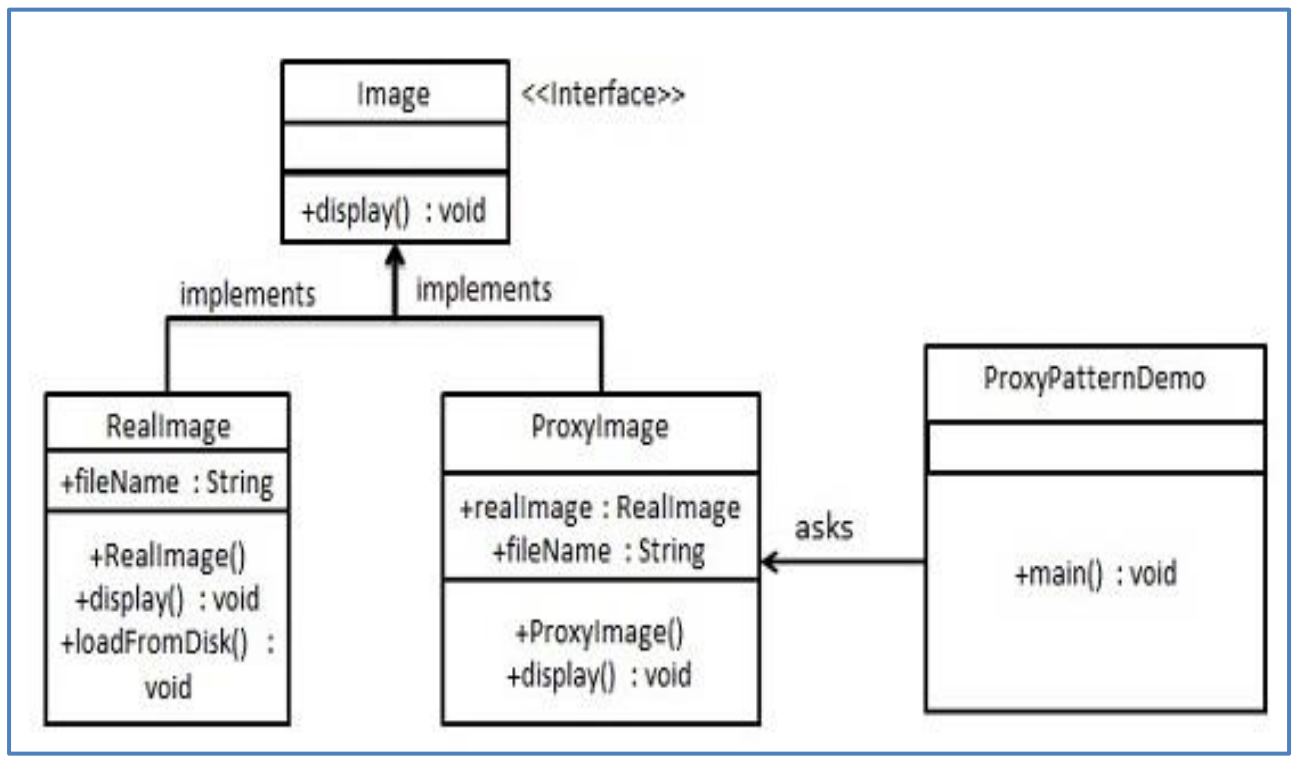

Figure 63 – Proxy Pattern Structure

The above class explains how proxy pattern works. *Image* interface and concrete classes implementing the *Image* interface. *ProxyImage* is a a proxy class to reduce memory footprint of *RealImage* object loading.

```
public interface Image {
   void display();
}
```

```
public class ProxyImage implements Image{

   private RealImage realImage;
   private String fileName;

   public ProxyImage(String fileName){
      this.fileName = fileName;
   }

   @Override
   public void display() {
      if(realImage == null){
         realImage = new RealImage(fileName);
      }
      realImage.display();
   }
}
```

Example of Proxy Pattern in GIPSY

```
public interface IDemandWorker extends Runnable
{
void setTransportAgent(EDMFImplementation poDMFImp);
void setTransportAgent(ITransportAgent poTA);
void setTAExceptionHandler(TAExceptionHandler
poTAExceptionHandler);
void startWorker();
void stopWorker();
}
public class MARFCATDWT extends DWTWrapper
{

@Override
public void setTransportAgent(EDMFImplementation poDMFImp)
{
this.oDemandWorker.setTransportAgent(poDMFImp);
}
@Override
public void setTransportAgent(ITransportAgent poTA)
{
this.oDemandWorker.setTransportAgent(poTA);
}
@Override
public void startWorker()
{
this.oDemandWorker.startWorker();
this.bIsWorking = true;
}
@Override
public void stopWorker()
{
this.oDemandWorker.stopWorker();
this.bIsWorking = false;
}
@Override
public void setTAExceptionHandler(TAExceptionHandler
poTAExceptionHandler)
{
this.oDemandWorker.setTAExceptionHandler(poTAExceptionHandler);
}
}
```

Class diagram

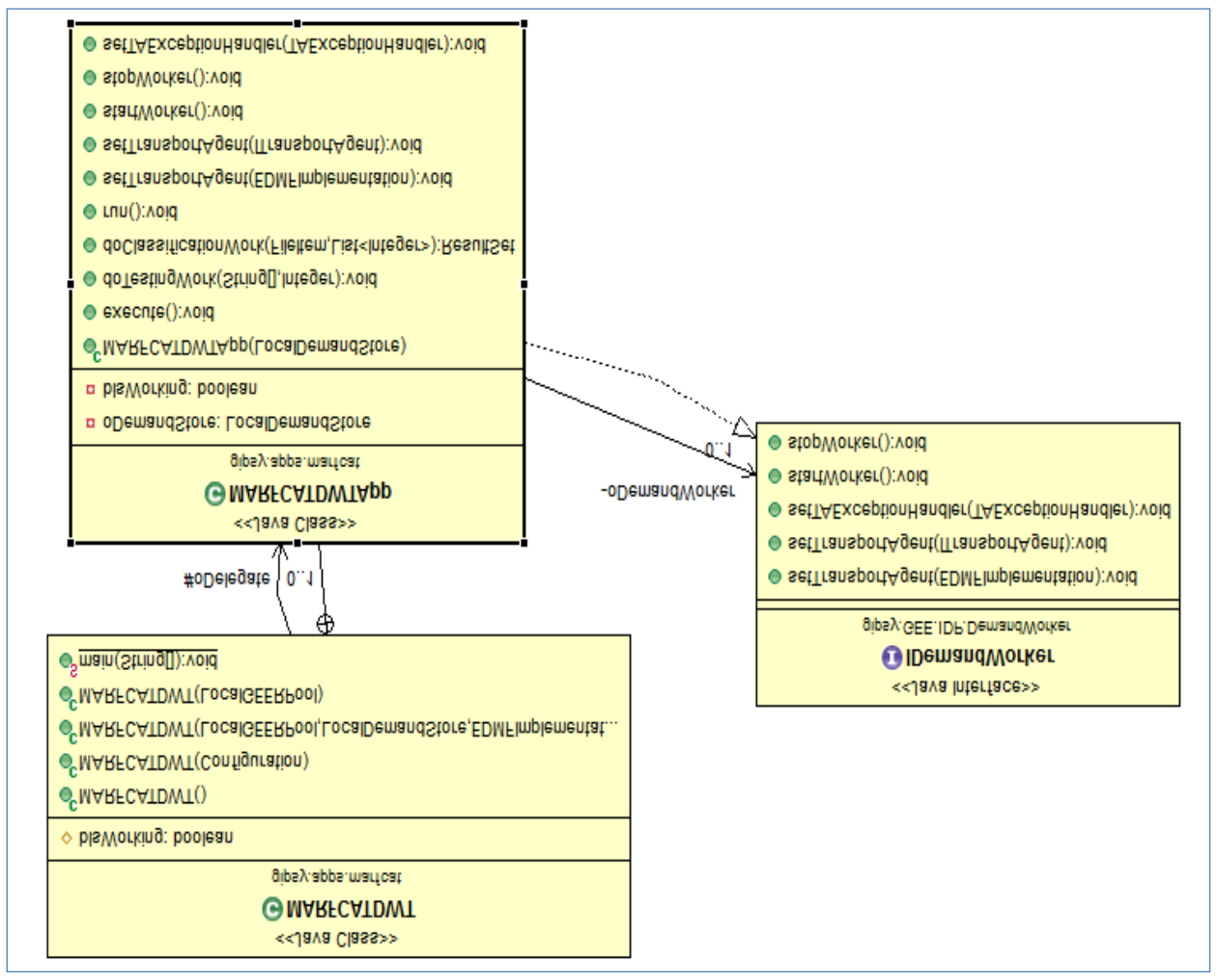

Figure 64 – IDeamandWorker Proxy Pattern Class Diagram

As shown in the class diagram, the class *MARFCATDWT* delegates the creation of objects regarding the transport of demands via transport agents to the class *MARFCATDWTApp*. The object initially calls the *IDemandWorker* interface to execute the operation of the methods as mentioned above. This helps the class to confine the use of heavy objects until it is actually needed. Thereby it provides the ability to control access to the objects.

## Factory pattern

In the Factory pattern [26], [27], [29], object is created without displaying logic to the client and then refers to the newly created object using a similar interface. It also defines a virtual constructor. The new operator is considered as a harmful. The super class specifies all the standard and generic behavior and works the developing details to the subclasses that are given by the client. It makes the design more personalize but bit complicated, whereas other design patterns need the classes, whereas factory method only needs need operations. [26], [27]

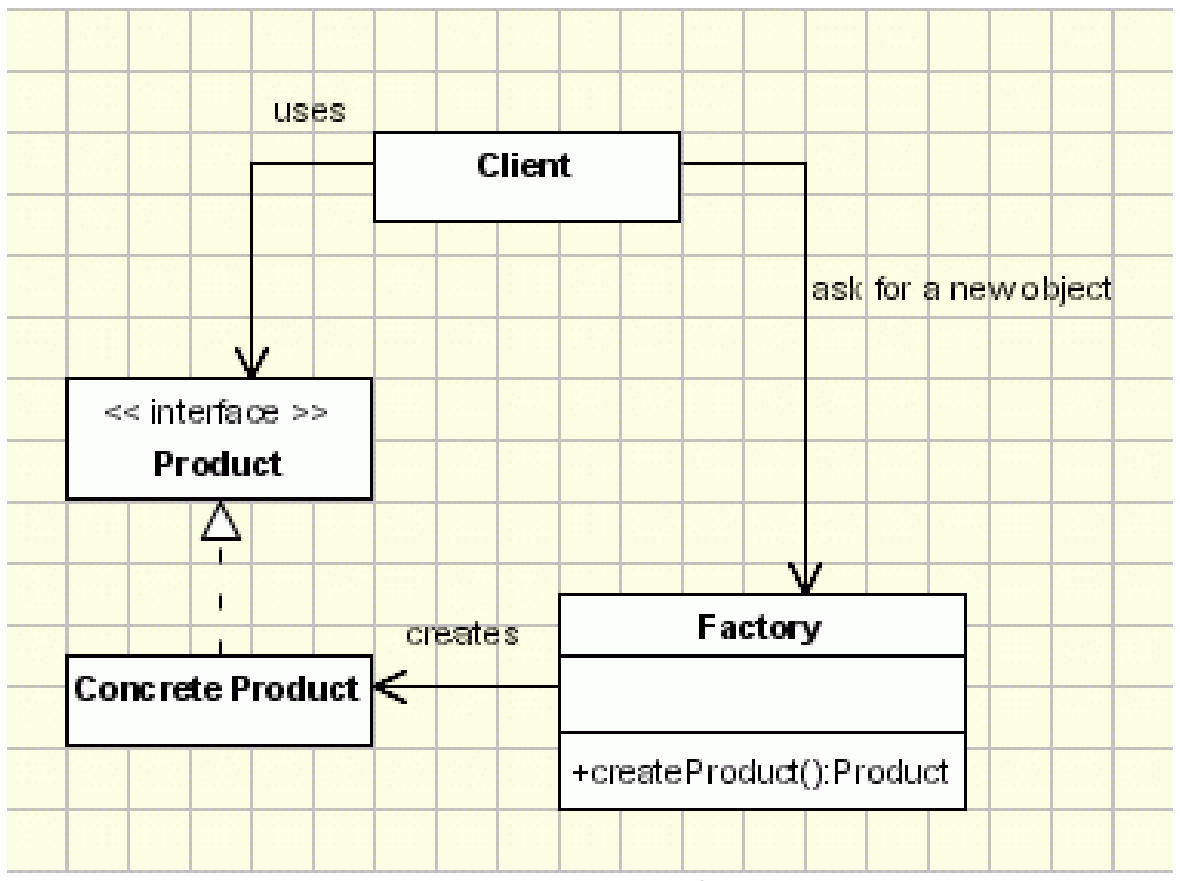

Figure 65 – Flow Diagram of Factory Pattern

Above class diagram explains how Factory Pattern works. Here the client asks the factory class to get the type of object it needs.

Class diagram

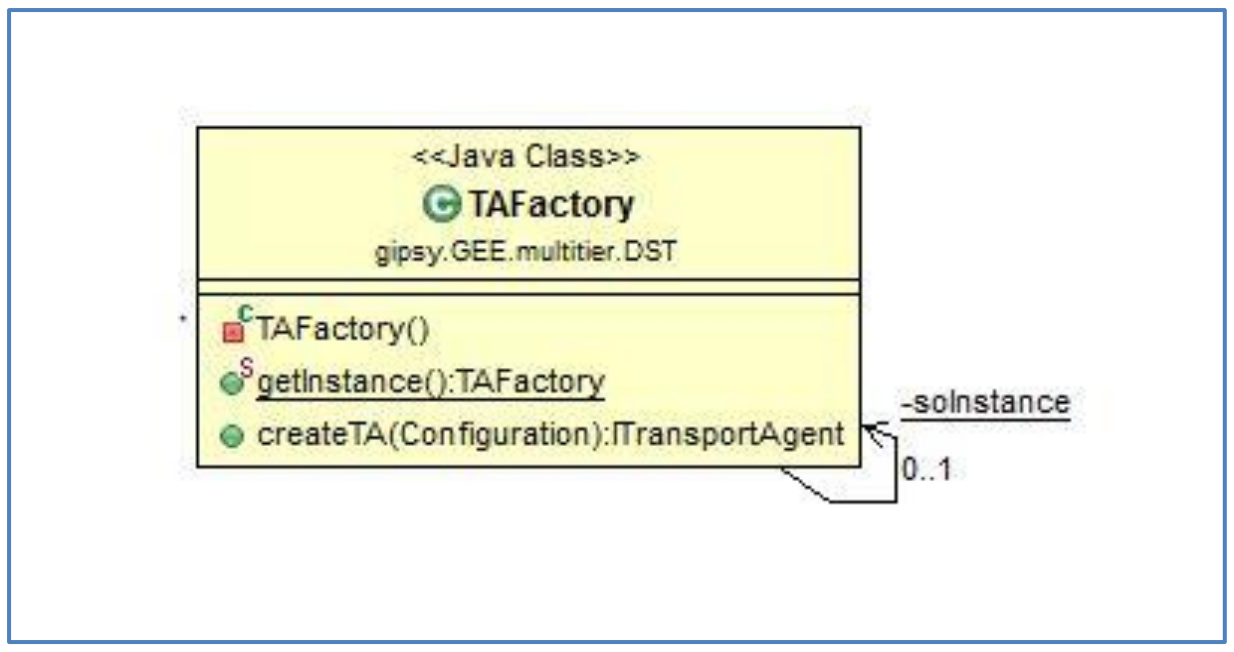

Figure 66 – TAFactory class Diagram

Example of Factory Pattern in GIPSY

```
publicclassTAFactory
{
…………………………………………………………
```

```
publicITransportAgentcreateTA(Configuration poConfiguration)
throwsMultiTierException
{
		try
{
String strTAImplClassName =
poConfiguration.getProperty(ITransportAgent.TA_IMPL_CLASS);
Class<?>oTAImplClass = Class.forName(strTAImplClassName);
Class<?>[] aoParamTypes = new Class[] { Configuration.class };
Constructor<?>oTAImplConstructor =
oTAImplClass.getConstructor(aoParamTypes);
Object[] aoArgs = new Object[]{poConfiguration};
ITransportAgentoTA =
(ITransportAgent)oTAImplConstructor.newInstance(aoArgs);
returnoTA;
}
catch(Exception oException)
{
if(Debug.isDebugOn())
{
oException.printStackTrace(System.err);
}
thrownewMultiTierException(oException);
				}
		}
}
```

### C. *Implementation of Refactoring*

All the classes and additional files are also available under the tarball file and CVS repository.

#### a) *Refactoring 1, DMARF - Replacement of DOM parser with JAXB.*

Motivation [14]

It is true that recursive solution is often more elegant and easier to understand than the iterative solution, but it should not abuse its usage. Recursion is often used without considering alternatives before using it. Complex Recursion that is hard to understand should be considered (as per Martin Flower) a "bad smell" in the code and a good candidate to be replaced with Iteration (usually in combination with some other refactoring). Moreover, iterative solutions are usually more efficient than recursive solutions as they don't incur the overhead of the multiple method calls. Recursion is used when a complex task must be performed that can be broken into the several subtasks. Recursion is implemented as a method that calls itself to solve subtasks. During the recursive call the values of the local fields of the method are placed on the method stack until the subtask performed by a recursive call is completed.

[14] Refactoring, Improving the Design of Existing Code, Fowler, M, and al, Addison-Wesley, 1999

Thus, whenever recursive method is called, local fields are put on the method stack and used again after the recursive call is completed. The general approach to refactoring could probably be to implement the alternative. Sometimes a recursive solution doesn't preserve any local fields during the recursive call, which makes refactoring much easier. This is the case with, so called, tail recursions. Tail recursions are recursions where the recursive call is the last line in the method. Tail recursions are generally considered a bad practice and should be replaced with iteration. This technique is often used in compiler implementations. A good compiler usually performs this refactoring on the fly.

<u>Mechanics</u>

- Determine the base case of the recursion. Base case, when reached, causes recursion to end. Every recursion must have a defined base case. In addition, each recursive call must make a progress towards the base case (otherwise recursive calls would be performed infinitely).
- Implement a loop that will iterate until the base case is reached.
- Make a progress towards the base case. Send the new arguments to the top of the loop instead to the recursive method.

The mechanics of some complicated refactoring other than tail recursion refactoring can also be refactored by Substitute Algorithm, which replaces the algorithm with one that’s clearer.

In the DMARF system, the *NeuralNetwork* class was modified and the method *createLinks* that originally was using DOM parser was replaced with a JAXB unmarshaller. Due to this refactoring, the following method can also be removed, since the JAXB variant takes in a file as a stream, although the validation of the XML is automatically performed by the JAXB framework, in such case the *pbValidateDTD* is replaced by validation against the XSD:

```
public final void initialize(final String pstrFilename, final boolean
pbValidateDTD)
```

Similar manner the following method can be refactored using JAXB, but the current study show only the refactoring of the *createLinks* method:

```
private final void buildNetwork(Node poNode)
```

XML file, NeuralNetwork.xml:

```
<?xml version="1.0" encoding="UTF-8"?>
<net>
        <layer type="input" index="1">
                <neuron index="1" thresh="0.5">
                        <output ref="1"/>
                        <output ref="2"/>
                </neuron>
                <neuron index="2" thresh="0.5">
        <output ref="2"/>
                </neuron>
                <neuron index="3" thresh="0.5">
        <output ref="1"/>
                        <output ref="2"/>
        <output ref="3"/>
        </neuron>
                <neuron index="4" thresh="0.5">
        <output ref="3"/>
                </neuron>
                <neuron index="5" thresh="0.5">
        <output ref="1"/>
        <output ref="3"/>
                </neuron>
        </layer>
        <layer type="hidden" index="2">
                <neuron index="1" thresh="0.5">
        <input ref="1" weight="0.42"/>
        <input ref="3" weight="0.42"/>
        <input ref="5" weight="0.42"/>
        <output ref="1"/>
        </neuron>
                <neuron index="2" thresh="0.5">
        <input ref="1" weight="0.42"/>
        <input ref="2" weight="0.42"/>
        <input ref="3" weight="0.42"/>
        <output ref="1"/>
        </neuron>
        <neuron index="3" thresh="0.5">
        <input ref="3" weight="0.42"/>
        <input ref="4" weight="0.42"/>
        <input ref="5" weight="0.42"/>
        <output ref="1"/>
        </neuron>
        </layer>
        <layer type="output" index="3">
        <neuron index="1" thresh="1.0">
        <input ref="1" weight="0.56"/>
        <input ref="2" weight="0.56"/>
        <input ref="3" weight="0.56"/>
        </neuron>
        </layer>
</net>
```

Schema generated, *NeuralNetwork.XSD:*

```
<?xml version="1.0" encoding="UTF-8"?>
<!--W3C Schema generated by XMLSpy v2011 sp1
(http://www.altova.com)-->
<xs:schema xmlns:xs="http://www.w3.org/2001/XMLSchema">
  <xs:element name="output">
        <xs:complexType>
                <xs:attribute name="ref" use="required">
                        <xs:simpleType>
                                <xs:restriction
base="xs:byte">
                                        <xs:enumeration
value="1"/>
                                        <xs:enumeration
value="2"/>
```

```
						<xs:enumeration value="3"/>
					</xs:restriction>
				</xs:simpleType>
			</xs:attribute>
		</xs:complexType>
	</xs:element>
	<xs:element name="neuron">
		<xs:complexType>
			<xs:choice>
				<xs:element ref="output" maxOccurs="unbounded"/>
				<xs:sequence>
					<xs:element ref="input" maxOccurs="unbounded"/>
					<xs:element ref="output" minOccurs="0"/>
				</xs:sequence>
			</xs:choice>
			<xs:attribute name="thresh" use="required">
				<xs:simpleType>
					<xs:restriction base="xs:decimal">
						<xs:enumeration value="0.5"/>
						<xs:enumeration value="1.0"/>
					</xs:restriction>
				</xs:simpleType>
			</xs:attribute>
			<xs:attribute name="index" use="required">
				<xs:simpleType>
					<xs:restriction base="xs:byte">
						<xs:enumeration value="1"/>
						<xs:enumeration value="2"/>
						<xs:enumeration value="3"/>
						<xs:enumeration value="4"/>
						<xs:enumeration value="5"/>
					</xs:restriction>
				</xs:simpleType>
			</xs:attribute>
		</xs:complexType>
	</xs:element>
	<xs:element name="net">
		<xs:complexType>
			<xs:sequence>
				<xs:element ref="layer" maxOccurs="unbounded"/>
			</xs:sequence>
		</xs:complexType>
	</xs:element>
	<xs:element name="layer">
		<xs:complexType>
			<xs:sequence>
				<xs:element ref="neuron" maxOccurs="unbounded"/>
			</xs:sequence>
			<xs:attribute name="type" use="required">
				<xs:simpleType>
					<xs:restriction base="xs:string">
						<xs:enumeration value="hidden"/>
						<xs:enumeration value="input"/>
```

```
						<xs:enumeration value="output"/>
					</xs:restriction>
				</xs:simpleType>
			</xs:attribute>
			<xs:attribute name="index" use="required">
				<xs:simpleType>
					<xs:restriction base="xs:byte">
						<xs:enumeration value="1"/>
						<xs:enumeration value="2"/>
						<xs:enumeration value="3"/>
					</xs:restriction>
				</xs:simpleType>
			</xs:attribute>
		</xs:complexType>
	</xs:element>
	<xs:element name="input">
		<xs:complexType>
			<xs:attribute name="weight" use="required">
				<xs:simpleType>
					<xs:restriction base="xs:decimal">
						<xs:enumeration value="0.42"/>
						<xs:enumeration value="0.56"/>
					</xs:restriction>
				</xs:simpleType>
			</xs:attribute>
			<xs:attribute name="ref" use="required">
				<xs:simpleType>
					<xs:restriction base="xs:byte">
						<xs:enumeration value="1"/>
						<xs:enumeration value="2"/>
						<xs:enumeration value="3"/>
						<xs:enumeration value="4"/>
						<xs:enumeration value="5"/>
					</xs:restriction>
				</xs:simpleType>
			</xs:attribute>
		</xs:complexType>
	</xs:element>
</xs:schema>
```

Classes generated from schema (*NeuralNetwork.XSD*), package *marf.Classification.NeuralNetwork.jaxb_generated*

*Input.java*
*Layer.java*
*Net.java*
*Neuron.java*
*Output.java*
*ObjectFactory.java*

```
/**
* Refactor of createLinks using JAXB The initialize method
needs to be refactored to use JAXB instead of the DOM model
*/
private void createLinks(File file) throws
ClassificationException {
BufferedInputStream bis = null;
try {
JAXBContext jc = JAXBContext.newInstance(Net.class);
Unmarshaller unmarshaller = jc.createUnmarshaller();
bis = new BufferedInputStream(new FileInputStream(file));
Net net = (Net) unmarshaller.unmarshal(bis);

// Next map the object obj to the instance variables, as on the
DOM
for (marf.Classification.NeuralNetwork.jaxb_generated.Layer
layer : net.getLayer()) {
switch (layer.getType()) {
case "input":
	this.oCurrentLayer = this.oInputs;
	this.iCurrenLayer = 0;
	break;
case "output":
	this.oCurrentLayer = this.oOutputs;
	this.iCurrenLayer = this.oLayers.size() - 1;
	break;
default:
	this.iCurrenLayer = ++this.iCurrLayerBuf;
	this.oCurrentLayer = (Layer)
this.oLayers.get(this.iCurrenLayer);
}
for (marf.Classification.NeuralNetwork.jaxb_generated.Neuron
neuron : layer.getNeuron()) {
this.oCurrNeuron =
this.oCurrentLayer.getNeuron(String.valueOf(neuron.getIndex()));
for (Object content : neuron.getContent()) {
	String strIndex = String.valueOf(neuron.getIndex());
	switch(content.getClass().getSimpleName()){
	// TODO: move the individual cases into distinct
private methods or merge them into one single method and provide the
class
	case "Input": Input input = (Input) content;
	strIndex = String.valueOf(input.getRef());
	double dWeight = input.getWeight().doubleValue();
	if(this.iCurrenLayer > 0){
			Neuron oNeuronToAdd =
((Layer)this.oLayers.get(this.iCurrenLayer - 1)).getNeuron(strIndex);
			if(oNeuronToAdd == null){
				throw new
ClassificationException("Cannot find neuron " + strIndex+ " in layer " +
(this.iCurrenLayer - 1));
			}

this.oCurrNeuron.addInput(oNeuronToAdd, dWeight);
		}
		else{
			throw new
ClassificationException("Input element not allowed in input layer");
		}

	case "Output": Output output = (Output)content;
	strIndex = String.valueOf(output.getRef());
	if(this.iCurrenLayer >= 0){
			Neuron oNeuronToAdd =
((Layer)this.oLayers.get(this.iCurrenLayer + 1)).getNeuron(strIndex);
			if(oNeuronToAdd == null){
				throw new
ClassificationException("Cannot find neuron " + strIndex+ " in layer " +
(this.iCurrenLayer + 1));
			}

this.oCurrNeuron.addOutput(oNeuronToAdd);
```

```
				}
			}
		}
		}
		}

		} catch (JAXBException | FileNotFoundException ex) {

		Logger.getLogger(NeuralNetwork.class.getName()).log(Level.S
EVERE, null, ex);
		} finally {
		if (bis != null) {
		try {
		bis.close();
		} catch (IOException ex) {

		Logger.getLogger(NeuralNetwork.class.getName()).log(Level.S
EVERE, null, ex);
		}
		}
		}
		}
```

Reduced from 147 to 72 lines (including comments) and can be reduced even more, plus the recursivety is eliminated and replaced by for loops.

The *NeuralNetwork.java.diff* file is attached to the CVS repository and tarball file and can be easily opened with TortoiseSVN client in order to have the changes displayed visually.

Simple JUnit test case was added to the DMARF project under dedicated folder:

```
<dmarf_root>/tests/marf/Classification/NeuralNetwork
```

```
NeuralNetwork.xml
NeuralNetworkTest.java
```

b) <u>Refactoring 2, GIPSY - Feature Envy refactoring gipsy.Configuration</u>

This is a simple example of refactoring for feature envy code smell for demonstration purposes, as identified by JDeodorant tool (brief description of the tool is available in the annex). The refactoring does not add much value since it is just a simple *MoveMethod* refactoring type, and more important “bad smells” have to be addressed first. The class *gipsy.GEE.multitier.DST.TAFactory::createTA(Configuration)* has FeatureEnvy code smell to *gipsy.Configuration*:

```
public ITransportAgent createTA(Configuration
poConfiguration)
```

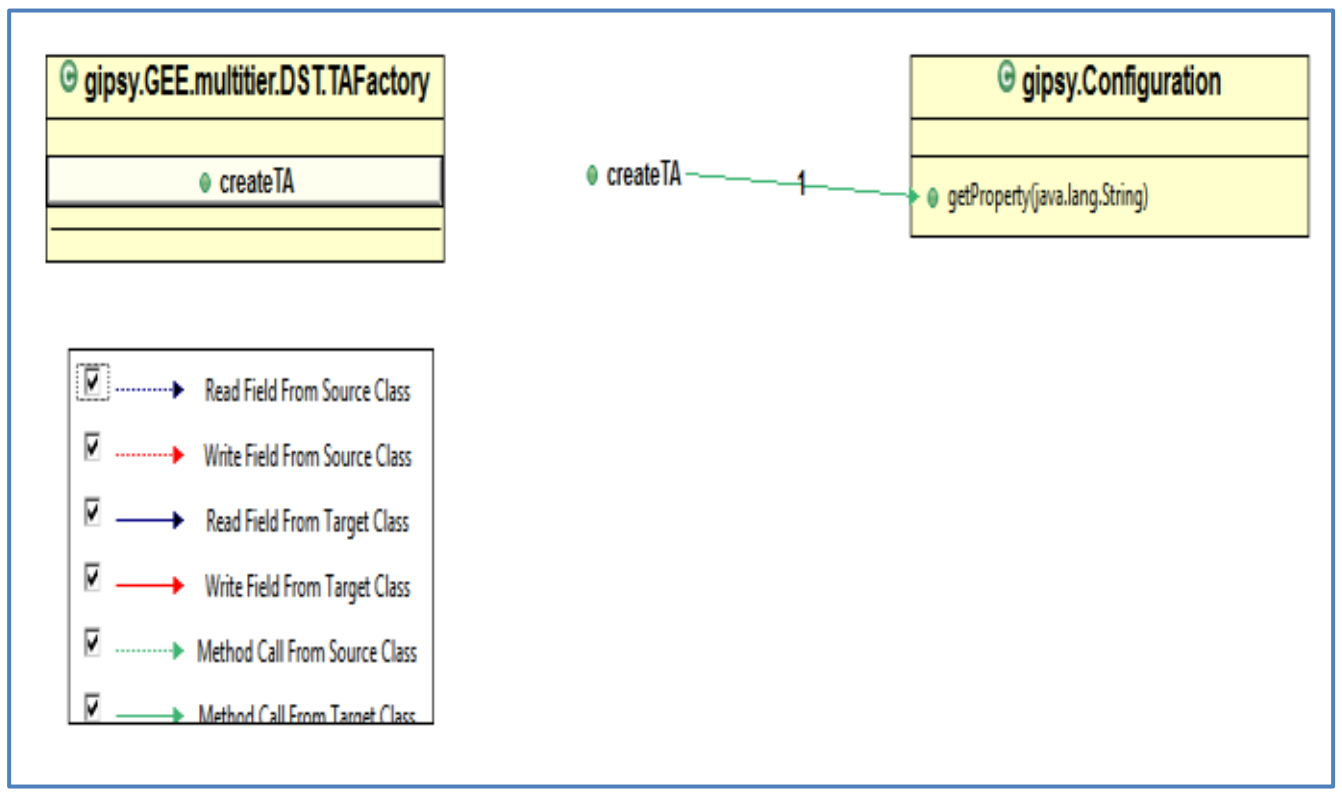

Figure 67 – FeatureEnvy refactoring

Original method source code in *gipsy.GEE.multiitier.DST.TAFactory*:

```
    /**
     * Create a TA instance based on configuration.
     *
     * @param poConfiguration - The TA configuration
     * @return the TA instance
     * @throws MultiTierException
     */
    public ITransportAgent createTA(Configuration poConfiguration)
    throws MultiTierException{
            try{
                    String strTAImplClassName =
poConfiguration.getProperty(ITransportAgent.TA_IMPL_CLASS);

                    Class<?> oTAImplClass =
Class.forName(strTAImplClassName);
                    Class<?>[] aoParamTypes = new Class[] {
Configuration.class };
                    Constructor<?> oTAImplConstructor =
oTAImplClass.getConstructor(aoParamTypes);

                    Object[] aoArgs = new Object[]{poConfiguration};
                    ITransportAgent oTA =
(ITransportAgent)oTAImplConstructor.newInstance(aoArgs);
                    return oTA;
            }
            catch(Exception oException)     {
                    if(Debug.isDebugOn())                   {
                            oException.printStackTrace(System.err);
                    }
                            throw new
MultiTierException(oException);
            }
    }
}
```

Refactoring applied is a simple *MoveMethod* :

```
/**
     * Create a TA instance based on configuration.
     *
     * @param poConfiguration - The TA configuration
     * @return the TA instance
     * @throws MultiTierException
     */
    public ITransportAgent createTA(Configuration poConfiguration)
    throws MultiTierException{
            return poConfiguration.createTA(poConfiguration);
    }
}
```

And in *Configuration.java*, the new method created:

```
     /**
     * Create a TA instance based on configuration.
     * @return  the TA instance
     * @throws MultiTierException
     */
    public ITransportAgent createTA(Configuration poConfiguration)
throws MultiTierException {
            try {
                    String strTAImplClassName =
getProperty(ITransportAgent.TA_IMPL_CLASS);
                    Class<?> oTAImplClass =
Class.forName(strTAImplClassName);
                    Class<?>[] aoParamTypes = new Class[] {
Configuration.class };
                    Constructor<?> oTAImplConstructor =
oTAImplClass.getConstructor(aoParamTypes);
                    Object[] aoArgs = new Object[] { poConfiguration
};
                    ITransportAgent oTA = (ITransportAgent)
oTAImplConstructor.newInstance(aoArgs);
                    return oTA;
            } catch (Exception oException) {
                    if (Debug.isDebugOn()) {
                            oException.printStackTrace(System.err);
                    }
                    throw new MultiTierException(oException);
            }
    }
```

The refactoring applied increased the cohesion in the class *Configuration.java*, reduced the coupling in *TAFactory.java* and eliminated the Feature Envy code bad smell.

Simple unit test have also been created, file *TAFactoryTest.java*:

```
    @Test
    public void testCreateTA() {

            TAFactory factory = TAFactory.getInstance();
            assertNotNull(factory);
            Configuration config = new Configuration();
            // manually set the property as the transport did not come on the
default settings (needs to investigate the default loader)
            // initialize to JMS transport agent
            config.setProperty("gipsy.GEE.TA.implementation",
"gipsy.GEE.IDP.DemandGenerator.jms.JMSTransportAgent");
            assertNotNull(config);
            config.list(System.out);

            // Having the agent config, create the connection
            try {
```

```
                    ITransportAgent ta = factory.createTA(config);
                    assertNotNull(ta);
                    ta.getDemand();
                    fail("Exception should have been thrown trying to
connect to localhost:7676 to aquire JMS connection");
            } catch (MultiTierException e) {
                    fail("Exception creating TA factory : " +
e.getMessage());
            } catch (DMSException e) {
                    assertTrue(e instanceof DMSException);
            }
    }
```

### c) Refactoring 3, GIPSY - God Class refactoring

An example of a God Class code smell have been identified in the following class described further down.

#### Motivation

When a class has become bloated with methods calling out to increased number of classes, then it is time to refactor that class. Developers kept pouring layers after layers into one class, and that became incredibly complex, low cohesion and highly coupled to other classes around it.

#### Mechanics

Refactoring the God Class has to first reduce it into a number of different classes, second, attaching some meaning to the classes the refactoring process has created:

- create utility classes;
- create test harness and decrease method visibility;
- look for candidates of pulling out into second class;
- focus on high cohesion and low coupling;
- avoid generalization, or delay until later iterations.

Here is an example of refactoring a god class in GIPSY system and how a second class has been extracted and the method was kept as a delegate of the first class:

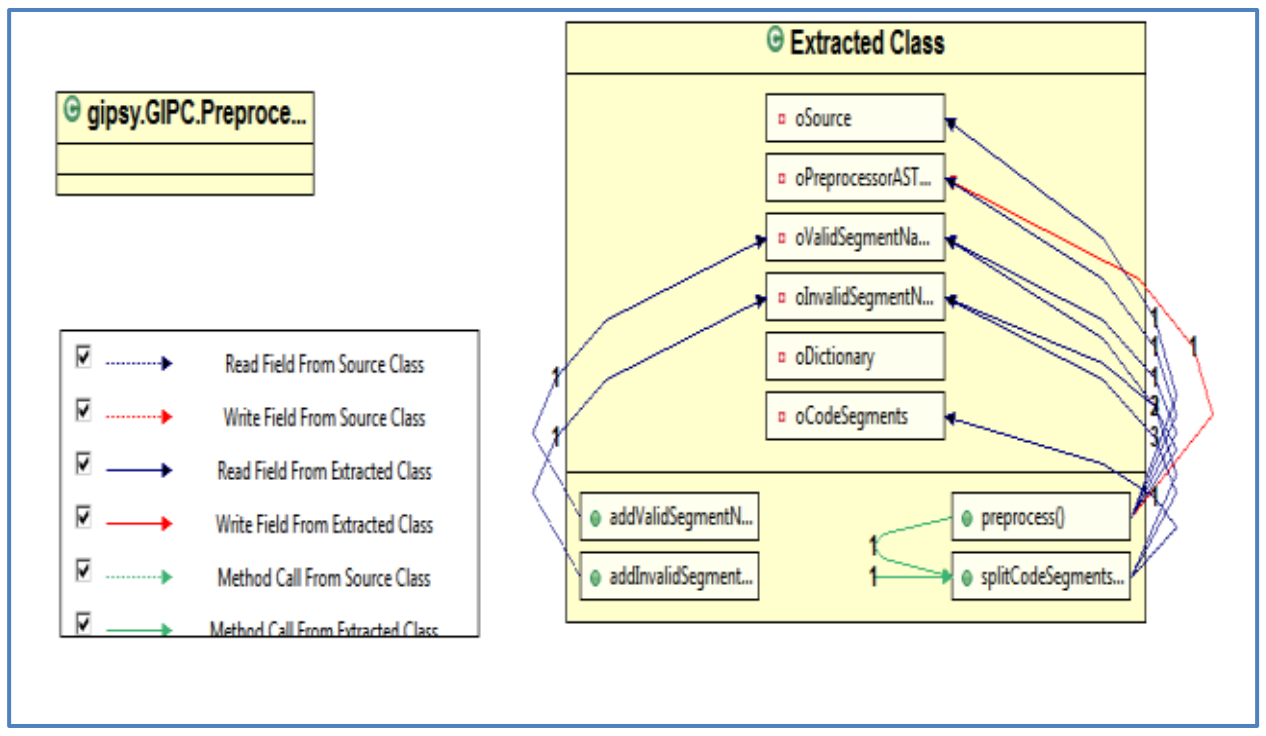

Figure 68 – God Class refactoring

Original attributes and methods, *gipsy.GIPC.Preprocessing.Preprocessor.java*

Attributes:

```
 /**
  * Source input stream.
  */
 protected InputStream oSource = null;

 /**
  * Root of the AST created by the PreprocessorParser.
  */
 protected SimpleNode oPreprocessorASTRoot = null;

 /**
  * List of references to the source code segments of
  * a GIPSY program.
  */
 protected Vector<CodeSegment> oCodeSegments = new
Vector<CodeSegment>();

 /**
  * List of valid segment names.
  */
 protected Vector<String> oValidSegmentNames = new
Vector<String>();

 /**
  * List of invalid segment names.
  */
 protected Vector<String> oInvalidSegmentNames = new
Vector<String>();

 /**
  * Embryo of the symbol dictionary.
  * Contains symbols from the #funcdecl and #typedecl
  * segments.
  */
 protected Dictionary oDictionary = new Dictionary();
```

Methods:

```
 /**
  * The body of the preprocessing.
  *
  * @throws GIPCException if there was a parsing or otherwise error.
  */
 public void preprocess()
 throws GIPCException
 {
         try
         {
                 PreprocessorParser oPreprocessorParser = new
PreprocessorParser(oSource);
                 oPreprocessorParser.parse();
                 this.oPreprocessorASTRoot =
oPreprocessorParser.getPreprocessorASTRoot();

                 // Valid takes precedence over invalid if specified
                 if(this.oInvalidSegmentNames.size() > 0 &&
this.oValidSegmentNames.size() > 0)
                 {
                         this.oInvalidSegmentNames.clear();
                 }

                 splitCodeSegments(this.oPreprocessorASTRoot);
```

```
//produceImperativeStubs(this.oPreprocessorASTRoot);
		}
		catch(NullPointerException e)
		{
			e.printStackTrace();
			throw new GIPCException
			(
				"Source InputStream has not been
initialized.\n" +
				"HINT: Make sure that the parameter to
the constructor or setSourceStream() is not null.",
				e
			);
		}
		catch(Exception e)
		{
			throw new GIPCException(e);
		}
	}
```

They were extracted in a new class and method:

Attributes:

```
private PreprocessorProduct preprocessorProduct = new
PreprocessorProduct();
```

Methods changed in the original class:

```
public Preprocessor(InputStream poGIPSYCode)  throws GIPCException {
		this.oSource = poGIPSYCode;
	}
```

To

```
public Preprocessor(InputStream poGIPSYCode) throws GIPCException  {
		preprocessorProduct.setOSource(poGIPSYCode);
	}
```

Then, from

```
public void preprocess()   throws GIPCException   {
		try {
			PreprocessorParser oPreprocessorParser = new
PreprocessorParser(oSource);
			oPreprocessorParser.parse();
			this.oPreprocessorASTRoot =
oPreprocessorParser.getPreprocessorASTRoot();

			// Valid takes precedence over invalid if specified
			if(this.oInvalidSegmentNames.size() > 0 &&
this.oValidSegmentNames.size() > 0)
			{
				this.oInvalidSegmentNames.clear();
			}

			splitCodeSegments(this.oPreprocessorASTRoot);

//produceImperativeStubs(this.oPreprocessorASTRoot);
		}
		catch(NullPointerException e)
		{
			e.printStackTrace();
			throw new GIPCException
			(
				"Source InputStream has not been
initialized.\n" +
				"HINT: Make sure that the parameter to
the constructor or setSourceStream() is not null.",
				e
			);
		}
		catch(Exception e)
		{
			throw new GIPCException(e);
		}
	}
```

To

```
public void preprocess()  throws GIPCException    {
		preprocessorProduct.preprocess();
	}
```

***Note***:
The list of the methods changed presented here is limited, the full list and the difference files are attached as annex to the study.

A new class was created, *PreprocessorProduct.java* that includes the delegation from the original class:

```
package gipsy.GIPC.Preprocessing;


import java.io.InputStream;
import java.util.Vector;
import gipsy.storage.Dictionary;
import gipsy.GIPC.GIPCException;
import marf.util.Debug;

public class PreprocessorProduct {
	private InputStream oSource = null;
	private SimpleNode oPreprocessorASTRoot = null;
	private Vector<String> oValidSegmentNames = new Vector<String>();
	private Vector<String> oInvalidSegmentNames = new
Vector<String>();
	private Dictionary oDictionary = new Dictionary();
	private Vector<CodeSegment> oCodeSegments = new
Vector<CodeSegment>();

	public InputStream getOSource() {
		return oSource;
	}

	public void setOSource(InputStream oSource) {
		this.oSource = oSource;
	}

	public SimpleNode getOPreprocessorASTRoot() {
		return oPreprocessorASTRoot;
	}

	public void setOPreprocessorASTRoot(SimpleNode
oPreprocessorASTRoot) {
		this.oPreprocessorASTRoot = oPreprocessorASTRoot;
	}
```

```
   public Vector<String> getOValidSegmentNames() {
            return oValidSegmentNames;
   }

   public void setOValidSegmentNames(Vector<String>
oValidSegmentNames) {
            this.oValidSegmentNames = oValidSegmentNames;
   }

   public Vector<String> getOInvalidSegmentNames() {
            return oInvalidSegmentNames;
   }

   public void setOInvalidSegmentNames(Vector<String>
oInvalidSegmentNames) {
            this.oInvalidSegmentNames = oInvalidSegmentNames;
   }

   public Dictionary getODictionary() {
            return oDictionary;
   }

   public void setODictionary(Dictionary oDictionary) {
            this.oDictionary = oDictionary;
   }

   public Vector<CodeSegment> getOCodeSegments() {
            return oCodeSegments;
   }

   public void setOCodeSegments(Vector<CodeSegment> oCodeSegments)
{
            this.oCodeSegments = oCodeSegments;
   }

   public void addValidSegmentName(String pstrName) {
            if (pstrName == null || pstrName.equals("")) {
                     throw new IllegalArgumentException("Segment
name cannot be null or empty string");
            }
            this.oValidSegmentNames.add(pstrName);
   }

   public void addInvalidSegmentName(String pstrName) {
            if (pstrName == null || pstrName.equals("")) {
                     throw new IllegalArgumentException("Segment
name cannot be null or empty string");
            }
            this.oInvalidSegmentNames.add(pstrName);
   }

   /**
    * The body of the preprocessing.
    * @throws GIPCException  if there was a parsing or otherwise error.
    */
   public void preprocess() throws GIPCException {
            try {
                     PreprocessorParser oPreprocessorParser = new
PreprocessorParser(oSource);
                     oPreprocessorParser.parse();
                     this.oPreprocessorASTRoot =
oPreprocessorParser.getPreprocessorASTRoot();
                     if (this.oInvalidSegmentNames.size() > 0 &&
this.oValidSegmentNames.size() > 0) {
                              this.oInvalidSegmentNames.clear();
                     }
                     splitCodeSegments(this.oPreprocessorASTRoot);
            } catch (NullPointerException e) {
                     e.printStackTrace();
                     throw new GIPCException("Source InputStream has
not been initialized.\n"
                                                  + "HINT: Make sure that the
parameter to the constructor or setSourceStream() is not null.", e);
            } catch (Exception e) {
                     throw new GIPCException(e);
            }
   }

   /**
    * Splits intensional and imperative code segments into separate code
pieces to be later fed to appropriate compilers.
    * @throws GIPCException
    */
   public void splitCodeSegments(SimpleNode poRoot) throws
GIPCException {
            SimpleNode oCurrentNode = poRoot;
            int i = oCurrentNode.jjtGetNumChildren();
            for (int c = 0; c < i; c++) {
                     SimpleNode oChild = (SimpleNode)
oCurrentNode.jjtGetChild(c);
                     switch (oChild.id) {
                     case
PreprocessorParserTreeConstants.JJTCODESEGMENT: {

Debug.debug("JJTCODESEGMENT!!!!!");
                              String strLanguage =
oChild.getLexeme().split("\\s")[0].substring(1);
                              if (this.oInvalidSegmentNames.size() !=
0 || this.oValidSegmentNames.size() != 0) {
                                       String strError = "Language
name " + strLanguage + "is not recognized as valid";
                                       if
(this.oValidSegmentNames.size() > 0) {
                                                if
(this.oValidSegmentNames.contains(strLanguage) == false) {
                                                         throw
new GIPCException(strError);
                                                }
                                       }
                                       if
(this.oInvalidSegmentNames.size() > 0) {
                                                if
(this.oInvalidSegmentNames.contains(strLanguage) == true) {
                                                         throw
new GIPCException(strError);
                                                }
                                       }
                              }
                              String strCode =
oChild.getLexeme().replaceFirst("#" + strLanguage, "").trim();
                              this.oCodeSegments.add(new
CodeSegment(strLanguage, strCode));
                              break;
                     }
                     case
PreprocessorParserTreeConstants.JJTPROTOTYPES: {
                              Debug.debug("PROTOTYPES: " +
oChild);
                              break;
                     }
                     case PreprocessorParserTreeConstants.JJTTYPES:
{
                              Debug.debug("TYPES: " + oChild);
                              break;
                     }
                     case
PreprocessorParserTreeConstants.JJTFUNCDECLS:
                     case
PreprocessorParserTreeConstants.JJTTYPEDECLS: {
                              Debug.debug("DECL: " + oChild);
                              break;
                     }
                     default: {
```

```
Debug.debug("Preprocessor.splitCodeSegments(): unhandled " +
PreprocessorParserTreeConstants.jjtNodeName[oChild.id]);
                        }
                    }
                    splitCodeSegments(oChild);
            }
    }
}
```

Two unit test cases were create in the test folder, *PreprocessorTest.java*, that runs the preprocessor component first with a valid Lucid file and second with an invalid file, expecting an exception:

```
    /*
     * Run the preprocessor with a valid
     */
    @Test
    public void testPreprocess1() {
            String testFileName = this.getClass().getResource("").getPath()
+ "/PreprocessorInputTest_good.ipl";
            InputStream bis = null;
            try {
                    bis = new FileInputStream(testFileName);
                    Preprocessor preprocessor = new Preprocessor(bis);
                    assertNotNull(preprocessor);
                    preprocessor.preprocess();
                    int code_segment_size = 2;

assertEquals(preprocessor.getCodeSegments().size(), code_segment_size);
            } catch (FileNotFoundException e) {
                    fail("Test file not found : " + e.getMessage());
            } catch (GIPCException e) {
                    fail("Preprocessor exception : " + e.getMessage());
            } finally {
                    try {
                            bis.close();
                    } catch (IOException e) {
                            // TODO Auto-generated catch block
                            fail("Exception closing the preprocessor
file stream : " + e.getMessage());
                    }
            }

    }

    /*
     * This test is expected to fail with TokenMrgError exception, since the
ipl file is not valid
     */
    @Test(expected = gipsy.GIPC.util.TokenMgrError.class)
    public void testPreprocess2() {
            String testFileName = this.getClass().getResource("").getPath()
+ "/PreprocessorInputTest_bad.ipl";
            InputStream bis = null;
            try {
                    bis = new FileInputStream(testFileName);
                    Preprocessor preprocessor = new Preprocessor(bis);
                    assertNotNull(preprocessor);
                    preprocessor.preprocess();
                    fail("Exception should have been thrown by
preprocessor...");
            } catch (FileNotFoundException e) {
                    fail("Test file not found : " + e.getMessage());
            } catch (GIPCException e) {
                    System.out.println(e.getCause().toString());
            } finally {
                    try {
                            bis.close();
                    } catch (IOException e) {
                            // TODO Auto-generated catch block
                            fail("Exception closing the preprocessor
file stream : " + e.getMessage());
                    }
            }

    }
```

d) *Refactoring 4, GIPSY - Facade refactoring*

Implementation of *Facade pattern* in the Transport Agent. Currently the implementation of the transport agent is provided by the method *setTransportAgent* which takes as parameter the type of the agent:

```
public void setTransportAgent(EDMFImplementation poDMFImp) {
        try {
                switch(poDMFImp) {
                        case JINI:
                        {
                                this.oTA = new JINITA();
                                break;
                        }
                        case JMS:
                        {
                                this.oTA = new
JMSTransportAgent();
                                break;
                        }
                        default:
                        {
                                this.oTA = null;
                                throw new
GIPSYRuntimeException("Unknown DMF Implementation Instance Type: "
+ poDMFImp);
                        }

                }
        } catch(Exception e) {
                e.printStackTrace(System.err);
        }
    }
```

Motivation

The client community should not be aware of the intricacies of the Transport Agent, and simple handler should be exposed, the implementation is hidden. The client community needs a simplified interface to the overall functionality of a complex subsystem. The Facade class is a simple facilitator, and should not become an all-knowing oracle or "god" object. It has to delegate the tasks throughout the subsystem which implementation hides.

Mechanism

Simple diagram showing the Facade pattern functionality is presented further down. In order to refactor the source code, few steps have been executed[15]:

1. Identified a simpler, unified interface for the subsystem or component;
2. Designed a 'wrapper' class that encapsulates the subsystem.
3. The facade/wrapper captures the complexity and collaborations of the component, and delegates to the appropriate methods.
4. The client uses (is coupled to) the Facade only.
5. Consider whether additional Facades would add value

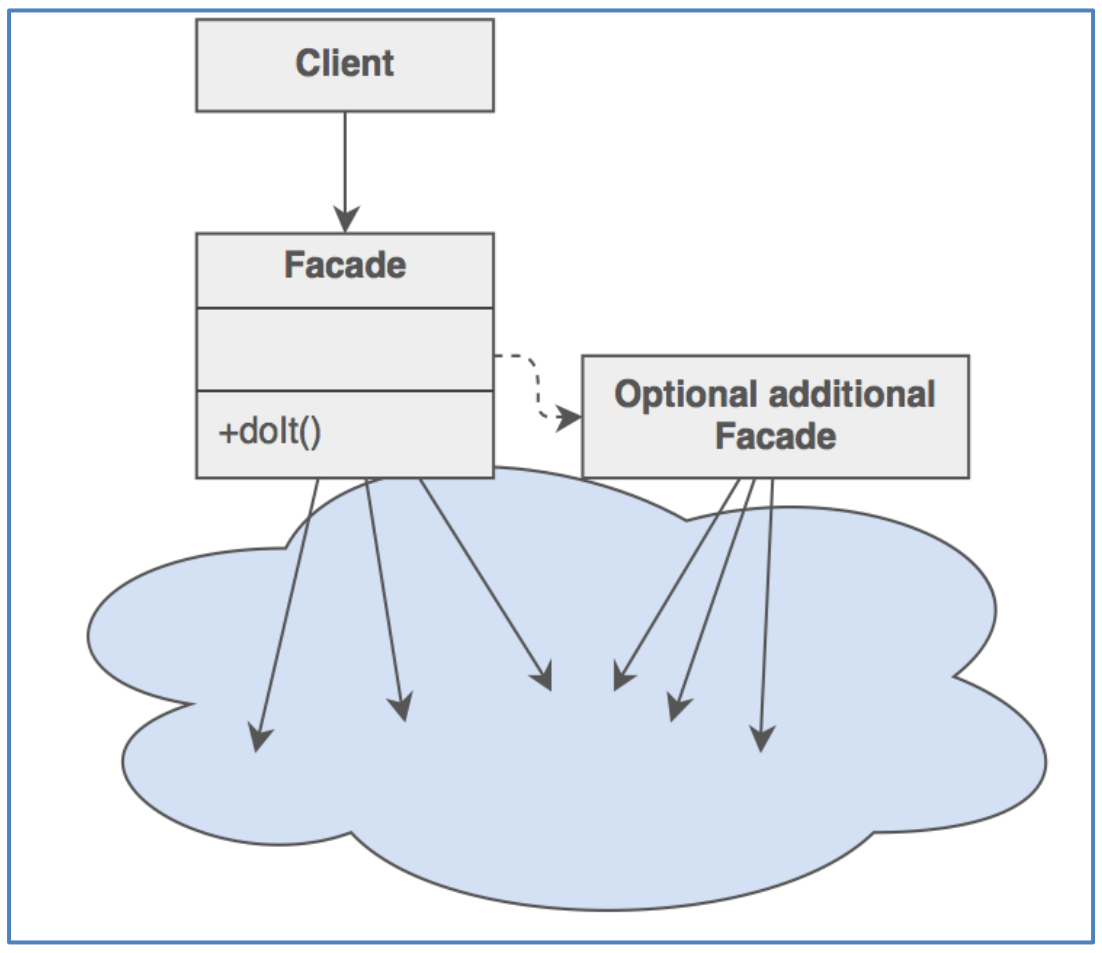

Figure 69 – Façade pattern on TransportAgent

Implementation is in progress and it is exposed as more details are added to the Facade object. However, might not be completely available at the moment of submission of the study due to the close deadline.

## VI. CONCLUSION

The paper focuses on the DMARF and GIPSY architecture and its implementation. The initial emphasis was on the extraction of the high level architectural aspects. These words are then directed towards identification of actors and stakeholders based on which the use cases are derived. Based on the concepts identified the architectural design and conceptual model was proposed. Forwarding into the implementation, bad code smells are identified using the tools (McCabe IQ, JDeodorant) and explained. Altova UModel and ObjectAid UML Explorer have been used as a reverse engineering tool to derive the architecture of the two case studies in order to compare the architectural design with the conceptual classes which have been refactored.

Besides, we have taken measurable steps to collaborate both the case studies and identified the fused architecture, in order to show the ability of DMARF with GIPSY architectures. We have used the JDeodorant and SonarQube to analyze the quality of the case studies. Finally, we have implemented four refactoring implementations for which supporting test cases and respective results have been provided as well.

The study actually gives the opportunity to apply good engineering practices in the software system architecture design. It started with an initial system implementation, identified the requirements, bad code smell and applied the refactoring in the system design and architecture.

The walkthrough of the design patterns and the joint usage of the architectural patterns upon the systems under study provide basic model of a systematic approach in the design and implementation. The refactoring, a continuous effort during the lifecycle of a system is an important factor of the quality of a system and its success to accomplish its scope.

The study exposed readers to certain approaches used to improve the quality of a software system.

## VII. METRICS

### A. Metrics definition

This section is concerned with the formal definition of the metrics used in this study. Basic metrics are collected from the research background studies, and the list along with the definition is presented further down.

- Number of Java files: Total number of Java files the software system is composed of;

- Number of classes: Total number of classes (OOP) written in Java;

- Number of methods: Total number of methods within the Java files;

- Number of lines of Java code: Total number of line of Java code of the software system source code.

[15] Refactoring, Improving the Design of Existing Code, Fowler, M, and al, Addison-Wesley, 1999

B. Metrics calculation/implementation details

In order to collect and calculate the metrics details, the SonarQube tool was used, which has proven to be an extremely good tool for gathering software metrics and performing basic software quality analysis. The computation results are depicted in the following section. Note that we only analyzed the Java source code strictly, although the case study project contained source code of other programming languages. The reports are presented in the appendix of the document.

**DMARF**

| METRIC | VALUE |
|---|---|
| Number of Java files | 1,024 |
| Number of Java classes | 1,054 |
| Number of methods | 7,152 |
| - accessors | 387 |
| Number of lines of code | 77,297 |

**GIPSY**

| METRIC | VALUE |
|---|---|
| Number of Java files | 602 |
| Number of Java classes | 666 |
| Number of methods | 6,262 |
| - accessors | 164 |
| Number of lines of code | 104,073 |

## VIII. CONTRIBUTION

### *A. Team members contribution*

| PM | Student | Chapter |
|---|---|---|
| 1 | i_iacob | Case study DMARF: Managing distributed MARF with SNMP. |
| 1 | i_iacob | Case study GIPSY: Towards a multi-tier runtime system for GIPSY. |
| 1 | g_mandap | Case study DMARF: Self-Optimization property in autonomic specification of distributed MARF with ASSL. |
| 1 | g_mandap | Case study GIPSY: Towards autonomic GIPSY. |
| 1 | p_tiruna | Case study DMARF: Towards security hardening of scientific distributed demand-driven and pipelined computing systems. |
| 1 | p_tiruna | Case study GIPSY: General architecture for demand migration in the GIPSY demand-driven execution engine. |
| 1 | a_gouris | Case Study DMARF: Towards Autonomic Specification of Distributed MARF with ASSL: Self-Healing. |
| 1 | a_gouris | Case Study GIPSY: Distributed eductive execution of hybrid intensional programs. |
| 1 | a_masna | Case Study DMARF: Autonomic speci cation of self-protection for Distributed MARF with ASSL |
| 1 | a_masna | Case Study GIPSY: The GIPSY architecture. |
| 1 | k_anthat | Case Study DMARF: Towards a self-forensics property in the ASSL toolset. |
| 1 | k_anthat | Case Study GIPSY : Using the General Intensional Programming System (GIPSY) for Evaluation of Higher-Order Intensional Logic (HOIL) |
| 1 | b_gujjul | Case Study DMARF: On Design and Implementation of Distributed Modular Audio Recognition Framework |
| 1 | b_gujjul | Case Study GIPSY: Advances in the Design and Implementation of a Multi-Tier Architecture in the GIPSY Environment with JAVA. |
| 1 | s_gaddam | Case Study DMARF: Distributed Modular Audio Recognition Framework (DMARF) and its Applications over Web Services. |
| 1 | s_gaddam | Case Study GIPSY: An Interactive Graph-Based Automation Assistant: A Case Study to Manage the GIPSY's Distributed Multi-tier Run-Time System. |
| 2 | i_iacob | Personas and stakeholders |
| 2 | i_iacob | Use cases |
| 2 | i_iacob | MARF over GIPSY architecture |
| 2 | g_mandap | DMARF Domain Models |
| 2 | p_tiruna | GIPSY Domain Models |
| 2 | a_gouris | MARF over GIPSY architecture |
| 2 | a_masna | Personas and stakeholders |
| 2 | k_anthat | DMARF Domain Models |
| 2 | b_gujjul | GIPSY Domain Models |
| 2 | s_gaddam | Use cases |
| 3 | i_iacob | Actual Architecture UML diagram |
| 3 | i_iacob | Refactoring – identification of code smells and system level refactoring |
| 3 | i_iacob | DMARF-Singleton pattern |
| 3 | P_tiruna | GIPSY Refactoring Suggestion |
| 3 | P_tiruna | DMARF- Factory Pattern |
| 3 | a_gouris | Refactoring – identification of code smells and system level refactoring |
| 3 | a_gouris | DMARF- Prototype Pattern |
| 3 | a_masna | DMARF Refactoring Suggestion |
| 3 | a_masna | GIPSY – Observer Pattern<br>GIPSY – Proxy Pattern |
| 3 | k_anthat | GIPSY-Decorator Pattern |
| 3 | k_anthat | Actual Architecture UML diagram |
| 3 | b_gujjul | GIPSY-Factory Pattern |
| 3 | b_gujjul | DMARF Refactoring Suggestion |
| 3 | s_gaddam | GIPSY Refactoring Suggestion |
| 3 | s_gaddam | DMARF-Strategy Pattern |
| 3 | g_mandap | Actual Architecture UML diagram |
| 3 | g_mandap | GIPSY-Singleton Pattern |
| 4 | i_iacob | Refactoring DMARF and GIPSY |
| 4 | i_iacob | JUnit tests |
| 4 | i_iacob | CVS commit of refactoring |
| 4 | g_mandap | JUnit tests |
| 4 | p_tiruna | Differences before and after refactoring CVS commit |
| 4 | a_gouris | JUnit tests, Refactoring DMARF and GIPSY |
| 4 | a_masna | Refactoring DMARF and GIPSY, JUnit tests |
| 4 | k_anthat | JUnit tests |
| 4 | b_gujjul | JUnit tests |
| 4 | s_gaddam | Differences before and after refactoring CVS commit |

*APPENDIX 1*

The metrics presented in the metrics section have been gathered using the SonarQube tool. Attached are the reports created by the tool. We based our metrics completely on SonarQube and assume these metrics are correct.

<u>DMARF report</u>

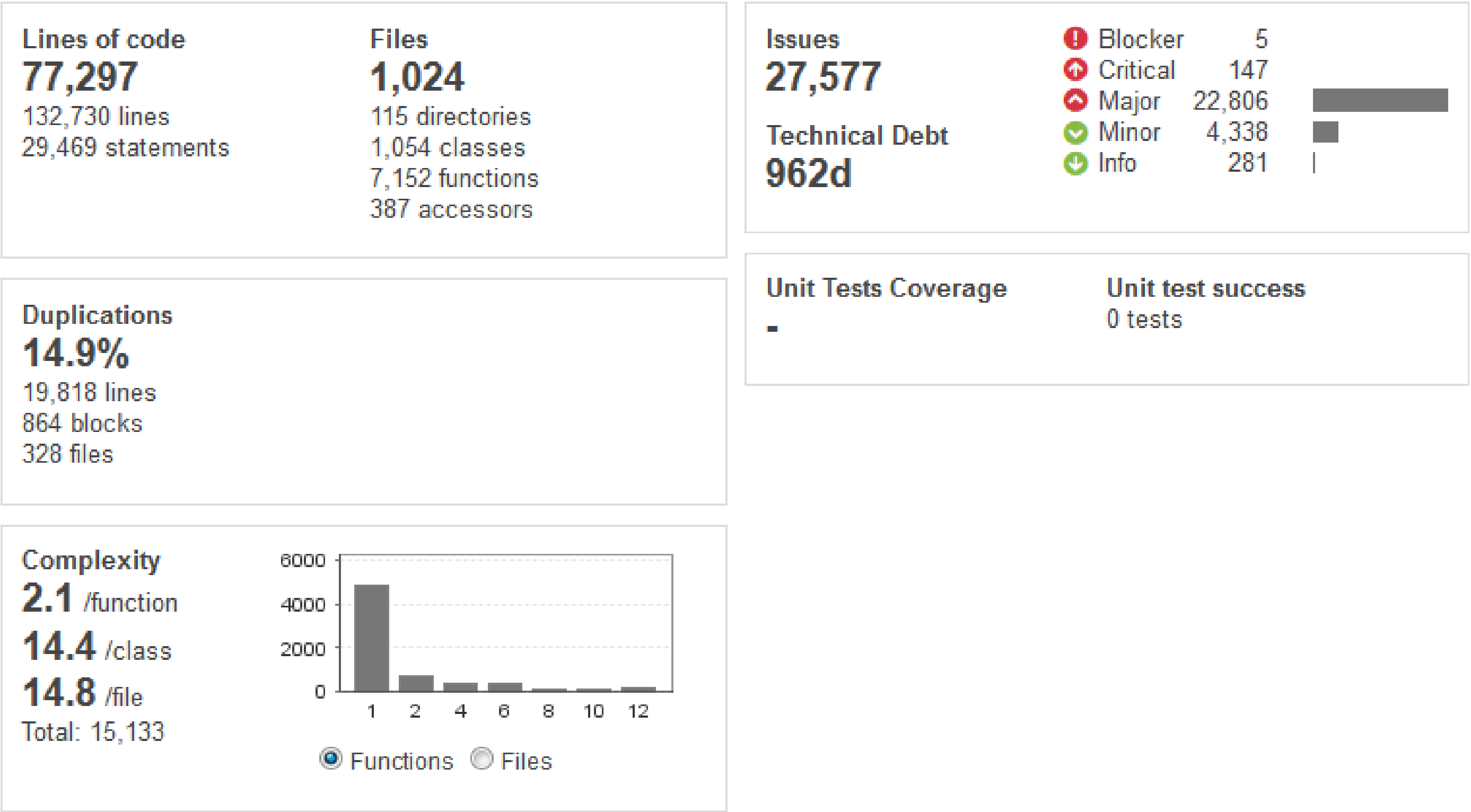

GIPSY report

**Lines of code**
**104,073**
140,250 lines
56,035 statements

**Files**
**602**
106 directories
666 classes
6,262 functions
164 accessors

**Issues**
**44,326**

**Technical Debt**
**1,057d**

| | |
|---|---|
| Blocker | 210 |
| Critical | 408 |
| Major | 38,069 |
| Minor | 5,364 |
| Info | 275 |

**Duplications**
**32.6%**
45,786 lines
2,870 blocks
106 files

**Unit Tests Coverage**
-

**Unit test success**
0 tests

**Complexity**
**5.1** /function
**47.6** /class
**52.7** /file
Total: 31,731

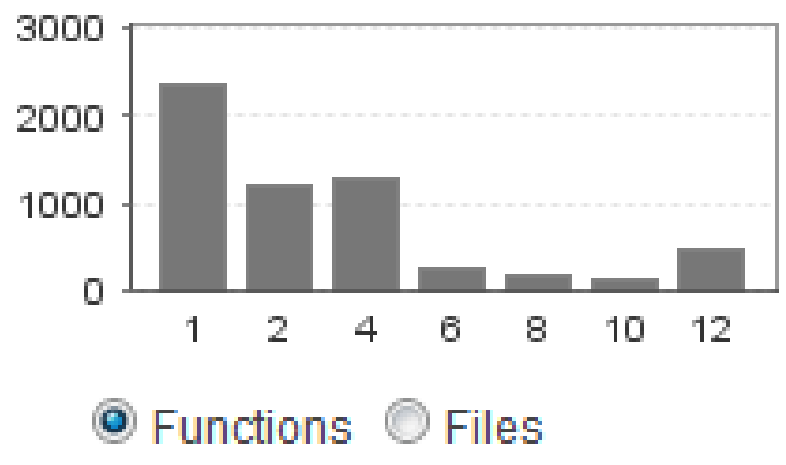

*APPENDIX 2*

The diagram in the following picture describes the architecture of MARF over GIPSY framework:

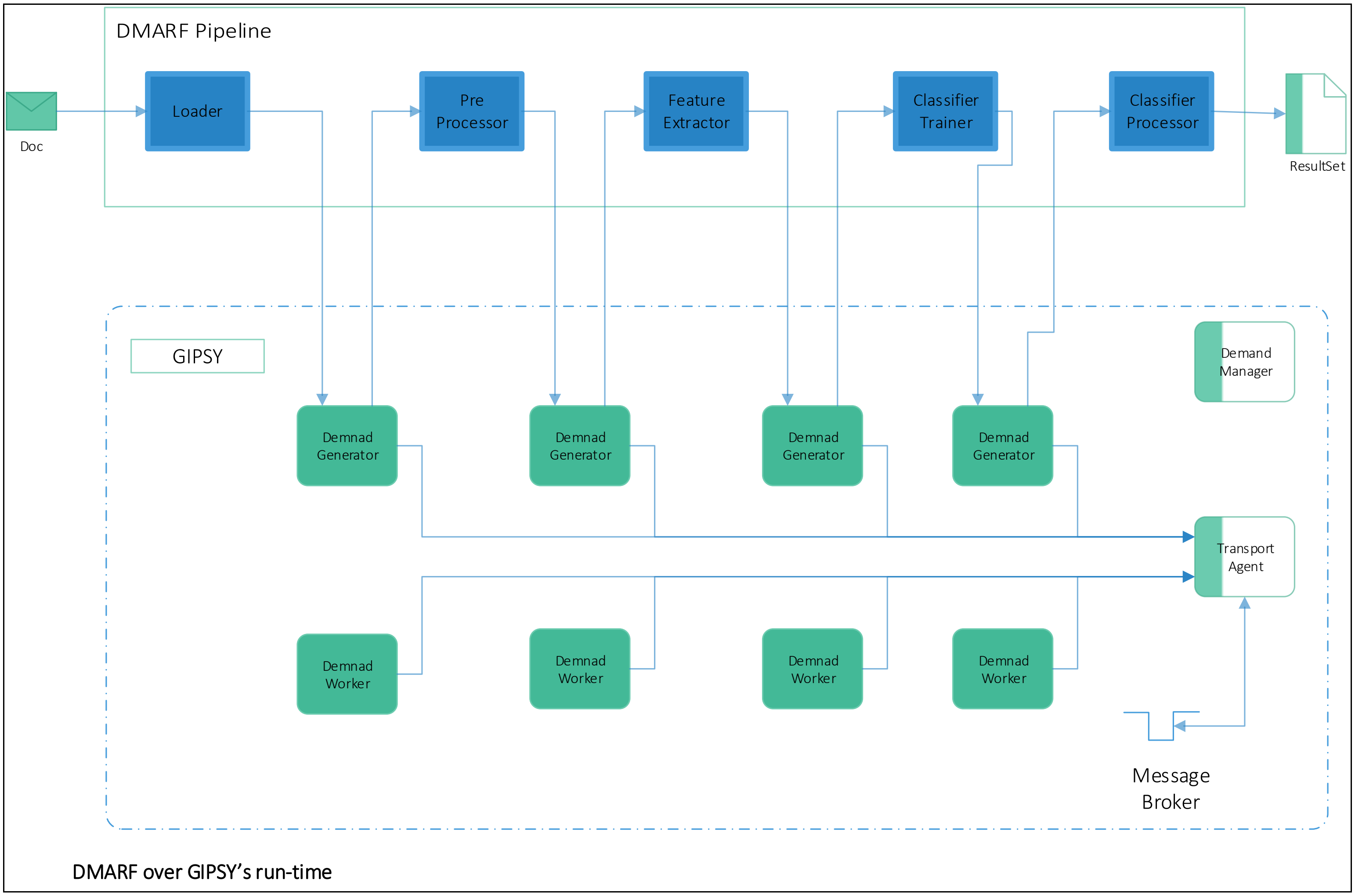

**JDeodorant** is an Eclipse plug-in that identifies design problems in software, known as bad smells and resolves them by applying appropriate refactoring. JDeodorant employs some novel methodologies in order to automatically identify bad smells. For the moment, the tool identifies two kinds of bad smells, namely "Feature Envy" and "Type Checking". "Feature Envy" problems are automatically resolved by "Move Method" and "Extract and Move Method" refactoring. "Type Checking" problems are automatically resolved by "Replace Conditional with Polymorphism" and "Replace Type code with State/Strategy" refactoring.

The tool is the outcome of the research effort in the Computational Systems and Software Engineering Lab, at the Department of Applied Informatics, University of Macedonia, Thessaloniki, Greece.

JDeodorant encompasses a number of innovative features:

- Transformation of expert knowledge to fully automated processes;
- Pre-Evaluation of the effect for each suggested solution;
- User guidance in comprehending the design problems;
- User friendliness (one-click approach in improving design quality).

JDeodorant is an Eclipse[16] plug-in used to identify the design problem in the software. These design problems are often referred as code smells (bad smells), can be resolved by applying appropriate refactoring. The tool can identify four kinds of bad smells, namely Feature Envy, Type Checking, Long Method, and God Class. Each bad smell can be resolved by using appropriate techniques as described below:

a) **Feature Envy**: Feature Envy problems occur when a method uses/invokes most of the other methods in other class. It does so to use the data in the other class, which is not present in the current class. Hence to resolve this problem, the method is moved to the other class where it mostly uses the data. Thus the method is called as Move Method refactoring. One other solution to overcome this problem is to use the Extract method.
b) **Long Method**: Long method is the method that is often difficult to change, understand and extend because of its size in terms of not only lines of code but in number of operators and operands which causes complexity and can be noted down based on the Cyclomatic complexity and Halstead complexity measures. The solution for this problem is to use Extract method refactoring. JDeodorant uses slicing technique to decide if extract method refactoring can be applied.
c) **Type Checking**: Type checking problem occurs when the conditional statements and any kind of objects or variables under late binding. This can be resolved by replacing the conditional statements using polymorphism and the code with strategic changes.
d) **God Class**: A god class is an object that controls other objects in the system and overpowered the system logic and became equivalent to the class that does everything in the system. God class is a threat to oop programming and will troubles in maintenance and evaluation phases. JDeodorant uses agglomerative clustering technique in order to find the distance between the entities of the class and then apply the clustering algorithms.

JDeodorant is a research program performed in the Department of computer science and software engineering at Concordia University, Canada and the Computational Systems and Software Engineering Lab at Department of Applied Informatics, University of Macedonia, Thessaloniki, Greece.

16 https://www.eclipse.org/ - An IDE used to develop applications in various programming languages. Mostly written in java, Eclipse can run other programming languages by using some additional plug-ins.